\documentclass[onecolumn]{emulateapj}
\usepackage{csvsimple}
\usepackage{adjustbox}
\usepackage{amsmath}
\usepackage{graphicx}
\usepackage[toc,page]{appendix}

\shorttitle{Molecular Gas Distributions in Lupus Disks}
\shortauthors{Arulanantham et al.}

\begin{document}

\title{Probing UV-Sensitive Pathways for CN \& HCN Formation in Protoplanetary Disks with the \emph{Hubble Space Telescope}}

\author{Nicole Arulanantham}
\affil{Laboratory for Atmospheric and Space Physics, University of Colorado, 392 UCB, Boulder, CO 80303, USA}
\author{K. France}
\affil{Laboratory for Atmospheric and Space Physics, University of Colorado, 392 UCB, Boulder, CO 80303, USA}
\author{P. Cazzoletti}
\affil{Max-Planck-Institute for Extraterrestrial Physics (MPE), Giessenbachstr. 1, 85748, Garching, Germany}
\author{A. Miotello}
\affil{European Southern Observatory, Karl-Schwarzschild-Str. 2, D-85748 Garching bei M\"{u}nchen, Germany}
\author{C. F. Manara}
\affil{European Southern Observatory, Karl-Schwarzschild-Str. 2, D-85748 Garching bei M\"{u}nchen, Germany}
\author{P. C. Schneider}
\affil{Hamburger Sternwarte, Gojenbergsweg 112, 21029 Hamburg, Germany}
\author{K. Hoadley}
\affil{Department of Astronomy, California Institute of Technology, 1200 East California Blvd., Pasadena, CA 91125, USA}
\author{E. F. van Dishoeck}
\affil{Leiden Observatory, Leiden University, PO Box 9513, 2300 RA Leiden, The Netherlands}
\affil{Max-Planck-Institute for Extraterrestrial Physics (MPE), Giessenbachstr. 1, 85748, Garching, Germany}
\author{H. M. G{\"u}nther}
\affil{MIT, Kavli Institute for Astrophysics and Space Research, 77 Massachusetts Ave., Cambridge, MA 02139, USA}

\begin{abstract}

The UV radiation field is a critical regulator of gas-phase chemistry in surface layers of disks around young stars. In an effort to understand the relationship between photocatalyzing UV radiation fields and gas emission observed at infrared and sub-mm wavelengths, we present an analysis of new and archival \emph{HST}, \emph{Spitzer}, ALMA, IRAM, and SMA data for five targets in the Lupus cloud complex and 14 systems in Taurus-Auriga. The \emph{HST} spectra were used to measure Ly$\alpha$ and FUV continuum fluxes reaching the disk surface, which are responsible for dissociating relevant molecular species (e.g. HCN, N$_2$). Semi-forbidden C II] $\lambda$2325 and UV-fluorescent H$_2$ emission were also measured to constrain inner disk populations of C$^+$ and vibrationally excited H$_2$. We find a significant positive correlation between 14 $\mu$m HCN emission and fluxes from the FUV continuum and C II] $\lambda$2325, consistent with model predictions requiring N$_2$ photodissociation and carbon ionization to trigger the main CN/HCN formation pathways. We also report significant negative correlations between sub-mm CN emission and both C II] and FUV continuum fluxes, implying that CN is also more readily dissociated in disks with stronger FUV irradiation. No clear relationships are detected between either CN or HCN and Ly$\alpha$ or UV-H$_2$ emission. This is attributed to the spatial stratification of the various molecular species, which span several vertical layers and radii across the inner and outer disk. We expect that future observations with \emph{JWST} will build on this work by enabling more sensitive IR surveys than were possible with \emph{Spitzer}.

\end{abstract}


\section{Introduction}

Multi-wavelength observations of gas- and dust-rich disks around young stars have allowed us to develop rudimentary maps of the composition and structure of planet-forming material. Infrared surveys with \emph{Spitzer} \citep{Oberg2008, Pontoppidan2010, Bottinelli2010, Salyk2011Spitzer, Pascucci2013} and \emph{Herschel} \citep{Dent2013} provided important constraints on warm molecular gas in surface layers of the inner disks $\left(r < 10 \, \text{au} \right)$, contributing column densities and temperatures of critical molecular gas species (e.g. H$_2$O, CO$_2$). Sub-mm observations of star-forming regions with ALMA have revealed the structure of cold gas in the outer disks with unprecedented sensitivity and angular resolution \citep{Ansdell2016, Ansdell2017, Barenfeld2016, Pascucci2016, Miotello2017, Tazzari2017, Long2018_Chamaeleon, Long2019_Taurus, vanTerwisga2019, Cazzoletti2019, Williams2019}, showing statistically significant trends in mass and radial extent as a function of cluster age and initial conditions of the parent cloud (e.g. angular momentum, temperature; \citealt{Cazzoletti2019}). Both observational and theoretical work demonstrate that the chemical evolution of molecular gas is strongly dependent on the ultraviolet radiation field reaching the surface of the disk \citep{Aikawa2002, Chapillon2012, Walsh2012, Walsh2015, Cazzoletti2018, Cleeves2018, Agundez2018, Bergner2019, Miotello2019}. However, the effects of disk geometry and varying UV flux are degenerate in physical-chemical models of the gas distributions (see e.g. \citealt{Cazzoletti2018}), making it difficult to trace the precise locations of critical species within the disk in the absence of high angular resolution datasets.

Observational constraints on the UV flux reaching the disk surface are available from \emph{HST} surveys of young stars with circumstellar disks (see e.g. \citealt{Yang2012, France2012, France2014}). The wavelength range available with the Cosmic Origins Spectrograph (\emph{HST}-COS) includes Ly$\alpha$ emission $\left( \lambda = 1215.67 \, \text{\AA} \right)$ and a portion of the FUV continuum $\left( \lambda > 1090 \, \text{\AA} \right)$, providing estimates of key photochemical ingredients in the molecular gas disk \citep{Bergin2003, Li2013}. In addition to these direct tracers of the UV radiation field, emission lines from electronic transitions of H$_2$ are also detected in \emph{HST}-COS and \emph{HST}-STIS spectra \citep{Herczeg2002, France2012}. These features originate in surface layers close to the star \citep{Hoadley2015}, providing an independent way to estimate the UV flux reaching the innermost regions of the disk \citep{Herczeg2004, Schindhelm2012}.           

In this work, we present new and archival \emph{HST}-COS and \emph{HST}-STIS observations of a sample of five young systems in the $\sim$3 Myr Lupus complex. We interpret our measurements of the UV radiation field and molecular gas features in the context of sub-mm CN observations and infrared HCN emission from \emph{Spitzer}, including disks in Taurus-Auriga for comparison. We focus on these two molecules because of the strong dependence of their physical distributions on the UV radation field \citep{Agundez2008, Agundez2018, Cazzoletti2018, Bergner2019, Greenwood2019}. Our dataset therefore allows us to observationally examine the theoretical relationships between molecular gas emission and UV radiation fields, in turn constraining physical-chemical models that map abundances of volatile elements (C, N, and H). This information may be particularly useful in disk regions where gas-phase oxygen is depleted and emission from more abundant species (e.g. CO) is fainter than expected \citep{Miotello2017, Schwarz2018}. The radial distributions of these molecules can then inform us about the composition of material available for in situ protoplanetary accretion, setting important initial conditions for atmospheric chemistry (see e.g. \citealt{Madhusudhan2011}). To this end, we discuss the relationships between spectral tracers of the UV radiation field and integrated fluxes from CN and HCN, with particular consideration given to the impact of disk geometry and optical depth of the molecular gas.   

\section{Targets \& Observations}

\subsection{A Sample of Young Disks in the Lupus Complex}

Our sample consists of five young stars with circumstellar disks in the nearby ($d \sim 160$ pc; \citealt{Gaia}) Lupus cloud complex: RY Lupi, RU Lupi, MY Lupi, Sz 68, and J1608-3070. Table \ref{stellar_props} lists the properties of each target from \citet{Alcala2017}, including stellar mass, disk inclination, and visual extinction $\left(A_V \right)$. Interstellar $A_V$ is low along the line of sight to the Lupus clouds \citep{Alcala2017}, making the region well-suited for UV observations. This group of young systems shows a broad range of outer disk morphologies in ALMA observations of their gas and dust distributions. At the time of the new \emph{HST} observations, two targets (Sz 68, RU Lupi) were categorized as full, primordial disks and three (MY Lupi, RY Lupi, J1608-3070) were identified as transition disks from sub-mm images ($r_{cav} = 25, 50, 75$ au; \citealt{vanderMarel2018}) under the traditional classification scheme (see e.g. \citealt{Strom1989, Skrutskie1990}). However, RY Lupi differs from MY Lupi and J1608-3070 in that it has strong 10 $\mu$m silicate emission \citep{KS2006} from warm grains close to the central star and undergoes periodic optical variability attributed to occultations by a warped inner disk \citep{Manset2009}. These signatures are not typical of depleted transition disks, indicating that the clearing of material seen inside 50 au is a gap, rather than a cavity \citep{Arulanantham2018, vanderMarel2018}. RU Lupi and MY Lupi were observed at high resolution ($\sim$5 au) with ALMA as part of the Disk Substructures at High Angular Resolution Project (DSHARP; \citealt{Andrews2018}), which revealed multiple rings of 1.25 mm continuum emission within each disk \citep{Huang2018_annular}. Sz 68, a triple system that was also included in DSHARP \citep{Andrews2018}, consists of a close binary and a distant third companion. The \emph{HST} and ALMA observations presented here include emission from both binary components. However, the secondary star (component B) and its disk are much smaller and fainter than the circumprimary disk ($F_{star \, B} / F_{star \, A} = 0.17$, \citealt{Ghez1997}; $I_{peak, disk B} / I_{peak, disk A} = 0.23$, \citealt{Kurtovic2018}), indicating that the bulk of the UV emission comes from the primary component.

\begin{deluxetable}{cccccccc}[b!]
\tablewidth{0.75\linewidth}
\tabletypesize{\scriptsize}
\tablecaption{Stellar \& Disk Properties \label{stellar_props}}
\tablehead{
\colhead{Target} & \colhead{Distance} & \colhead{$M_{\ast}$} & \colhead{$A_V$} & \colhead{$i$} & \colhead{$r_{cav,dust}$} & \colhead{$r_{cav,gas}$} & \colhead{References} \\ 
 & \colhead{[pc]} & \colhead{$\left[ M_{\odot} \right]$} & \colhead{[mag]} & \colhead{$^{\circ}$} & \colhead{[au]} & \colhead{[au]} & \\ 
}
\startdata
RU Lupi\tablenotemark{*} & 159 & 0.8 & 0.07 & $\sim$18.5$^{\circ}$ & 14, 17, 21, 24, 29.1, 34, 42, 50 & \nodata & \tablenotemark{a} \tablenotemark{b} \tablenotemark{c} \\
RY Lupi & 158 & 1.47 & 0.1 & 68 & 50 & 50 & \tablenotemark{a} \tablenotemark{d} \\
MY Lupi \tablenotemark{$\ast$} & 156 & 1.02 & 0.04 & 73 & 8, 20, 30, 40 & 25 & \tablenotemark{a} \tablenotemark{b} \tablenotemark{d} \\
Sz 68 & 154 & 2.13 & 0.15 & 34 & \nodata & \nodata & \tablenotemark{a} \tablenotemark{e} \\
J1608-3070 & 155 & 1.81 & 0.055 & 74 & 75 & 60 & \tablenotemark{a} \tablenotemark{d} \\ 
\enddata
\tablenotetext{*}{High-resolution ALMA images of RU Lupi have revealed a series of rings inside $\sim$50 au \citep{Huang2018_annular}. The two rings with constrained values have inclinations of 20$^{\circ}$ and 17$^{\circ}$, so we use an average of the two. For both RU Lupi and MY Lupi, the $r_{cav,dust}$ radii are locations of the dust rings resolved by \citep{Huang2018_annular}.}
\tablenotetext{a}{\citealt{Gaia}}
\tablenotetext{b}{\citealt{Huang2018_annular}}
\tablenotetext{c}{\citealt{vanderMarel2018}}
\tablenotetext{d}{\citealt{Ansdell2016}}
\end{deluxetable} 

These systems were selected for follow-up with \emph{HST} after a large ALMA survey of the Lupus clouds identified them as hosts to some of the most massive dust disks in the region \citep{Ansdell2016, Ansdell2018}. However, physical-chemical models of the $^{13}$CO and C$^{18}$O emission demonstrate that the total gas masses are unexpectedly low \citep{Miotello2016, Miotello2017}, which can be attributed to either shorter timescales than predicted for removing gas from the disk (e.g. via external photoevaporation) or chemical pathways that trap carbon in larger molecules with higher freeze-out temperatures. Since UV photons are critical regulators of chemical processes in disk environments (see e.g. \citealt{Aikawa2002, Bergin2003, Bethell2011, Walsh2012, Walsh2015, Cazzoletti2018, Visser2018}), the \emph{HST} data we present here provide currently missing observational constraints on the levels of irradiation at the surface of the gas disk. 

\subsection{Observations}
All five systems were observed with the \emph{Hubble Space Telescope (HST)}, using both the Cosmic Origins Spectrograph (COS; \citealt{Green2012}) and the Space Telescope Imaging Spectrograph (STIS; \citealt{Woodgate1997_overview, Woodgate1997_firstresults}). Table \ref{observing_props} lists exposure times and program IDs for the observations. The FUV spectra for three of our targets were previously presented in \citealt{France2012} (RU Lupi), \citealt{Arulanantham2018} (RY Lupi), and \citealt{Alcala2019} (MY Lupi). 

\begin{deluxetable}{ccccccc}
\tablewidth{0.75 \linewidth}
\tabletypesize{\scriptsize}
\tablecaption{Observations of Young Systems in Lupus \label{observing_props}}
\tablehead{
 & \colhead{RU Lupi\tablenotemark{*}} & \colhead{RY Lupi} & \colhead{MY Lupi} & \colhead{Sz 68} & \colhead{J1608-3070} \\ 
}
\startdata
\textbf{Program ID} & 12036, 8157 & 14469 & 14604 & 14604 & 14604 \rule{0pt}{1.2\normalbaselineskip}\\
\hline
\rule{0pt}{1.2\normalbaselineskip}
 & & & \textbf{Exposure Time [s]} &  & \rule{0pt}{1.2\normalbaselineskip}\\
 & & & \textbf{Observation Date} &  & \rule{0pt}{1.2\normalbaselineskip}\\
\hline
\rule{0pt}{1.2\normalbaselineskip}
\textbf{\emph{HST}-COS G140L} & \nodata & 2448 & 5658 & 5538 & 5574 \\
($\lambda$1280; $R \sim 1500$) & \nodata & 2016 Jun 16 & 2018 Sep 8 & 2018 Jul 27 & 2018 Jul 30 \\
\hline
\rule{0pt}{1.2\normalbaselineskip}
\textbf{\emph{HST}-COS G130M} & 1686 & 654 & 1242 & 1212 & 1218 \\
($\lambda$1291; $R \sim 16000$) & 2012 Jul 20 & 2016 Jun 16 & 2018 Sep 8 & 2018 Jul 27 & 2018 Jul 30 \\
\hline
\rule{0pt}{1.2\normalbaselineskip}
\textbf{\emph{HST}-COS G160M} & 1938 & 648 & 1296 & 1266 & 1278 \\
($\lambda$1577; $R \sim 16000$) & 2012 Jul 21 & 2016 Jun 16 & 2018 Sep 8 & 2018 Jul 27 & 2018 Jul 30 \\ 
\hline
\rule{0pt}{1.2\normalbaselineskip}
\textbf{\emph{HST}-STIS G230L} & \nodata & 2400 & 2028 & 1968 & 1986 \\
($\lambda$2375; $R \sim 1000$) & \nodata & 2016 Jun 16 & 2018 Sep 8 & 2018 Jul 26 & 2018 Jul 30 \\ 
\hline
\rule{0pt}{1.2\normalbaselineskip}
\textbf{\emph{HST}-STIS G430L} & 120 & 60 & 60 & 60 & 60 \\
($\lambda$4300; $R \sim 1000$) & 2001 Jul 12 & 2016 Jun 16 & 2018 Sep 8 & 2018 Jul 26 & 2018 Jul 30 \\ 
\enddata
\tablenotetext{*}{The G160M $\lambda$1589 setting was used for RU Lupi instead of $\lambda$1577.}
\end{deluxetable}
\medskip

\begin{figure}
%Figure made with Full_Lupus/Scripts/panchromatic_spectra_plots.py
\centering
\includegraphics[width=1.0\linewidth]
{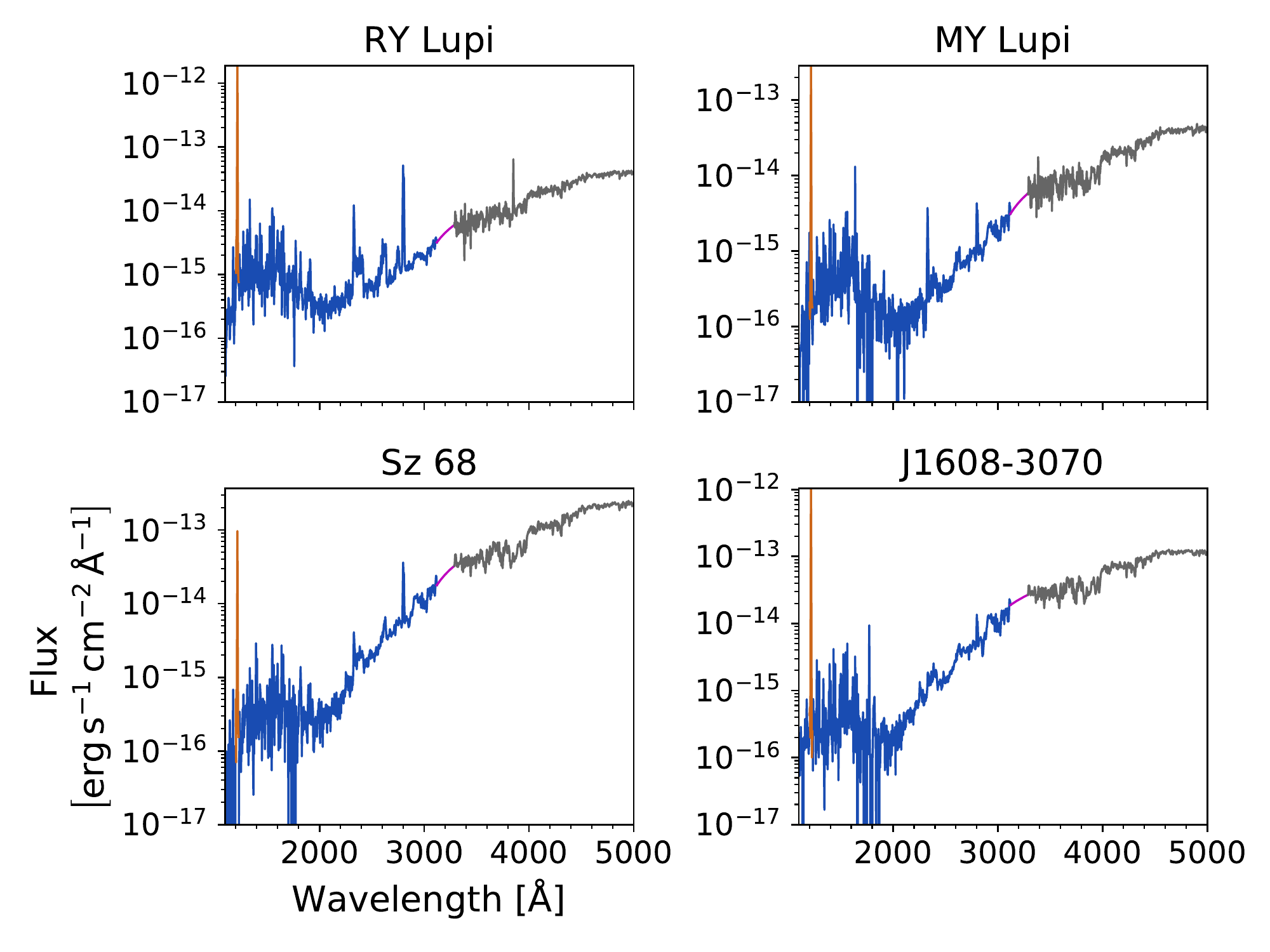}
\caption{A panchromatic spectrum was produced for four disks in Lupus by stitching together new data from five different observing modes of \emph{HST}-COS (blue) and \emph{HST}-STIS (gray). We include reconstructed model Ly$\alpha$ profiles (orange; see e.g. \citealt{Schindhelm2012}) in place of the observed features, which are contaminated by interstellar absorption and geocoronal emission. Interpolated fluxes spanning the overlap between gratings are shown in magenta. The spectra from RU Lupi and the disks in Taurus-Auriga were previously presented in \citet{Herczeg2005} and \citet{France2014}.}
\label{panchromatic_spectra}
\end{figure}

Five different modes of \emph{HST}-COS and \emph{HST}-STIS were used to observe four of our systems, to cover wavelengths from 1100-5000 \AA. NUV coverage is not included for the fifth star, RU Lupi, which was observed as part of a different program. These data were used to extrapolate the FUV continuum down to the hydrogen ionization edge at 912 \AA, which is a critical region for photodissociation of abundant molecular species (e.g. H$_2$ and CO) and gas-phase chemistry but not readily accessible with available UV facilities \citep{France2014}. All five spectral modes were then stitched together to produce a SED for each system using the methods outlined in \citet{France2014} and \citet{Arulanantham2018}. Figure \ref{panchromatic_spectra} shows an overview of the SEDs, highlighting the strong contribution from the accretion-dominated NUV continuum $\left( \lambda \sim 3000-4000 \text{ \AA} \right)$. Accretion processes enhance the FUV continuum as well, but the total flux at $\lambda < 2000$ \AA \, is dominated by line emission from hot, atomic gas (e.g. Ly$\alpha$, C IV, C II). Our resulting library of radiation fields, which encompasses far-ultraviolet to optical wavelengths, is available to the community\footnote[1]{http://cos.colorado.edu/$\sim$kevinf/ctts\_fuvfield.html}. We anticipate that the data will be used in gas-phase chemical modeling efforts that require an understanding of stellar irradiation at the disk surface.

Sub-mm CN fluxes for this work were taken from the literature, providing measurements of gas in the cold, outer regions of the disk. The data were acquired with the SMA \citep{Oberg2010, Oberg2011}, the IRAM 30-m telescope \citep{Guilloteau2013}, and ALMA \citep{vanTerwisga2019}, causing the full sample to span several different observing configurations and spectral features. While the Lupus ALMA survey measured fluxes from the strongest $N = 3-2, J = 7/2-5/2$ hyperfine CN transitions \citep{vanTerwisga2019}, $N=2-1, J=5/2-3/2$ features are reported from targets in Taurus-Auriga \citep{Guilloteau2013, Oberg2011}. To convert the fluxes from different datasets to comparable quantities, we adopted the methodology of \citet{vanTerwisga2019} and used the ratio
\begin{equation}
\frac{N = 2-1, J = 5/2-3/2}{N = 3-2, J = 7/2-5/2} = 1.6
\end{equation}
to obtain CN $N=2-1, J=5/2-3/2$ flux estimates for the Lupus disks. This scaling accounts for the gas temperature distribution within the disk, under the assumption that the CN emission is optically thin. The value chosen by \citet{vanTerwisga2019}, which we use here, is the median from the grid of models studied by \citet{Cazzoletti2018}, which spanned a range from 1.1-2.1. 

The total $N=2-1, J=5/2-3/2$ flux in all 19 hyperfine transitions listed by \citet{Guilloteau2013} was estimated by scaling the flux in the three strongest features (226.874, 226.887, and 226.892 GHz) by a factor of 1.67, as calculated by those authors from the measured flux ratios. We applied the same scaling to the 226.874 GHz fluxes from \citet{Oberg2010, Oberg2011}; however, these measurements are likely slightly underestimated, since they do not include the weaker emission lines at 226.887 and 226.892 GHz, and we depict them as lower limits in the figures presented here. We also note that different methods were used to derive the CN fluxes, with \citet{vanTerwisga2019} using aperture photometry and \citet{Guilloteau2013} and \citet{Oberg2010, Oberg2011} using the integrated spectral lines. \citet{Guilloteau2013} account for beam dilution in their flux measurements by incorporating a ``beam filling factor" in their CN line fitting procedure, while \citet{Oberg2010, Oberg2011} estimate synthesized beam sizes. We emphasize that the differences in observing methodologies outlined here introduce systematic uncertainties into our analysis that may contribute $\sim$30\% to the Lupus CN fluxes and $\sim$50\% to the \citet{Oberg2010, Oberg2011} measurements that are scaled and presented here. 

Infrared HCN features, originating in the warmer inner disk, were measured directly from observations with the InfraRed Spectrograph (IRS) onboard the \emph{Spitzer Space Telescope} \citep{Houck2004}. All targets except J1608-3070 were observed in the high-resolution mode $\left(R \sim 600 \right)$ over the course of several different observing programs. The data were retrieved from the Combined Atlas of Sources with Spitzer IRS Spectra (CASSIS; \citealt{Lebouteiller2011, Lebouteiller2015}), which provides a complete catalog of \emph{Spitzer}/IRS observations. Fluxes from the 14 $\mu$m HCN $v_{2}$ band were measured over the wavelength range defined for the feature in \citet{Najita2013}. That work used slab models of the molecular gas disk to identify line-free spectral windows for each target, which we used as a reference for continuum subtraction. Although the measurement errors from the \emph{Spitzer} spectra are $\sim5$\%, the continuum subtraction procedure is likely the largest source of uncertainty in the final HCN fluxes. We quantify this by measuring the fluxes using three different sets of continuum regions defined by \citet{Pascucci2009, Teske2011, Najita2013}, finding that the resulting fluxes are consistent to within $\sim$20\%. The average and standard deviation of the three measurements for each target are reported in Table \ref{CN_HCN_props}, along with observing program IDs and PIs.   

\begin{deluxetable}{cccccc}
\tablecaption{\emph{Spitzer}/IRS and sub-mm CN PIDs and Fluxes \label{CN_HCN_props}
}
\tablewidth{0 pt}
\tabletypesize{\scriptsize}
\tablehead{
\colhead{Target} & \colhead{Program ID} & \colhead{PI} & \colhead{Observation Date} & \colhead{HCN Fluxes} & \colhead{CN Reference} \\
 & & & \colhead{[mm/dd/yyyy]} & \colhead{[$10^{-14}$ erg s$^{-1}$ cm$^{-2}$]} & \\
}
\startdata
AA Tau & 20363 & J. Carr & 2005 Oct 15 & $4.9 \pm 0.7$ & \citealt{Guilloteau2013} \\
BP Tau & 20363 & J. Carr & 2006 Mar 19 & $6 \pm 1$ & \citealt{Guilloteau2013} \\
DE Tau & 50641 & J. Carr & 2008 Oct 8 & $3 \pm 1$ & \citealt{Guilloteau2013} \\
DM Tau & 30300 & J. Najita & 2007 Mar 24 & $2 \pm 1$ & \citealt{Guilloteau2013} \\
DR Tau & 50641 & J. Carr & 2008 Oct 8 & $18 \pm 3$ & \citealt{Guilloteau2013} \\
DS Tau & 50498 & J. Houck & 2008 Nov 9 & $9 \pm 2$ & \citealt{Guilloteau2013} \\
GM Aur & 30300 & J. Najita & 2007 Mar 14 & $2 \pm 1.5$ & \citealt{Oberg2010} \\
HN Tau A & 50641 & J. Carr & 2008 Oct 1 & $2.4 \pm 0.4$ & \nodata \\
LkCa 15 & 40338 & J. Najita & 2008 Nov 5 & $2.8 \pm 0.6$ & \citealt{Oberg2010} \\
MY Lupi & 20611 & C. Wright & 2005 Aug 8 & $1.4 \pm 0.2$ & \citealt{vanTerwisga2019} \\
RU Lupi & 172 & N. Evans & 2004 Aug 30 & $12 \pm 3$ & \citealt{vanTerwisga2019} \\
RY Lupi & 172 & N. Evans & 2004 Aug 30 & $3 \pm 2$ & \citealt{vanTerwisga2019} \\
SU Aur & 50641 & J. Carr & 2008 Nov 5 & $8 \pm 3$ & \citealt{Guilloteau2013} \\
Sz 68 & 172 & N. Evans & 2004 Aug 30 & $8 \pm 4$ & \citealt{vanTerwisga2019} \\
T Tau & 40113 & F. Lahuis & 2008 Oct 1 & $130 \pm 80$ & \citealt{Guilloteau2013} \\
J1608-3070 & \nodata & \nodata & \nodata & \nodata & \citealt{vanTerwisga2019} \\
V4046 Sgr & 3580 & M. Honda & 2005 Apr 19 & $4.2 \pm 0.7$ & \citealt{Oberg2011} \\
\enddata
\end{deluxetable}

\subsection{Archival \emph{HST} Data from YSOs in Taurus-Auriga}

To increase the sample size of this study and compare properties between different star-forming regions, we include all Taurus-Auriga sources from the literature with both sub-mm CN fluxes and UV spectra from \emph{HST}-STIS and/or \emph{HST}-COS. Disks in Taurus are roughly equivalent in age to the Lupus systems ($\sim$1-3 Myr; \citealt{Andrews2013, Alcala2014}) and have similar dust mass distributions \citep{Ansdell2016, Pascucci2016}, making it easier to isolate the impact of the UV radiation field on the molecular gas distributions. The UV observations were carried out as part of the COS Guaranteed Time program (PI: J. Green; PIDs: 11533, 12036), The Disks, Accretion, and Outflows (DAO) of T Tau stars program (PI: G. Herczeg; PID: 11616), and Project WHIPS (Warm H$_2$ In Protoplanetary Systems; PI: K. France; PID: 12876). \emph{HST}-COS observations were acquired with both the G130M and G160M gratings for all systems, providing wavelength coverage over the same range of FUV wavelengths ($\sim$1100-1700 \AA) as the spectra obtained for the five Lupus disks. Spectral features from these data (e.g. emission line strengths, accretion rates) have previously been analyzed in a number of papers, including \citet{Yang2012}, \citet{Ardila2013}, \citet{Ingleby2013}, \citet{France2011_FUVII, France2012, France2014, France2017}, and references therein. We mirror the techniques described in those works to identify various properties in the new Lupus spectra, reporting uniform measurements over the entire Lupus/Taurus-Auriga disk sample.  

\subsection{Uncertainty in Literature Measurements of $A_V$}

Observations at FUV wavelengths are highly affected by the amount of dust and gas along the line of sight to a disk $\left(A_V \right)$, making accurate reddening corrections critical in interpreting the spectral features. \citet{France2017} report a statistically significant positive correlation between $A_V$ (derived from e.g. broadband color excesses, deviations from stellar photospheric templates; \citealt{Kenyon1995, Hartigan2003}) and the total luminosity from UV-fluorescent H$_2$. The trend implies that more molecular gas emission is seen from disks with more intervening material; however, there is no physical or chemical process that would produce such a relationship between circumstellar and interstellar material, making systematic overestimates in the $A_V$ measurements a more likely driver. This effect must be removed in order to accurately assess relationships between the UV spectral features. 

To provide an estimate of the line-of-sight interstellar extinction that is less sensitive to the circumstellar dust properties, \citet{McJunkin2014} used observed Ly$\alpha$ profiles to directly measure the amount of neutral hydrogen (H I) along the line of sight. $A_V$ values were then calculated as $A_V / R_V = N(\mathrm{H \, I}) / \left(4.8 \times 10^{21} \text{ atoms cm}^{-2} \text{ mag}^{-1} \right)$ \citep{Bohlin1978}. \citet{France2017} find that the correlation between $A_V$ and $L \left(\mathrm{H_2} \right)$ becomes statistically insignificant when the \citet{McJunkin2014} reddening values are adopted, so we adopt the H I-derived $A_V$ values for the analysis presented here.  

\begin{figure}[t!]
	\begin{minipage}{0.5\textwidth}
	\centering
	\includegraphics[width=\linewidth]{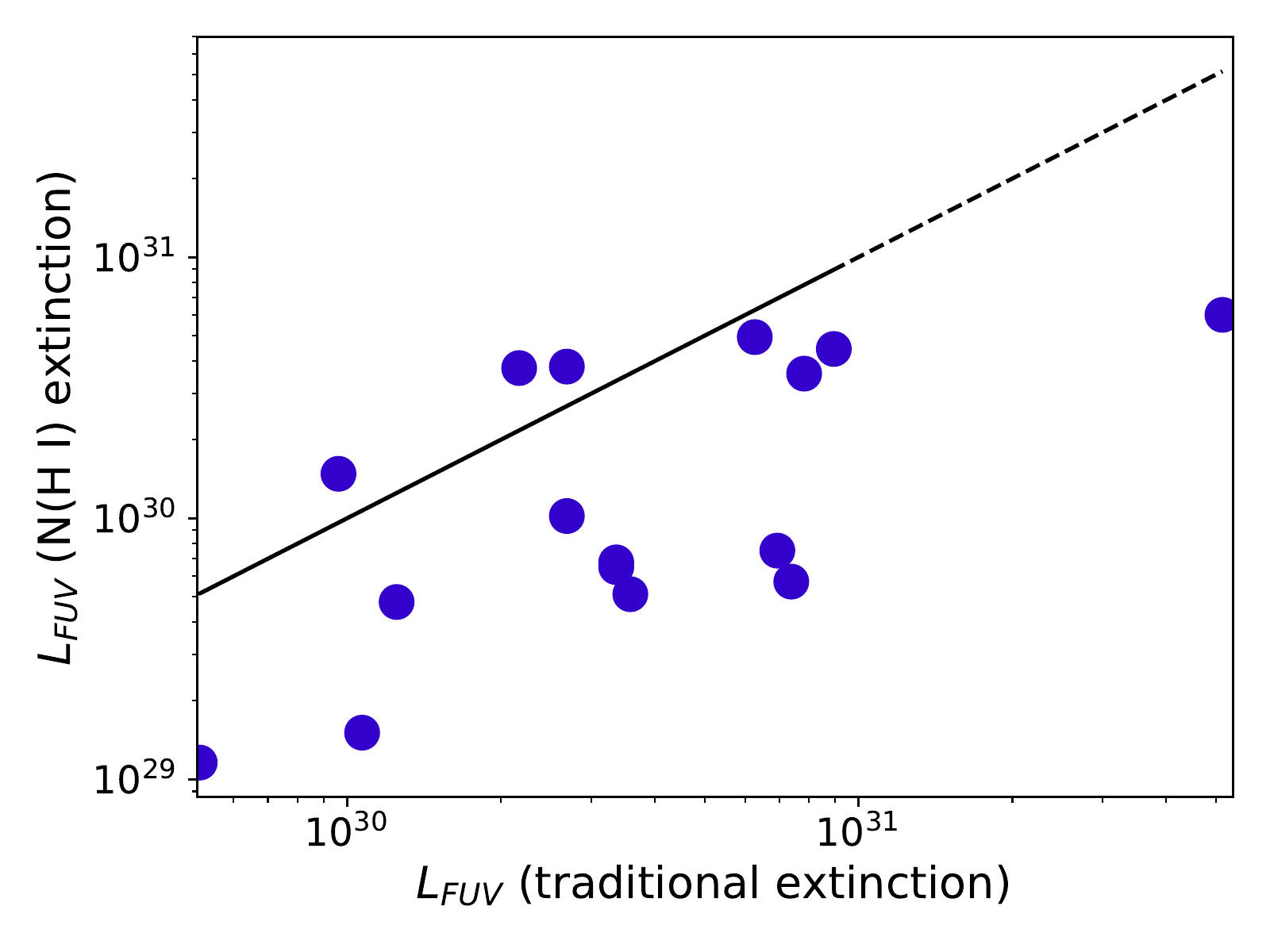}
	\end{minipage}
	\begin{minipage}{0.5\textwidth}
	\centering
	\includegraphics[width=\linewidth]{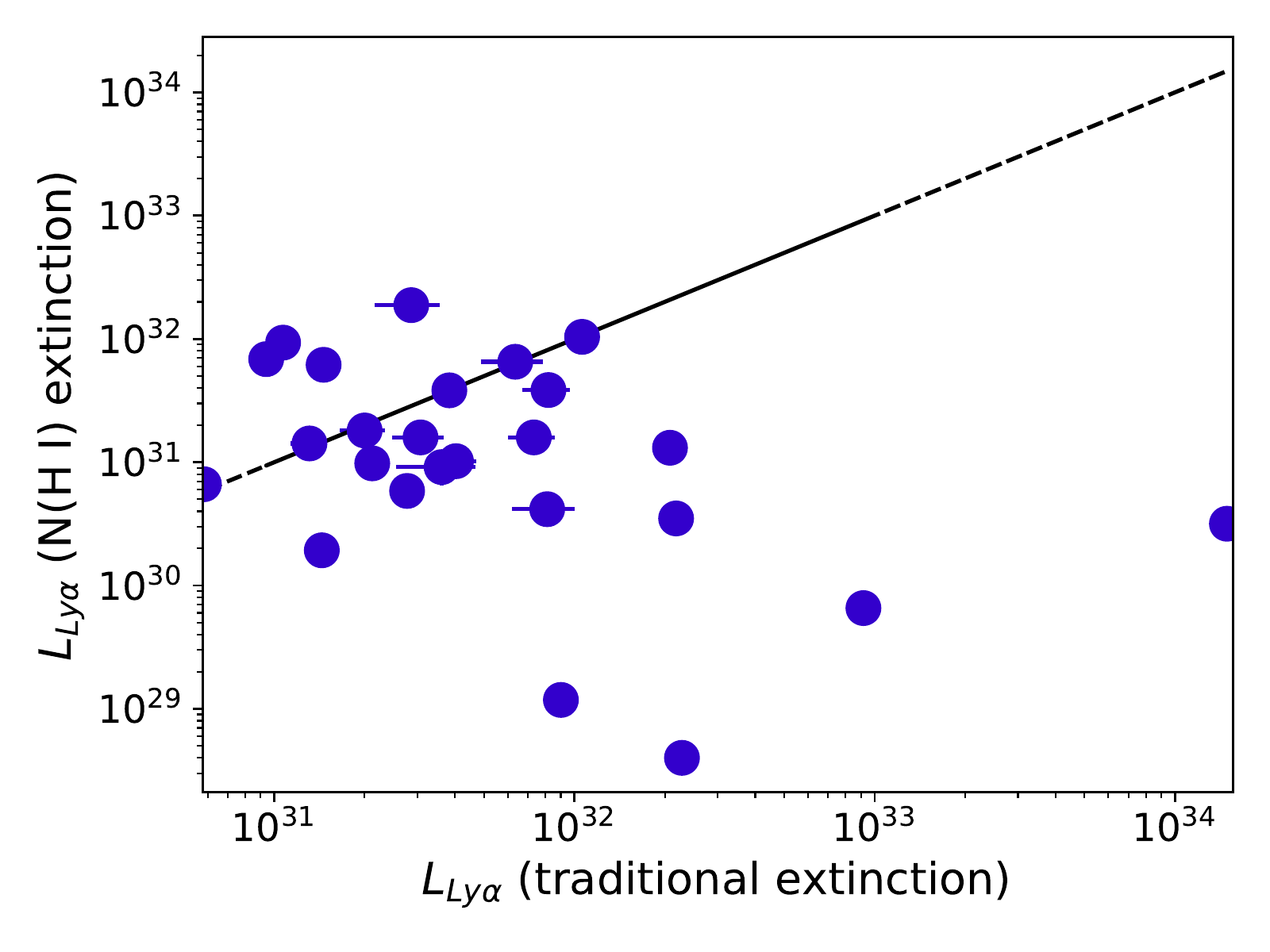}
	\end{minipage}
\caption{Comparison of luminosities from the FUV continuum (left) and Ly$\alpha$ (right), calculated using different values of $A_V$ to deredden the spectrum. The method described in \citet{McJunkin2014}, which uses the Ly$\alpha$ wings to estimate $N \left(H I \right)$ along the line of sight, typically yields smaller $A_V$ and lower luminosities than A$_V$ derived from broadband color excesses or comparisons to stellar photospheric templates (1-to-1 relationship traced by black, dashed lines). The major outliers in the right panel are disks with $A_V$ values that changed by $\Delta A_V > 0.8$. Since the $N \left(H I \right)$-based measurements alleviate the correlation between $A_V$ and total flux from UV-H$_2$ \citep{France2017}, we adopt those extinctions for the analysis presented here. } 
\label{ext_comp}
\end{figure}

Extinctions derived using the \citet{McJunkin2014} method are typically significantly lower than reported in the literature, since the measurements isolate interstellar $\mathrm{N (H \, I)}$ from circumstellar material that may significantly increase $A_V$ estimates (e.g. DR Tau; $\Delta A_V \sim 2.7$). Figure \ref{ext_comp} compares the FUV continuum luminosities calculated with $N \left( \mathrm{H \, I} \right)$-based $A_V$ values to the measurements using more traditional methods to estimate $A_V$, using targets from \citet{McJunkin2014} and \citet{France2017}. We find that the two luminosities are roughly consistent for most disks, demonstrating that the trends presented later in this work are independent of the choice of $A_V$. Figure \ref{ext_comp} also presents the same comparison for Ly$\alpha$ luminosities measured using the two different $A_V$ values, showing just five major outliers that have the largest changes in $A_V$ as well $\left(\Delta A_V > 0.8 \right)$. Of these outliers, only DR Tau is included in the sample we analyze in this work, and all other targets have $\Delta A_V < 0.8$. We therefore conclude that the trends in Ly$\alpha$ emission presented here are also not driven by the choice of $A_V$. 

\subsection{Normalized UV Luminosities}

Models of UV-sensitive molecular gas distributions have included a broad range of stellar blackbody temperatures, UV excesses due to accretion, and total UV luminosities, allowing the authors to isolate the impact of UV irradiation from disk geometric properties such as disk mass, flaring angle, and pressure scale height (see e.g. \citealt{vanZadelhoff2003, Walsh2015, Cazzoletti2018, Bergner2019}). Other observational work has split the targets by spectral type, allowing the authors to roughly correct for stellar mass and temperature \citep{Pascucci2009}. However, the sample we present here is too small to meaningfully bin the data by both disk and stellar properties. Instead, we divide the individual UV components by the total UV luminosity $\left(L_{UV, total} \right)$, defined as the sum of fluxes from the FUV continuum, Ly$\alpha$, UV-H$_2$, C IV, and C II].   

Since both $L_{UV, total}$ and fluxes from the individual UV components are impacted by target properties in the same way (e.g. $\mathbf{i_d}$, $\mathbf{M_{\ast}}$), this correction allows us to evaluate the entire sample without binning the data based on disk or stellar parameters. The normalization also reduces the impact of systematic uncertainties in measuring $A_V$, as described in Section 2.4, although we note that resonant scattering effects may preferentially enhance Ly$\alpha$ fluxes in deeper layers of the disk (see e.g. \citealt{Bethell2011}), relative to emission at other FUV and NUV wavelengths. When the data were analyzed before accounting for the diversity of disk and stellar parameters, no correlations were observed between any of the UV tracers and the 14 $\mu$m HCN or sub-mm CN fluxes. We therefore conclude that normalizing the UV luminosities allows us to provide more physically meaningful information about UV-dependent gas-phase chemistry than using the raw fluxes alone. All plots also differentiate between targets with resolved dust sub-structure (open markers) and full, primordial disks (filled markers), demonstrating that the results are roughly independent of evolutionary phase.

\section{Results}

The \emph{HST}-COS and \emph{HST}-STIS spectra described above provide direct measurements of the UV radiation field, which we use to estimate the UV flux reaching the surface of the gas disk. Here we focus on the role of UV photons in producing CN, which has been detected in ALMA Band 7 observations of a large sample of disks in the Lupus clouds \citep{vanTerwisga2019}, including the five studied here. We present the following results under the assumption that the dominant reaction pathway for CN production in disks is
\begin{equation}
\begin{aligned}
\mathrm{N + H_2^{\ast}} &\longrightarrow \mathrm{NH + H} \\
\mathrm{C^+ + NH} &\longrightarrow \mathrm{CN^+ + H} \\
\mathrm{CN^+ + H} &\longrightarrow \mathrm{CN + H^+}, 
\end{aligned}
\end{equation}
where H$_2^{\ast}$ is gas that has been pumped into excited vibrational states by FUV photons \citep{Walsh2015, Heays2017, Cazzoletti2018}. HCN can then be formed via reactions with H$_2$ and CH$_4$ 
\begin{equation}
\begin{aligned}
\mathrm{CN + H_2} &\longrightarrow \mathrm{HCN + H} \\
\mathrm{CN + CH_4} &\longrightarrow \mathrm{HCN + CH_3}
\end{aligned}
\end{equation}
\citep{Baulch1994, Walsh2015, Visser2018}. Destruction of HCN by UV photons also significantly influences the total abundance of CN in the disk \citep{Walsh2015, Cazzoletti2018, Pontoppidan2019}, with photodissociation occurring at a rate of $1.6 \times 10^{-9}$ s$^{-1}$ under a typical interstellar radiation field and a rate of $5.3 \times 10^{-9}$ s$^{-1}$ under a standard T Tauri Ly$\alpha$ profile \citep{vanDishoeck2006, Heays2017}. Although The UMIST Database for Astrochemistry \citep{UMIST2013} lists many other reactions for CN formation, physical-chemical models demonstrate that the H$_2^{\ast}$ pathway is the most important route (see e.g. \citealt{Visser2018}). This may be attributed to the high abundance of H$_2$ in disks, relative to molecular species like C$_2$H and OH that are required for alternate pathways \citep{Visser2018}.  

\subsection{Photodissociation of N$_2$ by the FUV Continuum}

Gas-phase chemical models of protoplanetary disks find that column densities of nitrogen-dependent molecular species (e.g. CN, HCN) vary with the total FUV flux at the disk surface \citep{Pascucci2009, Walsh2012, Walsh2015}. The trend is attributed to photoabsorption, since five excited electronic transitions of N$_2$ fall between 912-1100 \AA. Absorption at these wavelengths produces pre-dissociated N$_2$ via coupling to the continuum \citep{Li2013, Visser2018}. As shown in Eq. 1, the atomic nitrogen products react with H$_2^{\ast}$ to produce NH, therefore catalyzing formation of molecules like CN and HCN \citep{Walsh2015, Cazzoletti2018}. 

The predicted relationship between molecular abundances and UV irradiation is corroborated by observational studies at IR wavelengths, which find that disks around cool M stars show less emission from nitrogen bearing molecules than disks around hotter solar-type stars \citep{Pascucci2009, Pascucci2013, Najita2013}. The difference is attributed to varying UV photon production rates, which are expected to be lower in cooler stars \citep{vanZadelhoff2003, Walsh2012, Walsh2015}. To further investigate this relationship from an observational perspective, we compare CN fluxes measured from ALMA observations \citep{vanTerwisga2019} to FUV fluxes from the \emph{HST}-COS and \emph{HST}-STIS spectra presented here. 

\begin{figure}[t!]
	\begin{minipage}{0.5\textwidth}
	\centering
	\includegraphics[width=\linewidth]{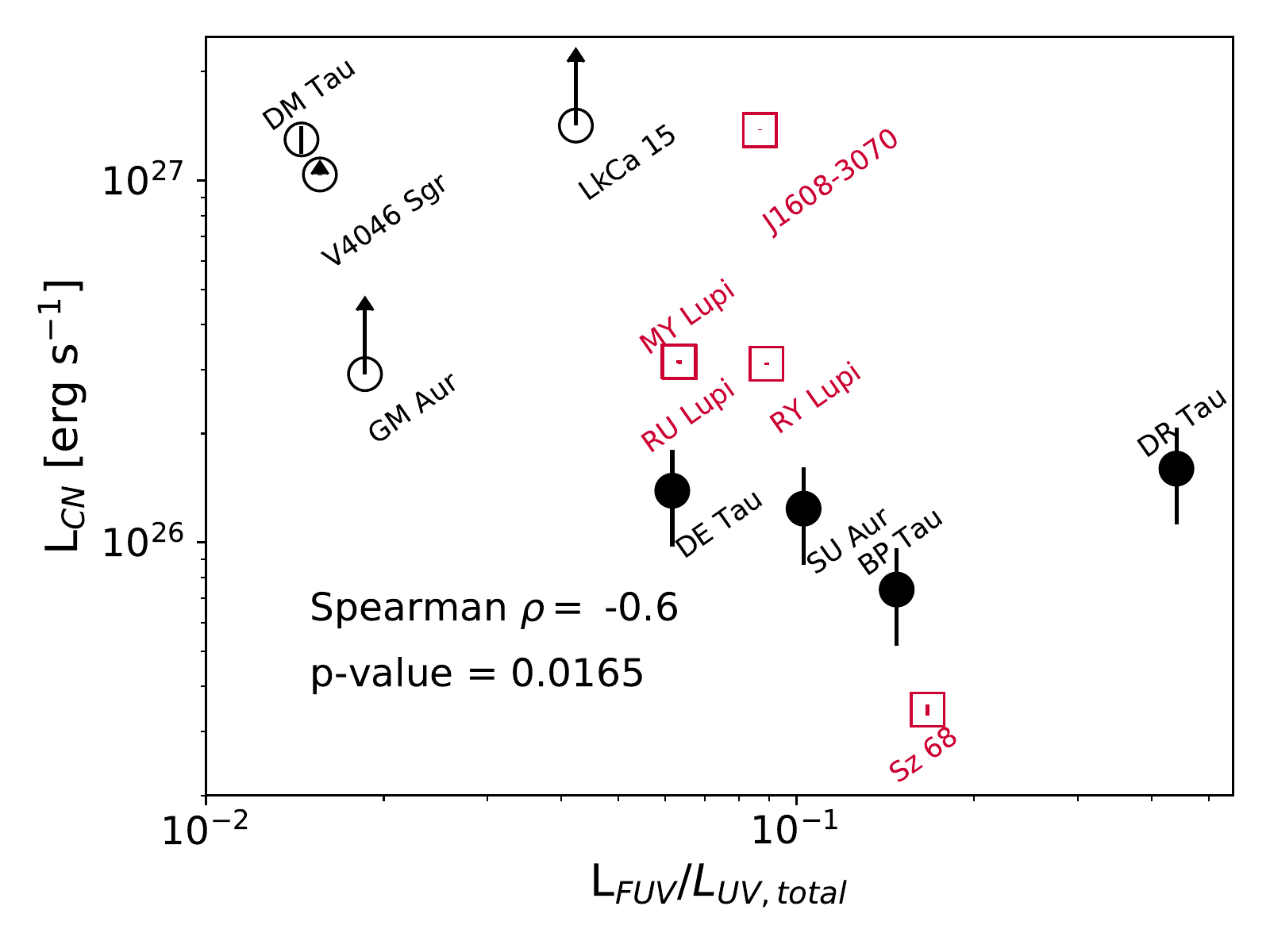}
	\end{minipage}
	\begin{minipage}{0.5\textwidth}
	\centering
	\includegraphics[width=\linewidth]{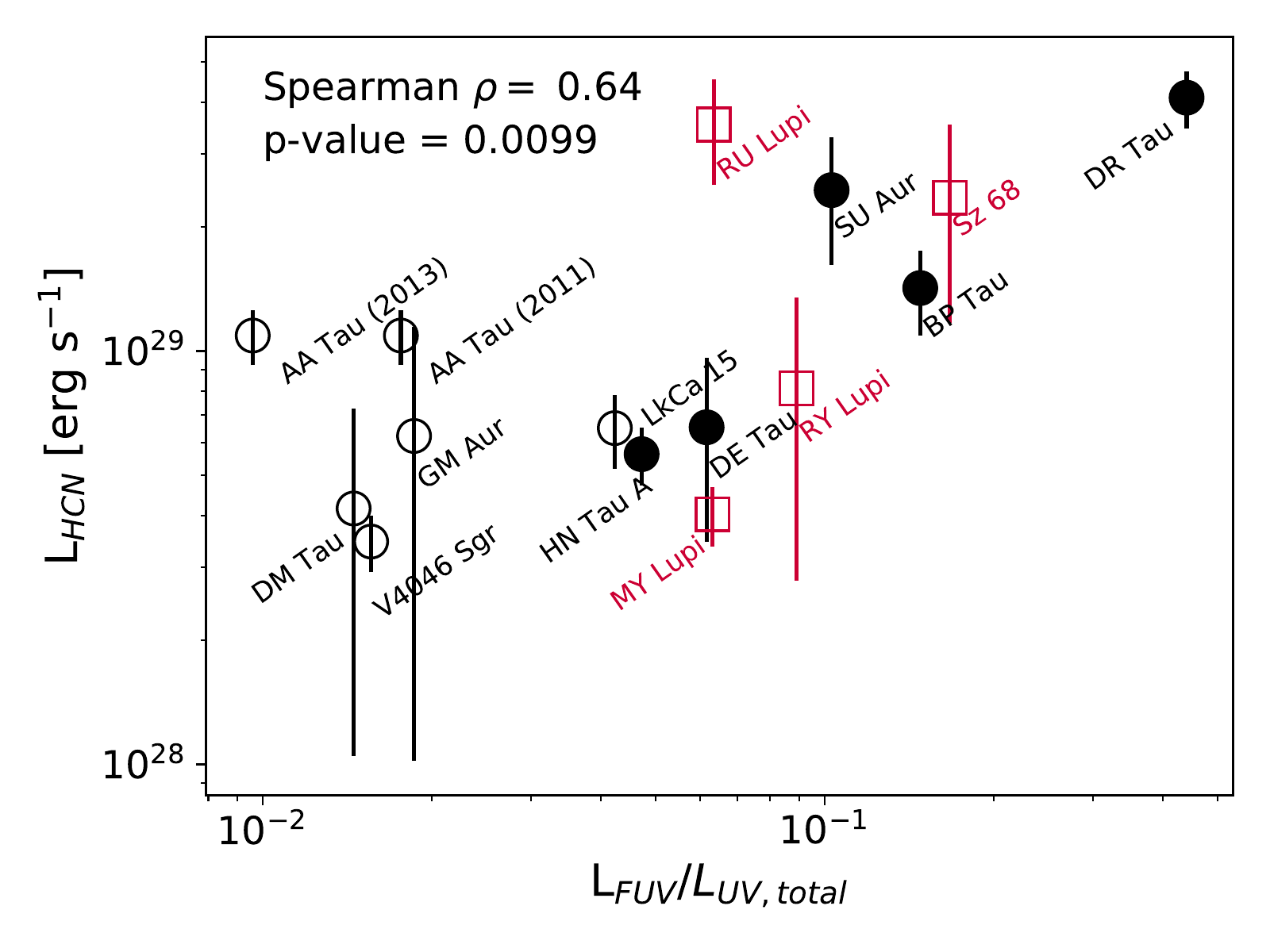}	
	\end{minipage}
\caption{Sub-mm CN (left) and 14 $\mu$m HCN (right) luminosities versus fractional flux from the FUV continuum (integrated from 912-1650 \AA). The five Lupus systems are shown as red squares and the subset of disks from \citet{France2017} as black circles, with open markers representing disks with resolved dust substructure. N$_2$ molecules are readily predissociated by photons between 912-1000 \AA \, \citep{Li2013}, and the atomic nitrogen then reacts with H$_2^{\ast}$ as a first step in CN formation \citep{Walsh2015, Cazzoletti2018}. The two relationships demonstrate that while N$_2$ photodissociation may proceed more efficiently in disks that are more strongly irradiated by FUV photons relative to the rest of the UV spectrum, CN photodissociation may increase as well. Measured FUV luminosities in this plot are accurate to within a factor of $\sim$1.5-2, with the choice of $A_V$ dominating the uncertainties.} 
\label{LFUV_vs_LCN}
\end{figure}

Although our \emph{HST} data are truncated below 1000 \AA, the method from \citet{France2014} was used to estimate the FUV continuum at shorter wavelengths. The extrapolation was performed on binned fluxes, which were calculated in 210 line-free regions from the longer wavelength data. The continua were fitted with a second order polynomial and extended down to 912 \AA. For uniformity with the larger sample presented in \citet{France2014}, we report the total integrated flux from wavelengths $<$1650 \AA \, and compare these FUV continuum luminosities to both CN and HCN emission (see Figure \ref{LFUV_vs_LCN}).

We report a significant positive correlation between the FUV and IR HCN fluxes $\left( \rho = 0.64; p = 0.01 \right)$, a trend that is consistent with model predictions. An analysis of the H-leverage values and studentized residuals for a linear regression model of the form $L_{HCN} = m \times L_{FUV} / L_{UV, total} + b$ shows that the DR Tau and AA Tau (2013) spectra are highly influential data points. When these outliers are removed from the correlation, the Spearman rank coefficient increases to $\rho = 0.66$, although $p$ remains at 0.01 because of the reduction in sample size. Both targets are included in Figure \ref{LFUV_vs_LCN}, but we note that the AA Tau (2013) spectrum in particular is impacted by an inner disk warp \citep{Schneider2015, Hoadley2015, Loomis2017} that is attenuating the FUV flux. In contrast with both the HCN results and modeling work, we find a significant negative correlation between the FUV and sub-mm CN fluxes $\left(\rho = -0.6; p = 0.02 \right)$ that points to increased photodissociation of CN in disks with strong FUV irradiation. Both DR Tau and AA Tau are once again the most influential data points.  

The degeneracy between gas column densities and temperature of the emitting region makes it difficult to conclusively determine whether the observed trends are due to efficient molecule formation/destruction pathways or increased emission from warmer disks. The negative correlation between sub-mm CN emission and the FUV continuum implies that this trend at least is dominated by photodissociation, since increased UV irradiation could lead to a warmer disk and therefore an opposite relationship. However, the trend of increased IR HCN emission in systems with stronger FUV fluxes could be attributed to either scenario, with both a larger number of HCN molecules and higher temperatures leading to enhanced populations in the upper state of the 14 $\mu$m transition. This result will become easier to interpret as more measurements of sub-mm HCN emission (see e.g. \citealt{Oberg2010, Oberg2011, Bergner2019}) are acquired with ALMA in coming years. Despite the uncertainty in the driving force behind the HCN trend, the negative and positive correlations between the FUV continuum and CN and HCN, respectively, demonstrate that the amount of FUV flux in a given disk helps control the balance of formation and destruction pathways that determine abundances of UV-sensitive species. 

\subsection{Ly$\alpha$ as a Regulator of Disk Chemistry}

Ly$\alpha$ emission is by far the strongest component of the UV radiation field, comprising roughly 75-95\% of the total flux from $\sim 912-1700$ \AA \, in typical accreting T Tauri systems \citep{Bergin2003, Herczeg2004, Schindhelm2012, France2014}. Its broad width in wavelength space encompasses transitions of a variety of molecules, including electronic transitions of H$_2$ \citep{Herczeg2002, Herczeg2004, Hoadley2017} and CO \citep{France2011_FUVII, Schindhelm2012_CO} and photodissociation energies of HCN, C$_2$H$_2$ \citep{Bergin2003, Walsh2015, Heays2017}, and H$_2$O \citep{France2017}. Destruction via Ly$\alpha$ photons is not necessarily more efficient than dissociative transitions at other wavelengths, but this is mitigated by the large number of Ly$\alpha$ photons relative to other regions of the UV spectrum. Unfortunately, the observed Ly$\alpha$ feature is always contaminated by geocoronal emission and ISM absorption along the line-of-sight, making direct measurements difficult. Instead, we use the method from \citet{Schindhelm2012} to reconstruct the Ly$\alpha$ profiles for the five Lupus disks from observations of UV-fluorescent H$_2$.  

UV-H$_2$ fluorescence is activated when a population of hot, vibrationally excited H$_2$ ($T > 1500$ K; \citealt{Black1987, Adamkovics2016}) is ``pumped" from the ground electronic state $\left(X^1 \Sigma_{g}^+ \right)$ into the first and second dipole-allowed excited electronic state $\left(B^1 \Sigma_{u}^+, C^1 \Pi_u \right)$ by photons with energies that fall along the Ly$\alpha$ line profile \citep{Herczeg2002, Herczeg2004, France2011_FUVI}. A cascade of UV emission lines is then observed as the molecular gas population transitions back to the ground electronic state. The features can be divided into groups called progressions, where a single progression, denoted $\left[ \nu', J' \right]$, consists of all transitions out of the same upper electronic level with vibrational state $\nu'$ and rotational state $J'$ \citep{Herczeg2002, Herczeg2004}. We measured fluxes from the strongest emission lines in 12 progressions (see Table 2 of \citealt{France2012}) by integrating over models of a Gaussian profile convolved with the \emph{HST}-COS line-spread function (LSF) and superimposed on a linear continuum. Upper limits on features indistinguishable from the continuum were calculated as the RMS flux within a 3 \AA \,  range across the expected line center. The total progression fluxes $\left(F_{\rm{H_2}, i} \right)$ were then used as Ly$\alpha$ ``data'' $\left(y \right)$, where the $x$ value for each data point is the Ly$\alpha$ pumping wavelength for the electronic transition. We then fit a model Ly$\alpha$ profile to the $\left(x_i, y_i \right) = \left( \lambda_{LyA, i}, F_{H_2, i} \right)$ data points.    

\begin{figure}
\centering
\includegraphics[width=0.75\linewidth]
{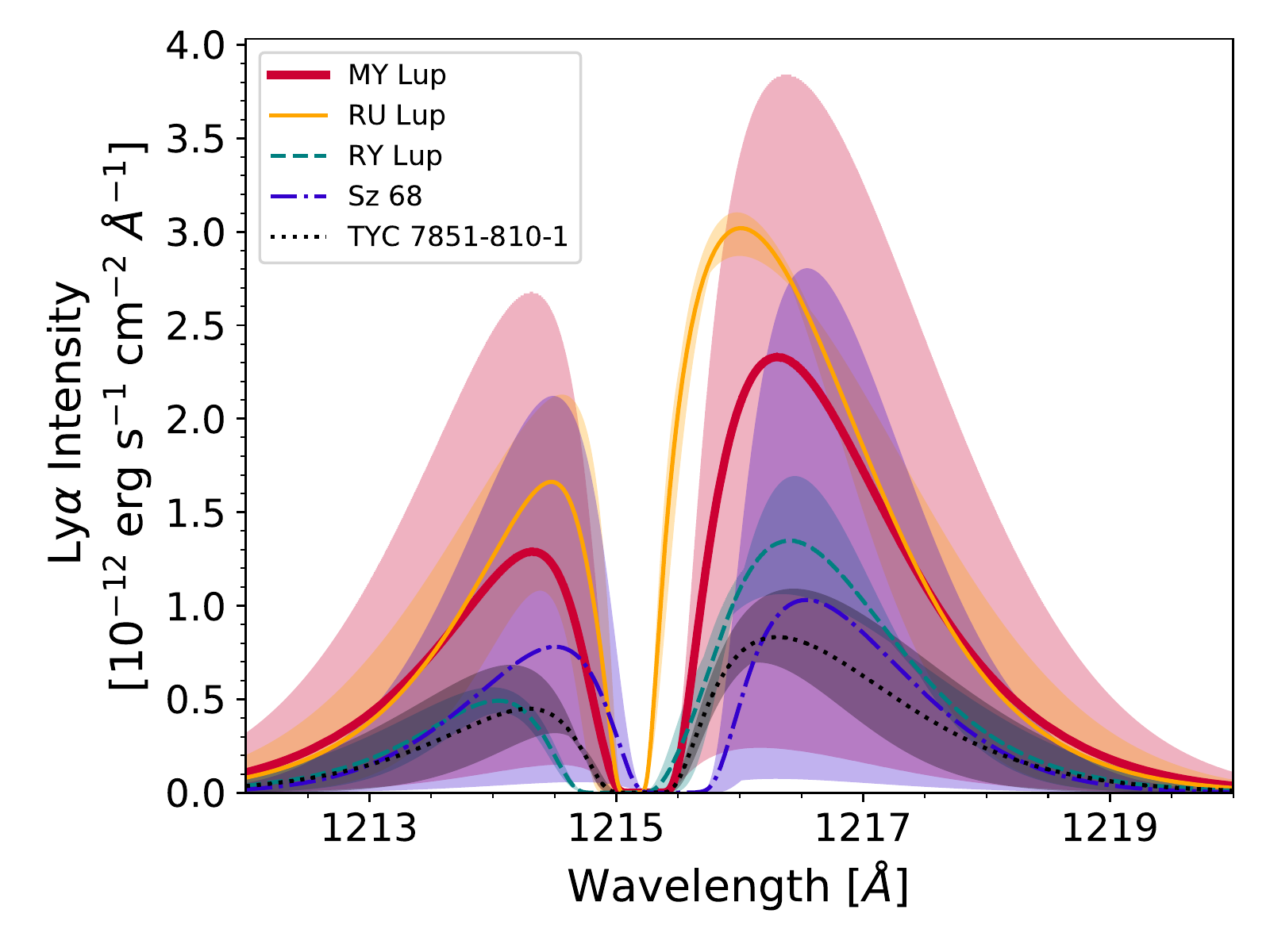}
\caption{A comparison of the reconstructed Ly$\alpha$ profiles at the disk surface for all five Lupus targets in our sample, with colored contours showing rough uncertainties associated with the modeling procedure. The reconstruction is done using observed UV-H$_2$ emission lines as data points \citep{Schindhelm2012}, since molecules are pumped into these excited electronic states by Ly$\alpha$ photons. The absorption seen on the blue side of the line profiles is due to an atomic outflow between the star and disk, rather than interstellar H I along the line-of-sight.}
\label{LyA_comp}
\end{figure}

The Ly$\alpha$ model consists of an initial ``intrinsic" Gaussian emission line, an H I outflow between the star/accretion shock and the molecular gas disk, and a population of H$_2$ that absorbs the Ly$\alpha$ photons. We allow five model parameters to vary: the amplitude of the intrinsic profile $\left(I_{Ly\alpha} \right)$, the velocity and column density of the outflowing H I $\left(v_{out}, \, N_{out} \right)$, and the temperature and column density of the absorbing H$_2$ $\left(T_{H_2}, \, N_{H_2} \right)$. The FWHM of the intrinsic profile for each system was fixed to the average, maximum, and minimum values from \citet{Schindhelm2012}, resulting in three model profiles for each target. Posterior distributions for the model parameters were constructed using MCMC sampling \citep{ForemanMackey2013} within the bounds defined by \citet{Schindhelm2012}. However, we find that the model uncertainties are better captured by the variations in the average, maximum, and minimum FWHM profiles. Figure \ref{LyA_comp} shows the median Ly$\alpha$ profile at the disk surface for all five Lupus systems, with colored contours representing the bounds set by the three FWHM values (see Table \ref{LyA_prop_table}). 

\begin{deluxetable}{cccccc}
\tablecaption{Best-Fit Parameters\tablenotemark{*} for Ly$\alpha$ Reconstruction \label{LyA_prop_table}
}
\tablewidth{0.5\linewidth}
\tabletypesize{\scriptsize}
\tablehead{
\colhead{Target} & \colhead{$I_{Ly\alpha}$} &  \colhead{$v_{out}$} & \colhead{$N_{out}$} & \colhead{$T_{H_2}$} & \colhead{$N_{H_2}$} \\ 
 & \colhead{[erg s$^{-1}$ cm$^{-2}$ \AA$^{-1}$]} & \colhead{[km s$^{-1}$]} & \colhead{[dex]} & \colhead{[K]} & \colhead{[dex]} \\
}
\startdata
RU Lupi & $\left(3.5 \pm 0.5 \right) \times 10^{-12}$ & $-138^{+6}_{-18}$ & $18.2^{+0.4}_{-0.2}$ & $3100 \pm 900$ & $18.2^{+0.5}_{-0.2}$ \\
RY Lupi & $2^{+2}_{-1} \times 10^{-12}$ & $-162^{+12}_{-6}$ & $19.1^{+0.4}_{-0.2}$ & $2000 \pm 1000$ & $18.8^{+1.5}_{-0.5}$ \\
MY Lupi & $3^{+50}_{-30} \times 10^{-13}$ & $-108^{+50}_{-20}$ & $18.5^{+0.3}_{-1}$ & $3000^{+1500}_{-900}$ & $19.5 \pm 2$ \\
Sz 68 & $1^{+50}_{-4} \times 10^{-13}$ & $-60^{+30}_{-100}$ & $18 \pm 1$ & $4000 \pm 1000$ & $20.5^{+1.5}_{-3}$ \\
J1608-3070 & $1.1^{+0.4}_{-0.1} \times 10^{-12}$ & $-114^{+1}_{-11}$ & $18.9^{+0.9}_{-0.1}$ & $3100^{+400}_{-200}$ & $18^{+2}_{-1}$
\enddata
\tablenotetext{*}{Model parameters are: $I_{Ly\alpha} = $ amplitude of intrinsic Ly$\alpha$ profile, $v_{out} = $ velocity of intervening outflowing gas, $N_{out} = $ column density of intervening outflowing gas, $T_{H_2} = $ temperature of fluorescent H$_2$, $N_{H_2} = $ column density of fluorescent H$_2$. The FWHM of the intrinsic profile for each system was held constant in the model, set to the average (708 km s$^{-1}$) from \citealt{Schindhelm2012}. Error bounds on the parameters were estimated by running the model with the FWHM fixed to the minimum (573 km s$^{-1}$) and maximum (912 km s$^{-1}$) values from that work.} 
\end{deluxetable}

\begin{figure}[t!]
	\begin{minipage}{0.49\textwidth}
	\centering
	\includegraphics[width=\linewidth]{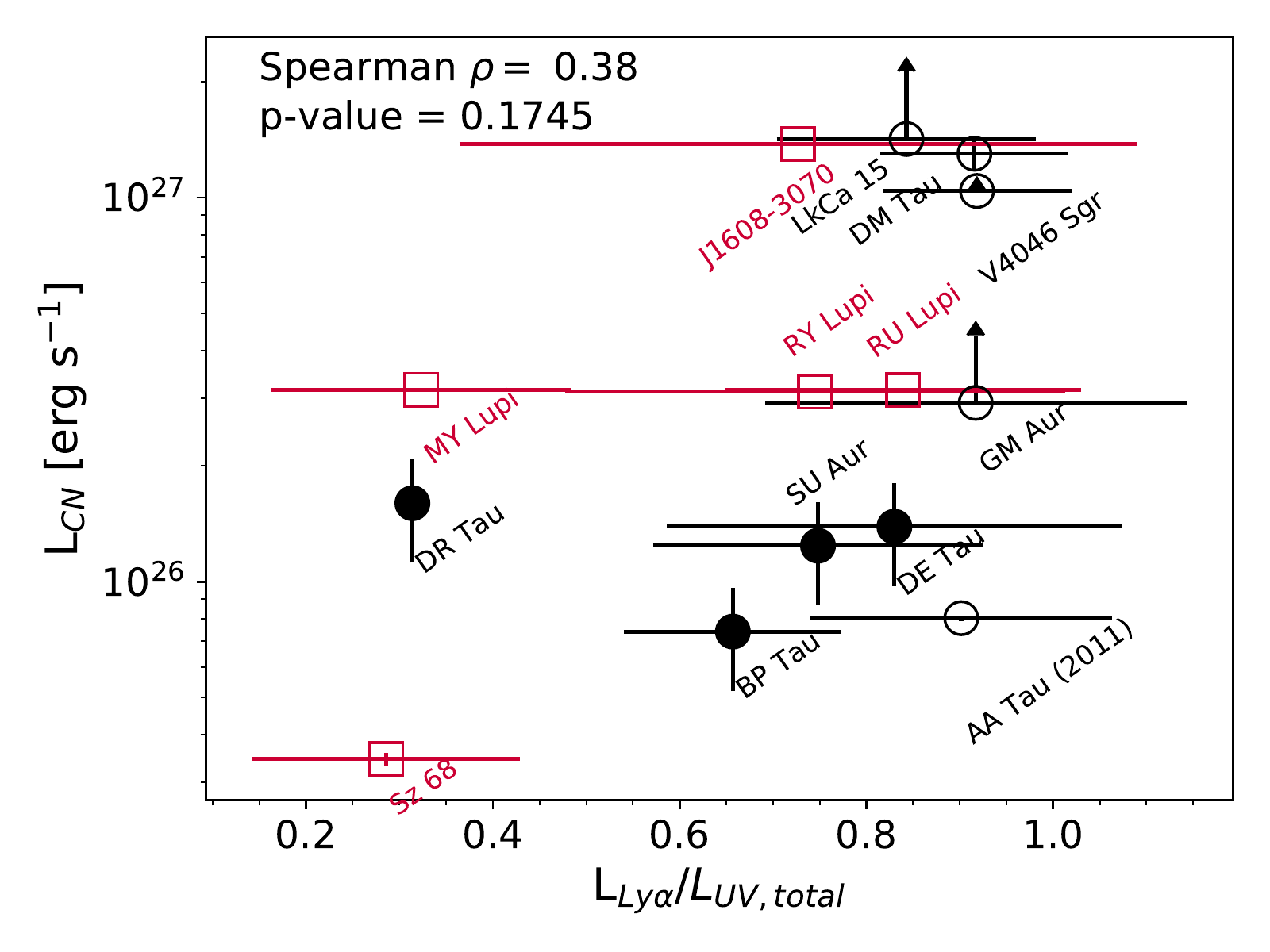}
	\end{minipage}
	\begin{minipage}{0.49\textwidth}
	\centering
	\includegraphics[width=\linewidth]{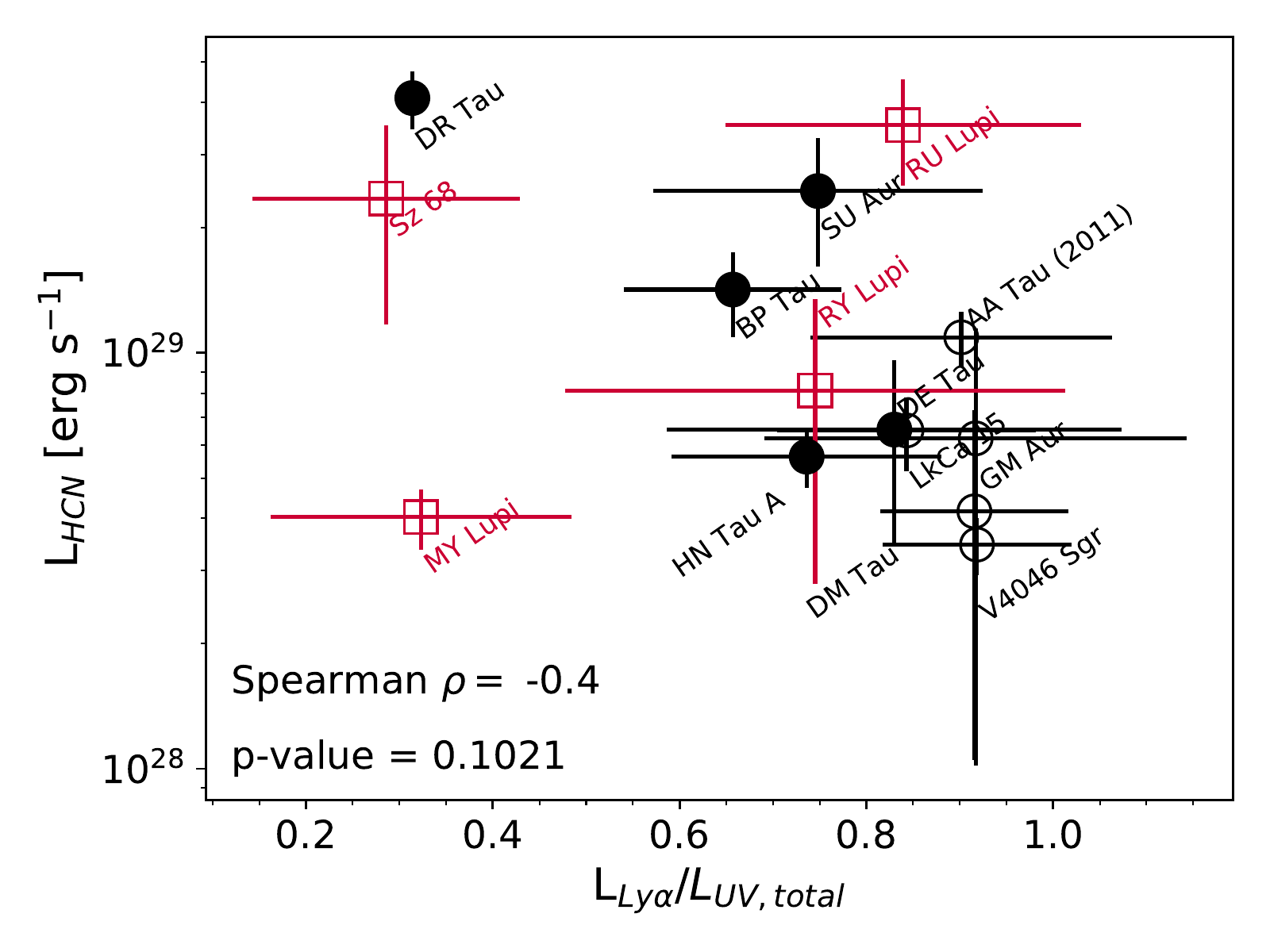}
	\end{minipage}
\caption{Sub-mm CN (left) and 14 $\mu$m HCN (right) emission versus fractional Ly$\alpha$ luminosity. The Lupus disks are shown as red squares and systems from \citet{France2017} as black circles, with open markers representing systems with resolved dust substructure. Neither species is significantly correlated with Ly$\alpha$ emission, but the Spearman rank coefficients are tentatively consistent with models predicting increased CN and decreased HCN abundances with increased Ly$\alpha$ irradiation. The 2013 AA Tau spectrum is omitted from these plots, since the Ly$\alpha$ fluxes were very similar to the spectrum from 2011.} 
\label{LyA_vs_CN}
\end{figure}

We compare the total luminosities from our reconstructed Ly$\alpha$ profiles to the the CN and HCN luminosities in Figure \ref{LyA_vs_CN}. Since CN molecules can also form as byproducts of HCN or CH$_3$CN photodissociation \citep{Walsh2015} via Ly$\alpha$ photons near 1216 \AA \, \citep{Nuth1982, Bergin2003}, we expect that increased Ly$\alpha$ irradiation of the disk surface will increase the significance of dissociative pathways in regulating both CN and HCN column densities. However, no statistically significant relationships are detected between Ly$\alpha$ and either HCN $\left( \rho = -0.4; p = 0.10 \right)$ or CN $\left(\rho = 0.38; p = 0.17 \right)$. We do note that the Spearman rank coefficients themselves $\left( \rho \right)$ are consistent with model predictions, with increased CN and decreased HCN emission observed from targets with stronger Ly$\alpha$ emission. 

Since the Ly$\alpha$ profiles are derived from UV-fluorescent H$_2^{\ast}$, the lack of correlation between CN and Ly$\alpha$ emission may be attributed to the radial stratification of the UV-H$_2$ and sub-mm CN. The UV-H$_2$ emission originates from gas in surface layers of the inner disk $\left(r < 10 \text{ au} \right)$, while the CN population extends to radii of $\sim$30-50 au in the outer disk \citep{vanTerwisga2019, Bergner2019}. Similarly, vertical stratification may be responsible for the null relationship between HCN emission and Ly$\alpha$ emission, with UV-H$_2$ emission originating from much closer to the disk surface than the 14 $\mu$m HCN features. A direct comparison between observed Ly$\alpha$, HCN, and CN luminosities likely requires more careful treatment of optical depth effects. Furthermore, a bootstrapping analysis of the data returns $\pm 1 \sigma$ confidence intervals on the Spearman rank coefficients of [-0.71, -0.29] and [0.2, 0.67] for the HCN and CN vs. $\rm{L\left(Ly\alpha \right) / L \left( UV, total \right)}$ correlations. Although the upper limits of the confidence intervals are consistent with robust linear relationships, it is possible that targets with larger uncertainties on the reconstructed Ly$\alpha$ profiles (e.g. MY Lupi, Sz 68) are masking underlying trends in the data. 

We also note that the \emph{HST}-COS, \emph{Spitzer}-IRS, and sub-mm CN observations were not conducted simultaneously (see Tables \ref{observing_props}, \ref{CN_HCN_props}), implying that the reconstructed Ly$\alpha$ profiles may not be representative of the flux reaching the disk surface at the time of the molecular gas observations. However, an \emph{HST} analysis of older K and M dwarfs has demonstrated that Ly$\alpha$ line strengths do not increase as much as other atomic features during flares (e.g. Si IV $\lambda$1400; \citealt{Loyd2018}), perhaps indicating that Ly$\alpha$ emission also remains steady during typical YSO variability. Although France et al. (2011) report a strong correlation between the FUV continuum and the C IV $\lambda$1550 fluxes that are often used as a proxy for mass accretion rate, multiple phased observations of a single target are likely required to determine whether the FUV continuum also fluctuates with accretion rate during periods of variability. Such observations will be acquired in the next few years through the Ultraviolet Legacy Library of Young Stars as Essential Standards (ULLYSES)\footnote[2]{http://www.stsci.edu/stsci-research/research-topics-and-programs/ullyses} Director's Discretionary program on \emph{HST}.  

\subsection{Mapping the population of H$_2^{\ast}$} 

\subsubsection{Ly$\alpha$-pumped H$_2$}

\begin{figure}
\centering
\includegraphics[width=0.75\linewidth]
{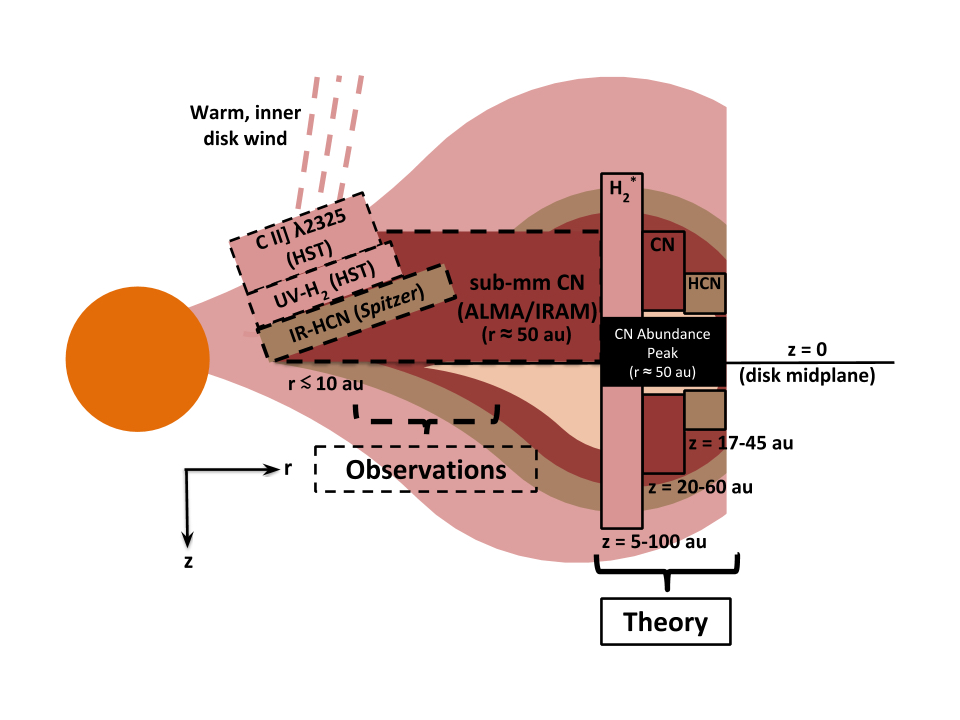}
\caption{Rough spatial locations of emitting gas that produces C II] $\lambda$2325, UV-H$_2$, IR-HCN, and sub-mm-CN emission lines, compared to radius where physical-chemical models predict peak abundances of H$_2^{\ast}$, CN, and HCN \citep{Cazzoletti2018}. The observed gas populations should overlap at radii close to the central star, although the sub-mm CN emission is the only component that generally extends across the full disk.}
\label{chem_cartoon}
\end{figure}

As introduced in Section 3.2, we detect a suite of emission lines from hot ($T \sim 1500-2500$ K; \citealt{Adamkovics2016}), fluorescent H$_2$ in the \emph{HST}-COS spectra of all five disks. Although our \emph{HST}-COS spectra are not spatially resolved, \emph{HST}-STIS spectra of the same features in the disk around TW Hya show that the H$_2$ emission must be within $\sim$2 au of the central star \citep{Herczeg2004}. The features may therefore provide constraints on the distribution of H$_2^{\ast}$ available for CN and HCN formation in the inner disk region. Figure \ref{chem_cartoon} provides a cartoon demonstrating the rough spatial locations of the emitting gas.

The UV-H$_2$ emission lines are spectrally resolved (i.e. broader than the \emph{HST}-COS resolution), allowing us to extract information about the spatial distribution of hot, fluorescent gas. Assuming that the material is in a Keplerian disk, the FWHMs of the emission lines can be mapped to an average radial location as 
\begin{equation}
\left< R_{H_2} \right> = G M_{\ast} \left( \frac{2 \sin i}{FWHM} \right)^2
\end{equation} 
\citep{Salyk2011CO, France2012}, where $M_{\ast}$ is the stellar mass and $i$ is the disk inclination. We average the FWHMs of the strongest emission lines from the $\left[\nu', J' \right] = $ [1,4], [1,7], [0,1], and [0,2] progressions and calculate a radius for each progression (see Table \ref{H2_radii}), finding that the bulk UV-H$_2$ emission originates inside $\sim$2 au for all five disks. The most distant emission radii in the [1,4] and [1,7] progressions are measured for J1608-3070, which has the largest sub-mm dust cavity $\left(r \sim 75 \text{ au; \citealt{vanderMarel2018}} \right)$. This is consistent with the results of \citet{Hoadley2015}, who found that disks with less advanced dust evolution typically have broad UV-H$_2$ features that are dominated by gas located close to the star. Although resonant scattering should allow pumping photons from the Ly$\alpha$ line wings to penetrate deeper into the disk than those from the line center \citep{Bethell2011}, we find no trends between the average emitting radius calculated for each progression and the Ly$\alpha$ pumping wavelength responsible for exciting the transitions.  

\begin{deluxetable}{ccccccccc}
\centering
\tablecaption{Average Emission Radii of Hot, Fluorescent H$_2$ \label{H2_radii}
}
\tablewidth{\linewidth}
\tabletypesize{\scriptsize}
\tablehead{ \colhead{Target} & \colhead{$FWHM_{\left[1,4\right]}$} & \colhead{$\left< R_{H_2} \right>_{\left[1,4\right]}$} & \colhead{$FWHM_{\left[1,7\right]}$} & \colhead{$\left< R_{H_2} \right>_{\left[1,7\right]}$} & \colhead{$FWHM_{\left[0,1\right]}$} & \colhead{$\left< R_{H_2} \right>_{\left[0,1\right]}$} & \colhead{$FWHM_{\left[0,2\right]}$} & \colhead{$\left< R_{H_2} \right>_{\left[0,2\right]}$} \\
 & \colhead{[km/s]} & \colhead{[au]} & \colhead{[km/s]} & \colhead{[au]} & \colhead{[km/s]} & \colhead{[au]} & \colhead{[km/s]} & \colhead{[au]} \\
}
\startdata
RU Lupi & $51 \pm 1$ & $0.10 \pm 0.03$ & $49 \pm 2$ & $0.11 \pm 0.03$ & $31 \pm 2$ & $0.29 \pm 0.08$ & $50 \pm 3$ & $0.11 \pm 0.03$ \\
RY Lupi\tablenotemark{*} & $53.3 \pm 0.7$ & $1.8 \pm 0.5$ & $51 \pm 1$ & $2.0 \pm 0.5$ & $63 \pm 10$ & $1 \pm 1$ & $45 \pm 4$ & $2.6 \pm 0.8$ \\
MY Lupi & $48 \pm 1$ & $1.4 \pm 0.2$ & $46 \pm 2$ & $1.5 \pm 0.2$ & $48 \pm 1$ & $1.4 \pm 0.2$ & $59 \pm 3$ & $0.9 \pm 0.1$ \\
Sz 68 & $57 \pm 2$ & $0.5 \pm 0.3$ & $57 \pm 5$ & $0.5 \pm 0.3$ & $57 \pm 5$ & $0.5 \pm 0.4$ & $43 \pm 3$ & $0.8 \pm 0.6$ \\
J1608-3070 & $47.1 \pm 0.7$ & $2 \pm 1$ & $42.5 \pm 0.7$ & $3 \pm 2$ & $77 \pm 10$ & $0.8 \pm 0.4$ & $75 \pm 10$ & $0.8 \pm 0.4$
\enddata
\tablenotetext{*}{The FWHMs listed here for RY Lupi are from single-component fits to the emission lines. \citet{Arulanantham2018} presents a more detailed analysis of the H$_2$ line shapes, showing that the strongest features in the [1,4] progression are better fit by a two-component model. We also adopt the disk inclination from \citet{vanderMarel2018} $\left( 68^{\circ} \right)$, instead of the scattered light inclination from \citet{Manset2009} $\left(85.6^{\circ} \right)$.} 
\end{deluxetable}

\begin{figure}
	\begin{minipage}{0.5\textwidth}
	\centering
	\includegraphics[width=\linewidth]{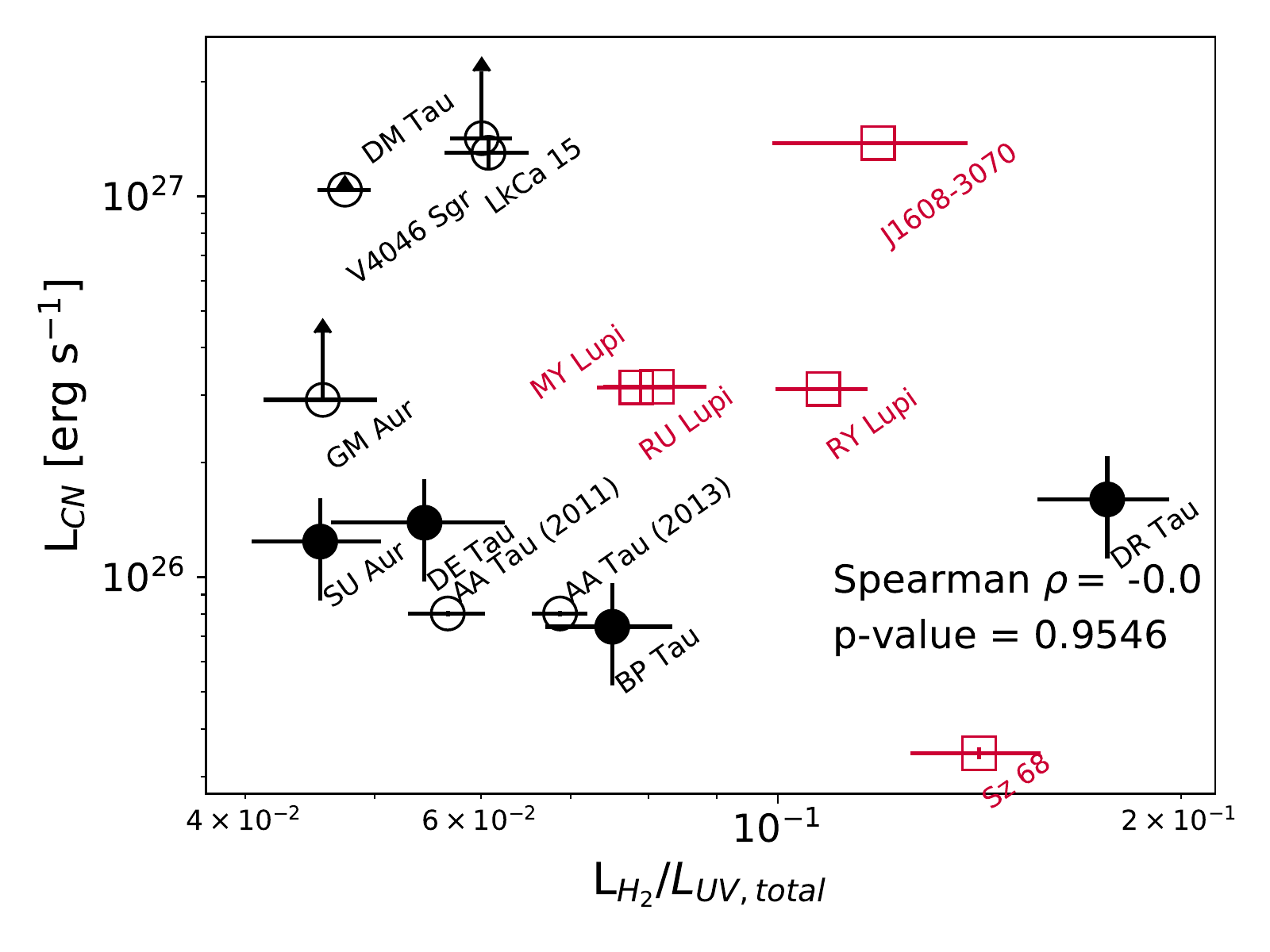}
	\end{minipage}
	\begin{minipage}{0.5\textwidth}
	\centering
	\includegraphics[width=\linewidth]{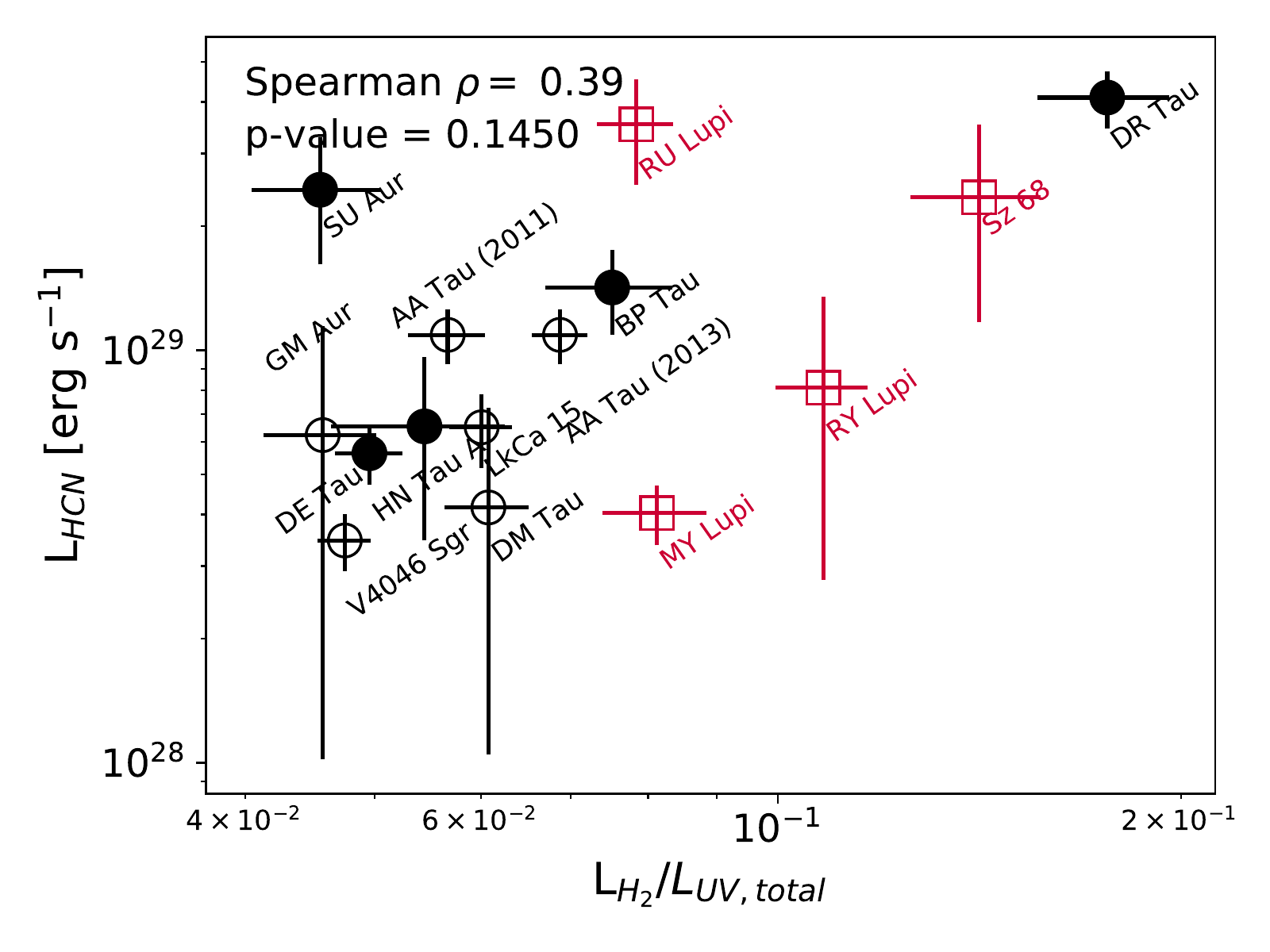}
	\end{minipage}
\caption{CN and HCN luminosities versus fractional luminosity from UV-pumped H$_2$, with open markers representing disks with resolved dust substructure (i.e. rings, gaps, or cavities). The Lupus disks are shown as red squares, while the black circles are systems from \citet{France2017}. Neither sub-mm CN or IR HCN is significantly correlated with UV-H$_2$, which is likely due to the disparate radial and vertical stratification of the three populations of emitting gas.}
\label{H2_vs_CN}
\end{figure}

Since H$_2^{\ast}$ is required to produce CN (see Eq. 1), the UV-H$_2$ features are a probe of the uppermost layer of available reactants located at the average radius of emitting gas $\left(R_{H_2} \right)$. We explore this relationship by comparing the CN and HCN luminosities to the total flux from fluorescent UV-H$_2$ emission lines (see Figure \ref{H2_vs_CN}), finding that neither species is correlated with the UV-H$_2$. In the case of HCN, the scatter can be attributed to the vertical stratification effects discussed in Section 3.2, since the H$_2^{\ast}$ and HCN abundances are expected to peak at different heights relative to the disk midplane \citep{Cazzoletti2018}. Ly$\alpha$ photons are only able to pump H$_2^{\ast}$ in a thin surface layer, so the UV-H$_2$ features do not contain information from vibrationally excited gas present deeper in the disk that would be co-located with the HCN. By contrast, the sub-mm emission traces CN molecules in the cold outer disk, where the population of H$_2^{\ast}$ declines due to extinction of pumping photons from the UV continuum \citep{Visser2018, Cazzoletti2018}. The radial distribution of UV-H$_2$ may provide estimates of either how far Ly$\alpha$ emission is able to travel in the disk or a rough boundary for the population of H$_2^{\ast}$. We explore this degeneracy further in Section 4 by using 2-D radiative transfer models to reproduce the UV-H$_2$ emission lines. 

\subsubsection{1600 \AA \, ``Bump'' as a Signature of H$_2$O Dissociation}

Previous studies of UV continuum emission from young stars with disks have identified an excess ``bump'' in the spectra around $\sim$1600 \AA \, \citep{Herczeg2004, Bergin2004}. The feature is attributed to continuum and line emission from Ly$\alpha$-pumped H$_2^{\ast}$, where the population of H$_2^{\ast}$ is indirectly produced during H$_2$O dissociation in the inner disk ($r < 2$ au; \citealt{France2017}). While the Ly$\alpha$ and bump luminosities are strongly correlated $\left( \rho = 0.74, \, p = 1.72 \times 10^{-3}; \, \text{\citealt{France2017}} \right)$, no relationship is observed between the bump and X-ray luminosities \citep{Espaillat2019}. This implies that Ly$\alpha$ photons may play a more prominent role than the X-ray radiation field in regulating the distribution of hot H$_2$O and vibrationally excited H$_2$ at radii close to the central star.

\begin{figure}[t!]
	\begin{minipage}{0.5\textwidth}
	\centering
	\includegraphics[width=\linewidth]{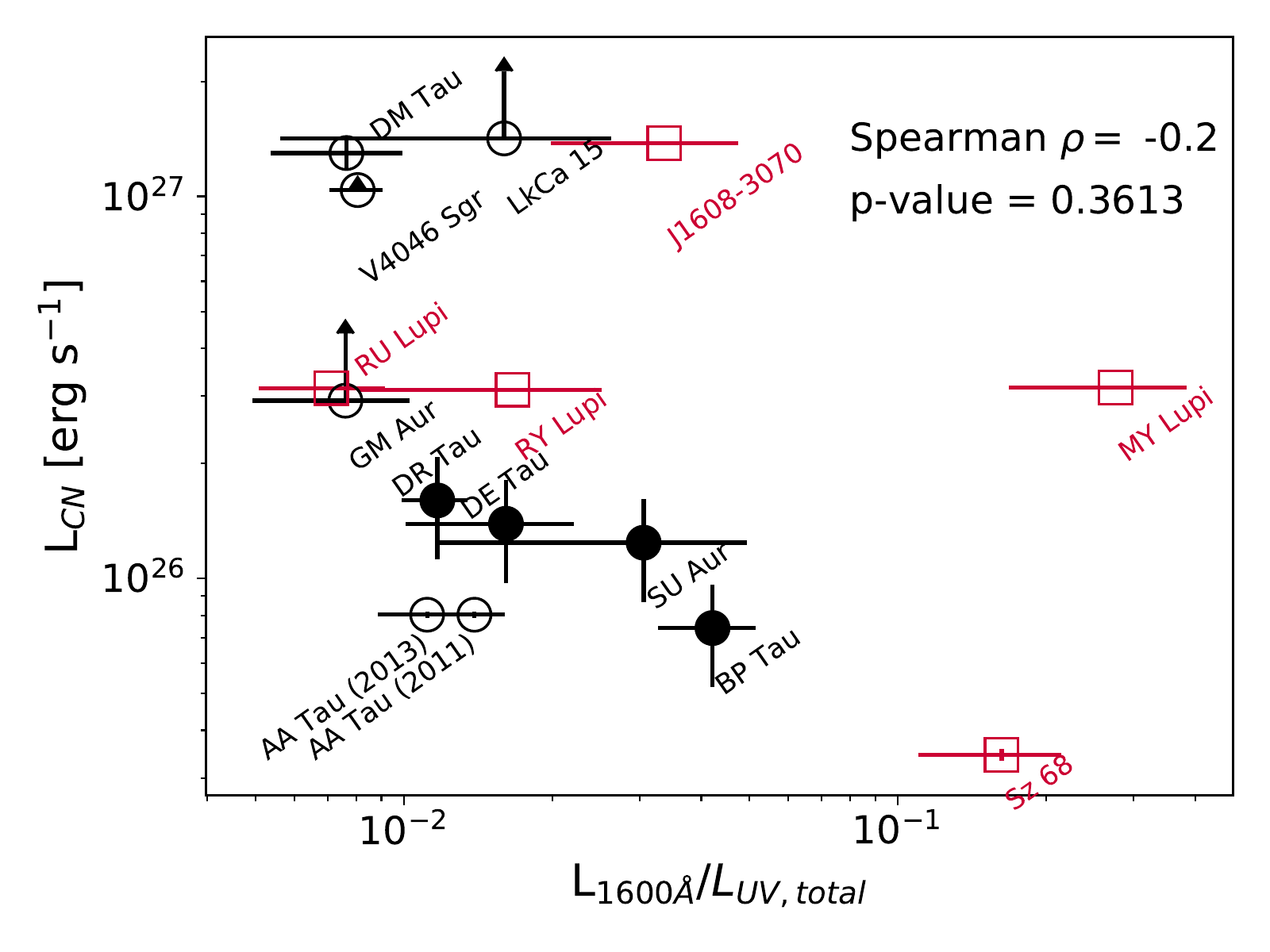}
	\end{minipage}
	\begin{minipage}{0.5\textwidth}
	\centering
	\includegraphics[width=\linewidth]{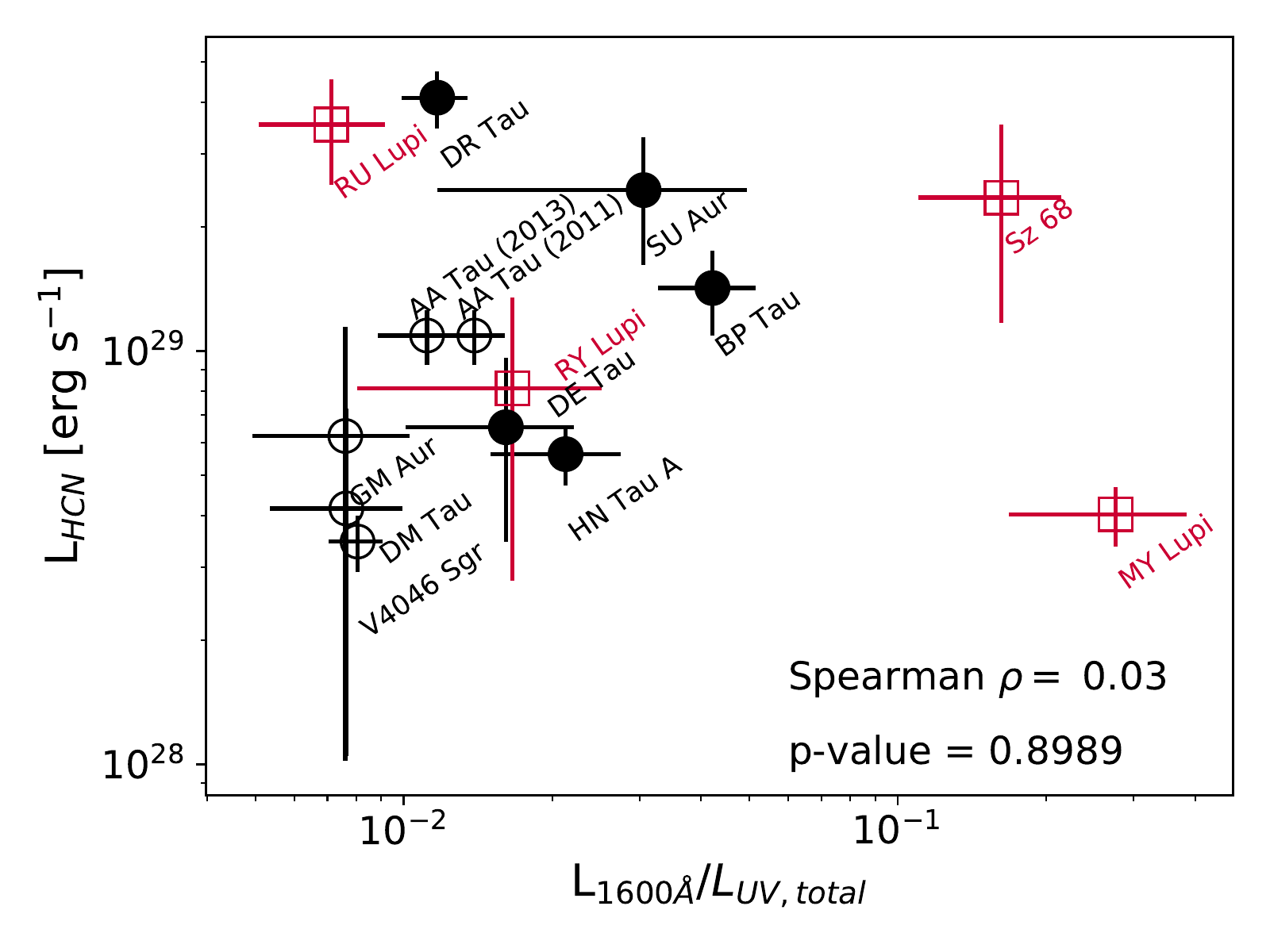}
	\end{minipage}
\caption{CN and HCN luminosities versus fractional luminosity from the 1600 \AA \, bump (produced by Ly$\alpha$-pumped H$_2^{\ast}$ left behind during H$_2$O photodissociation; \citealt{France2017}). The Lupus disks are shown as red squares, while the black circles are systems from \citet{France2017}. The strength of the bump is correlated with time-varying accretion luminosities \citep{Espaillat2019}, implying that the measurements shown here are snapshots of the population of hot, non-thermal gas and not necessarily reflective of equilibrium chemical conditions.} 
\label{1600_vs_LCN}
\end{figure}

We use the method described in \citet{France2017} to measure bump fluxes and calculate luminosities for the three disks in our sample that were not included in that work. A second order polynomial fit to the FUV continuum, representative of the underlying flux, was integrated from 1490-1690 \AA \, and subtracted from the total observed flux in the same wavelength region. The residual flux can be attributed to the bump alone (see Table \ref{lum_table}). Figure \ref{1600_vs_LCN} compares the CN/HCN and bump luminosities, again showing no clear linear trend between the spectral features. Similar to the Ly$\alpha$-pumped fluorescent gas, the H$_2^{\ast}$ responsible for producing the bump is therefore likely located higher in the disk surface than the 14 $\mu$m HCN and constrained to closer radii than the sub-mm CN.  

\begin{deluxetable}{cccccc}
\tablecaption{Hot Gas \& Stellar Luminosities \label{lum_table}
}
\tablewidth{0 pt}
\tabletypesize{\scriptsize}
\tablehead{
\colhead{Target} & \colhead{$L_{H_2}$} & \colhead{$L_{1600 \AA}$} & \colhead{$L_{Ly\alpha}$} & \colhead{$L_{C II]}$} & \colhead{$L_{FUV}$\tablenotemark{c}} \\ 
 & \colhead{[10$^{30}$ erg s$^{-1}$]} & \colhead{[10$^{29}$ erg s$^{-1}$]} & \colhead{[10$^{29}$ erg s$^{-1}$]} & \colhead{[10$^{29}$ erg s$^{-1}$]} & \colhead{[10$^{29}$ erg s$^{-1}$]} \\
}
\startdata
RU Lupi & $6.1 \pm 0.4$ & $5.54 \pm 1.59$\tablenotemark{a} & $170 \pm 70$ & \nodata & 49.4\tablenotemark{a} \\
RY Lupi & $1.01 \pm 0.08$ & $1.55 \pm 0.8$\tablenotemark{a} & $70 \pm 25$ & $2.18 \pm 0.02$ & 8.3 \\
MY Lupi & $0.45 \pm 0.04$ & $15.3 \pm 6$ & 18\tablenotemark{b} & $0.69 \pm 0.01$ & 3.5 \\
Sz 68 & $0.27 \pm 0.03$ & $3.1 \pm 1$ & 5\tablenotemark{b} & $0.73 \pm 0.01$ & 3.2 \\
J1608-3070 & $0.6 \pm 0.1$ & $1.7 \pm 0.7$ & 37\tablenotemark{b} & $0.18 \pm 0.01$ & 4.4 \\ 
\enddata
\tablenotetext{a}{Values from \citet{France2017}.}
\tablenotetext{b}{Reconstructed Ly$\alpha$ profiles have large uncertainties, due to noisy UV-H$_2$ emission lines.}
\tablenotetext{c}{FUV continuum luminosities are accurate to within a factor of $\sim$1.5-2.}
\end{deluxetable}

\subsection{C II] $\lambda$2325 Emission as a Tracer of Inner Disk C$^+$}

The semi-forbidden C II] emission lines at $\lambda$2325 are detected in all the Lupus and Taurus-Auriga targets with \emph{HST}-STIS spectra. Models of the feature in other young systems find that its shape is consistent with an origin at the base of a warm, inner disk ($r \sim 0.1-1$ au) wind, coincident with surface layers of the disk itself \citep{GdC2005} and with similar formation conditions to the [O I] 6300 \AA \, line \citep{Simon2016}. Studies of the [O I] line show that the profile is narrower in systems with dust-depleted inner disks \citep{Simon2016, Banzatti2019}, indicating that the base of the wind, and therefore all the emitting gas it contains, shifts to more distant radii within the disk. C II] therefore may also be an important observational proxy for the population of inner disk C$^+$.    

\citet{Simon2016} find that the low-velocity component of the [O I] profile is consistent with Keplerian rotation, with emitting gas originating in radially separated components located between 0.05 and 5 au. Only one target (AA Tau) in the sample presented here was observed at high enough spectral resolution and signal-to-noise to obtain kinematic information from the C II] emission, with three individual features resolved at $\lambda$2325.40, $\lambda$2326.93, and $\lambda2328.12$ \AA. We find that the central feature ($\lambda$2326.93 \AA) has a similar FWHM to the broad component of the [O I] 6300 \AA \, line (96 km s$^{-1}$; \citealt{Banzatti2019}), which likely implies that the two emission lines have similar inner disk origins. The C II] $\lambda$2325 emission is also correlated with the ionization fraction derived from carbon recombination lines (e.g. DR Tau; \citealt{McClure2019}), providing further support that the two features come from a roughly co-spatial population of inner disk gas.    

Ionized carbon (C$^+$) is a key reactant in the main formation pathways of C$_2$H \citep{Henning2010, Walsh2015, Miotello2019} and CN \citep{Walsh2015, Visser2018, Cazzoletti2018}. C$^+$ also plays an important role in CO$_2$ destruction, with physical-chemical models showing an enhancement in the CO$_2$ column density at $r \sim 10$ au, where gas self-shielding allows the C$^+$/C ratio to drop below unity \citep{Walsh2012}. We note that the FUV C II $\lambda$1335 resonant feature is also included in the datasets presented here. However, its line profiles are likely dominated by the accretion pre-shock region and protostellar chromosphere and consistently show detectable wind absorption signatures \citep{JKH2007}, requiring a more complicated method to untangle the associated population of C$^+$ that is directly involved in molecule formation and destruction than needed for the C II] $\lambda$2325 feature.

\begin{figure}[t!]
	\begin{minipage}{0.47\textwidth}
	\centering
	\includegraphics[width=\linewidth]{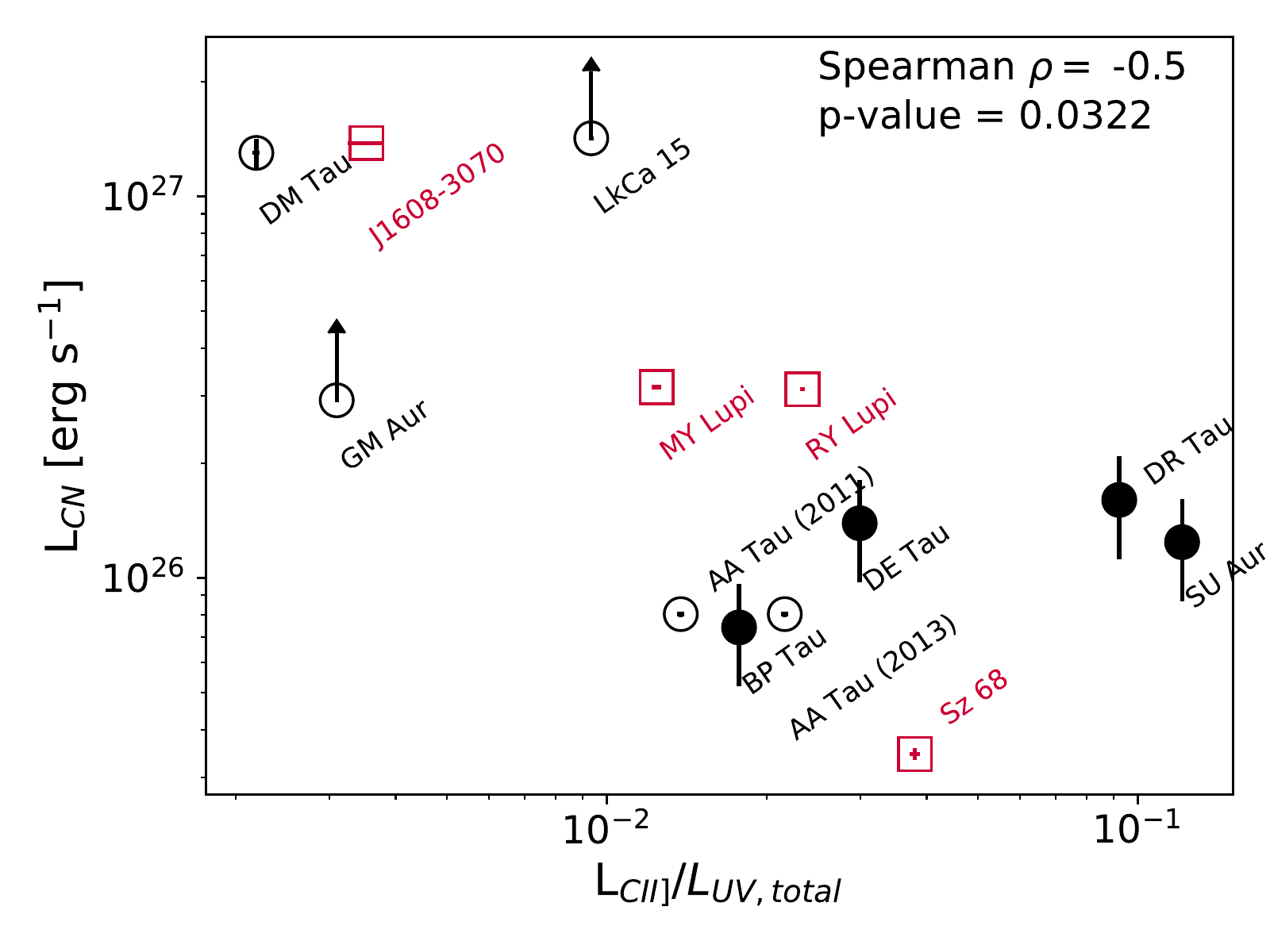}
	\end{minipage}
	\begin{minipage}{0.47\textwidth}
	\centering
	\includegraphics[width=\linewidth]{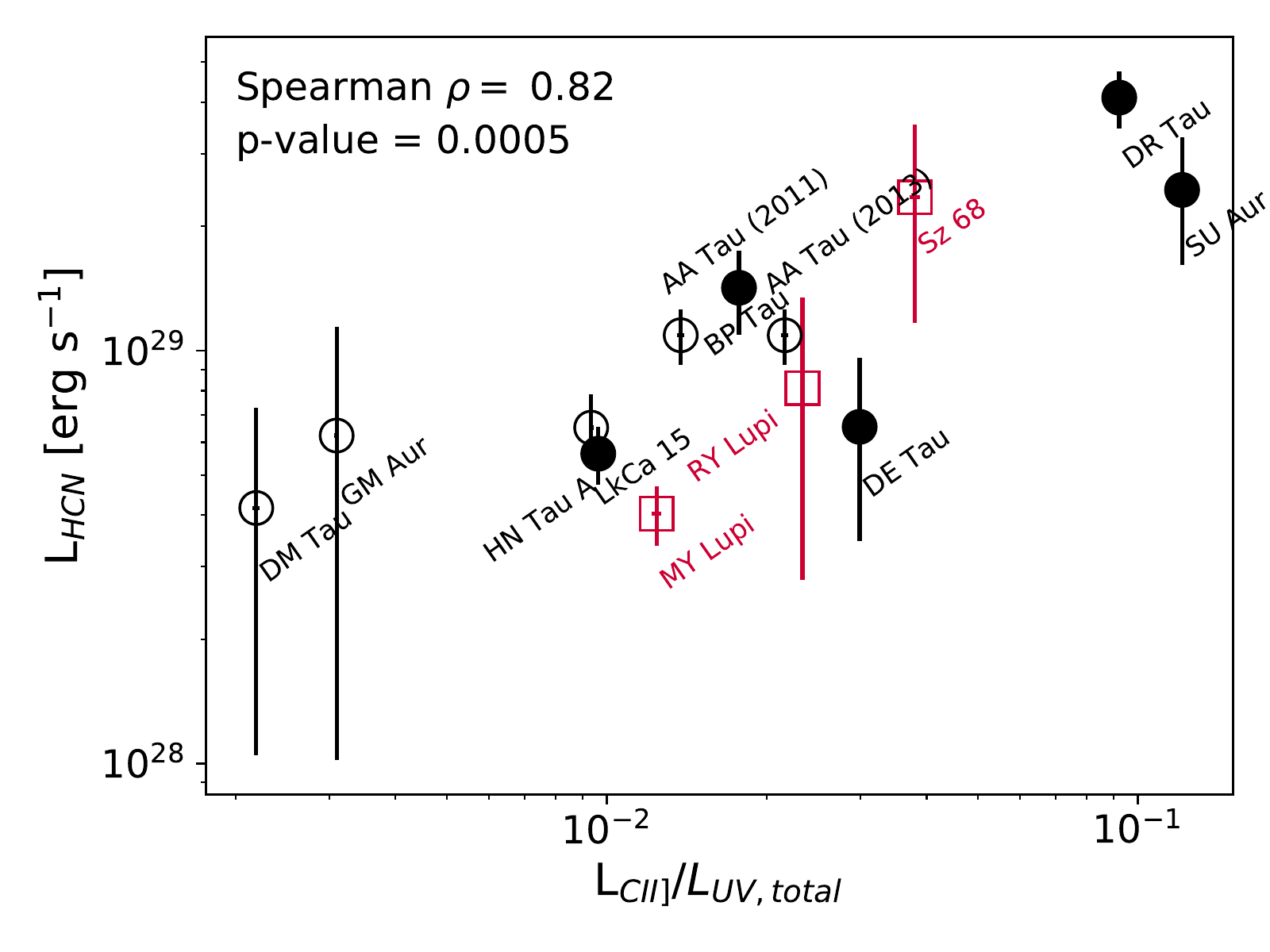}
	\end{minipage}
\caption{CN and HCN luminosities versus fractional C II] $\lambda$2325 emission. The five Lupus systems are shown as red squares and the subset of disks from \citet{France2017} as black circles, with open markers representing disks with resolved dust substructure. The sub-mm CN emission is negatively correlated with the C II] feature, and a strong positive correlation is seen with the IR HCN emission. The relationships can likely be attributed to accretion processes that enhance both C II] and HCN emission and dissociate CN.}
\label{LCII_vs_FCN}
\end{figure}

To explore whether the C II] emission is related to the CN and HCN observations, C II] line fluxes were measured for each target in our sample by integrating the dereddened \emph{HST}-STIS spectra over a wavelength range spanning all three C II] features and subtracting a linear continuum (see Table \ref{lum_table}). Figure \ref{LCII_vs_FCN} compares the CN, HCN, and C II] $\lambda$2325 luminosities, showing statistically significant relationships between the relative fluxes from C II] and emission from both molecular species. We find that HCN emission is stronger in systems with higher C II] fluxes $\left( \rho = 0.82; p = 0.0005 \right)$, although CN emission appears to decline as C II] fluxes increase $\left( \rho = -0.5; p = 0.03 \right)$. However, we note that the CN relationship is largely driven by GM Aur and SU Aur, which both have high H-leverage values and studentized residuals when included in a linear regression model of the form $L_{CN} = m \times L_{C II]} / L_{UV, total} + b$.

\begin{figure}
\centering
\includegraphics[width=0.7\linewidth]{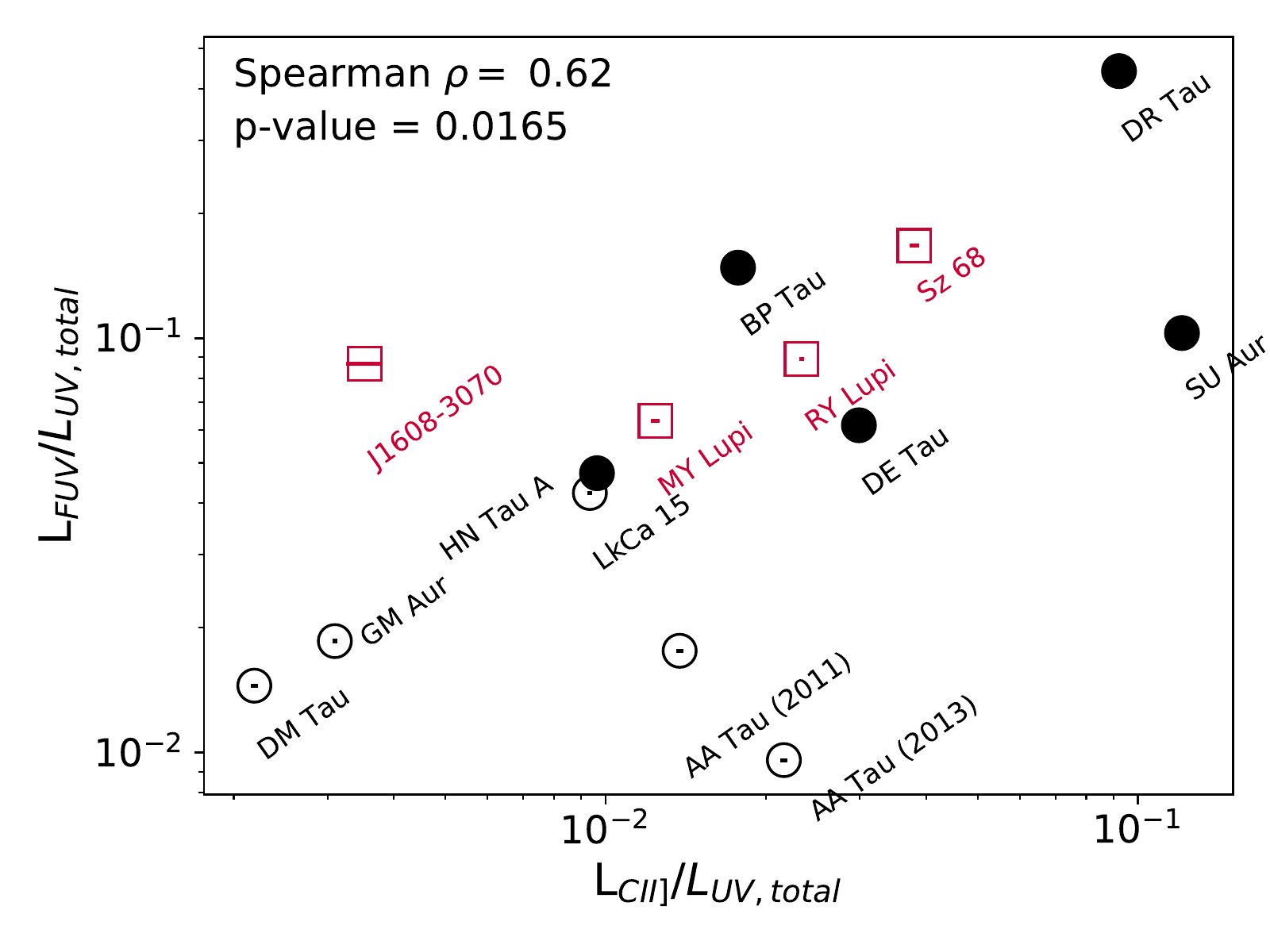}
\caption{Fractional FUV continuum emission versus fractional C II] $\lambda$2325 luminosities. Both quantities are significantly correlated with 14 $\mu$m HCN emission, implying that systems with stronger FUV and C II] fluxes are better able to produce HCN in the inner disk. A full kinematic analysis of the C II] line profiles is likely required to determine whether the C II] emission traces the C$^+$ population involved in gas-phase chemistry or accretion processes that enhance the strength of the feature.}
\label{CII_vs_FUV}
\end{figure}

The negative correlation between fractional C II] and sub-mm CN emission points to a relationship between C II] $\lambda$2325 emission and the FUV continuum (see Figure \ref{CII_vs_FUV}), implying that more CN is photodissociated in systems with stronger FUV fluxes and correspondingly larger C II] fluxes. Since the FUV continuum is strongly associated with accretion \citep{France2014}, we explore whether the population of C$^+$ is similarly driven. Although we find that the C II] emission is not significantly correlated with the mass accretion rate $\left( \rho = 0.33; p = 0.26 \right)$, the positive relationship between C II] and the FUV continuum may still be a byproduct of accretion-related processes that are difficult to trace without contemporaneous measurements of mass accretion rates and C II] fluxes. Since only a handful of targets presented here have high-resolution \emph{HST}-STIS observations of the C II] $\lambda$2325 feature, a more detailed kinematic analysis of the line properties is outside the scope of this work.   

\begin{figure}
\centering
\includegraphics[width=0.75\linewidth]{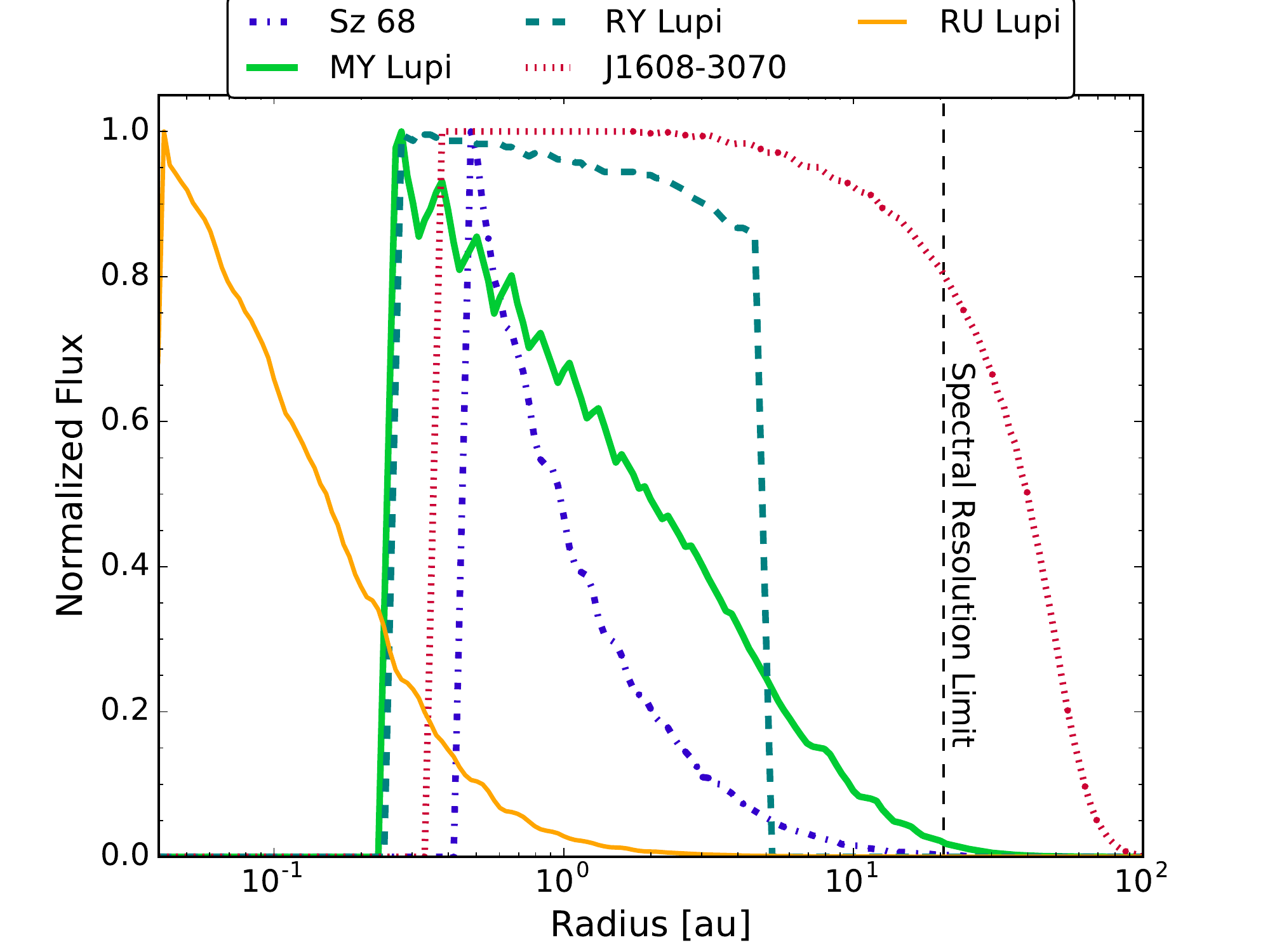}
\caption{Radial distributions of flux from hot, UV-fluorescent H$_2$ pumped by Ly$\alpha$ in the five Lupus disks presented here, obtained by fitting a 2-D radiative transfer code to individual emission lines. The flux distribution in J1608-3070 extends to much more distant radii than the other three systems, consistent with a depletion of small dust grains inside a large sub-mm cavity ($r_{cav} \sim 75$ au; \citealt{vanderMarel2018}). Although the flux distribution from MY Lupi spans the radii of its first two dust rings ($r \sim 8, 20$ au; \citealt{Huang2018_annular}), we detect no sign of a break in the population of hot gas. Finally, we note that the flux distribution from Sz 68 is sharply truncated around 10 au, implying that the UV-H$_2$ emission extends to the edge of the system's circumprimary disk \citep{Kurtovic2018}.}
\label{H2_model_results}
\end{figure}

\section{Discussion}

Physical-chemical models of disks have suggested that emission from Ly$\alpha$ and the FUV continuum directly impact molecular gas abundances, providing photons at the energies required for gas-phase reactions to proceed. Although we find significant correlations between both 14 $\mu$m HCN and sub-mm CN and the FUV continuum, neither species appears to be related to Ly$\alpha$ emission. In order to understand this discrepancy, we examine the impact of disk geometry on the observed spectra and consider whether the optical depth of the inner disk has a significant impact on our results. 

\subsection{Extent of UV-H$_2$ Emitting Region}

UV-H$_2$ features observed with \emph{HST}-COS are typically much broader than the instrument resolution $\left( \Delta v \sim 17 \text{ km s}^{-1} \right)$, implying that most of the detected flux is emitted at radii $\leq$10 au from the central star (see e.g. \citealt{France2012}). This is supported by \emph{HST}-STIS spectra of H$_2$ in the disk around TW Hya, which show that the UV-H$_2$ emission lines are not spatially extended and are therefore confined to the inner disk \citep{Herczeg2002}. However, physical-chemical models suggest that both Ly$\alpha$ and FUV photons reach large swaths of the outer disk (see e.g. \citealt{Cleeves2016}). The FUV photons can pump H$_2$ into vibrationally excited states at radii where the gas temperature is too low for thermal populations of H$_2^{\ast}$ to survive (see e.g. \citealt{Cazzoletti2018}). Ly$\alpha$ photons then act as a searchlight illuminating the vibrationally excited population. 

Since fluorescent emission is not detected from the outer disk (see Figure \ref{H2_model_results}), the radial extent of the UV-H$_2$ emitting region may be restricted by either: 
\begin{enumerate}

\item the extent to which Ly$\alpha$ pumping photons can travel into the disk (Ly$\alpha$-limited), or

\item the total abundance of H$_2^{\ast}$, excited thermally $\left(T > 1500 \text{ K} \right)$ and via FUV-pumping (H$_2^{\ast}$-limited). 

\end{enumerate}
The negative correlation between sub-mm CN fluxes and the FUV continuum (measured both directly and via C II] $\lambda$2325 emission) reported here implies that UV photons are able to reach the outer disk, perhaps enhancing the population of H$_2^{\ast}$ in those regions. With this in mind, we focus on a simple model of the Ly$\alpha$-limited scenario. A more advanced full disk H$_2^{\ast}$ distribution will be analyzed in a future paper on the H$_2^{\ast}$-limited scenario.   

\subsection{2-D Radiative Transfer Models of UV-H$_2$ Emission}

To investigate whether our UV-H$_2$ observations are Ly$\alpha$-limited, we use the 2-D radiative transfer model developed by \citet{Hoadley2015} to reproduce the distributions of fluorescent gas in 4/5 of the Lupus disks presented here. The model propagates Ly$\alpha$ photons into a Keplerian disk with a power-law temperature gradient of coefficient $q$, fixed temperature $T_{1 au}$ at a radial distance of 1 au, and a minimum $T = 1000$ K for UV-fluorescence,
\begin{equation}
T \left( r \right) = T_{1 au} \left( \frac{r}{1 \, \text{au}} \right)^{-q}
\end{equation}
a pressure scale height dependent on the radial temperature distribution $T \left(r \right)$ and stellar mass $M_{\ast}$, 
\begin{equation}
H_p \left(r \right) = \sqrt{ \frac{k T \left( r \right)}{\mu m_H} \frac{r^3}{G M_{\ast}}}
\end{equation}
and a surface density distribution with characteristic radius $r_c$, power-law coefficient $\gamma$, and normalization factor $\Sigma_c = M_{\rm{H_2}} \left(2 - \gamma \right) / \mu \rm{r_{char}} \left(2 \pi \right)^{3/2}$ 
\begin{equation}
\Sigma \left( r \right) = \Sigma_c \left( \frac{r}{r_c} \right)^{-\gamma} \exp \left[ - \left( \frac{r}{r_c} \right)^{2 - \gamma} \right]
\end{equation}
We calculate the mass density distribution at some height $z$ above the disk midplane $\left( \rho \left(r, z \right) \right)$ for the entire volume of H$_2$ gas and the corresponding number density $\left( n_{\left[ \nu, J \right]} \left(r, z \right) \right)$ and optical depth $\left( \tau_{\lambda} \left(r, z \right) \right)$ of molecules in the upper level of each progression. Once the physical structure of the underlying hot H$_2$ population has been derived, the distribution of UV-H$_2$ flux from each transition is calculated as 
\begin{equation}
F_{\lambda_{H_2}} \left( r, z \right) = \eta F_{\ast, Ly\alpha} \left( \frac{R_{\ast}^2}{r^2} \right) \left( \frac{\left( r \cos i_{\rm{disk}} \right)^2}{s \left(r, z \right)^2} \right) \times B_{mn} \sum^{\tau_{\lambda}'} \left( 1 - e^{-\tau_{\lambda}' \left(r, z \right)} \right),
\end{equation}    
where $\eta$ represents the geometric fraction of the disk exposed to Ly$\alpha$ photons (held constant at 0.25; \citealt{Herczeg2004}), F$_{\ast, Ly\alpha}$ is the Ly$\alpha$ flux reaching the gas, $B_{mn}$ is the branching ratio that describes the likelihood of a given transition relative to all other transitions from the same upper level, and $s \left(r, z \right)$ is the sightline from the observer to a gas parcel at position $\left(r, z \right)$ in the disk. The flux distribution is then summed over the entire disk, producing an emission line profile that we fit directly to the observed UV-H$_2$ spectra. The resulting model distribution of gas informs us about where in the disk the H$_2^{\ast}$ is exposed to Ly$\alpha$ radiation, providing radial constraints on the uppermost layer of reactants for producing CN molecules in the inner disk. 

To fit these models to the observed UV-H$_2$ features from the Lupus disks, we used the reconstructed Ly$\alpha$ profiles shown in Figure \ref{LyA_comp} to estimate $F_{\ast, Ly\alpha}$ for each progression. The disk inclinations $\left(i_{\rm{disk}} \right)$ and stellar masses $\left(M_{\ast} \right)$ were fixed to values from the literature (see Table \ref{stellar_props}). The parameters $z$, $\gamma$, $T_{1 \, \rm{au}}$, $q$, $r_{\rm{char}}$, and $M_{\rm{H_2}}$ were allowed to vary, and uncertainties on the best-fit models were estimated using MCMC re-sampling \citep{ForemanMackey2013} with uniform priors spanning the grid space defined by \citet{Hoadley2015}. The MCMC algorithm used 3000 walkers to explore the parameter space, finding no strong degeneracies between the six variables (see Figure \ref{RYLup_corner} for an example corner plot). The final distributions of UV-H$_2$ flux were most sensitive to the values of $T_{1 \, \rm{au}}$ and $q$ used to define the radial temperature structure defined in Equation 5. 

\subsection{Radial Distributions of Flux from UV-H$_2$}

Figure \ref{H2_model_results} shows the radial distributions of UV-H$_2$ flux that best reproduce the observed emission lines from the Lupus disks. The shapes of the gas distributions are generally correlated with the sub-mm dust distributions, in agreement with the results from \citet{Hoadley2015} that showed less UV-H$_2$ close to the star in disks with dust gaps or cavities. Sz 68, which is a close binary \citep{Ghez1997}, shows a distribution that is sharply truncated at 10 au. This is consistent with UV-H$_2$ emission from the primary component alone \citep{Kurtovic2018}, with Ly$\alpha$ photons reaching the gas surface layers out to the circumprimary disk edge. MY Lupi, which has two shallow gaps at 8 and 20 au \citep{Huang2018_annular}, shows no sign of breaks in the gas disk at those radii, although the flux distribution declines rapidly from its peak at $\sim$0.25 au. Finally, we report that the UV-H$_2$ emission in J1608-3070 extends to more distant radii than the other systems, with a flat distribution from $\sim$0.3-10 au followed by a slow decline to the outer disk regions. This is consistent with the observed dust depletion inside of 75 au \citep{vanderMarel2018}, which allows FUV photons to travel further in the disk and pump a more extended population of H$_2$ into vibrationally excited states. However, we note that the outer radius of the UV-H$_2$ distribution is limited by the \emph{HST}-COS spectral resolution $\left( \Delta v \sim 17 \text{ km s}^{-1} \right)$. Given the stellar mass and disk inclination of J1608-3070, this corresponds to a spatial scale maximum of $\sim$20 au.

Although the UV-H$_2$ lines do not originate from the same region as the sub-mm CN emission, the distributions of UV-H$_2$ flux provide constraints on either the radial extent of the population of H$_2^{\ast}$ in surface layers of the gas disk or the location where those surface layers become optically thick to Ly$\alpha$ photons. However, the 2-D radiative transfer models described above only include the thermal population of H$_2^{\ast}$ and do not account for FUV-pumped gas located at more distant radii. Distinguishing between the Ly$\alpha$-limited and H$_2^{\ast}$-limited scenarios for UV-H$_2$ emission, and subsequently identifying disk regions where H$_2^{\ast}$-driven CN formation pathways can proceed, will require a model that accounts for both the thermal and non-thermal gas populations. This relationship between the spatial distributions of H$_2^{\ast}$ and CN will therefore be explored in more detail in a forthcoming paper. 

\section{Summary \& Conclusions}

We have analyzed the ultraviolet spectral properties of 19 young stars in the Lupus and Taurus-Auriga associations, using spectra from \emph{HST}-COS and \emph{HST}-STIS to directly measure fluxes from Ly$\alpha$, the FUV continuum, semi-forbidden C II] $\lambda$2325, and UV-fluorescent H$_2$. Each of these is a potential tracer of the photochemical pathways responsible for producing CN and HCN molecules in disks. To investigate the formation chemistry of these two species, we compare the UV tracers to sub-mm CN and 14 $\mu$m HCN fluxes. We find that 
\begin{enumerate}
\item HCN fluxes are positively correlated with relative fluxes from the FUV continuum and C II] $\lambda$2325, implying that disks with strong continuum fluxes are more readily able to produce the atomic N and C$^+$ reactants required in the first step of the main HCN formation pathway. 
\item By contrast, CN fluxes are negatively correlated with relative fluxes from the FUV continuum and C II] $\lambda$2325. This result indicates that while molecule formation may proceed more efficiently when more atomic N is produced, CN destruction may increase accordingly as well. 
\item Neither CN or HCN emission is significantly correlated with Ly$\alpha$ emission. However, we report very tentative positive (CN) and negative (HCN) correlations that are consistent with modeling work that predicts increased photodissociation with stronger Ly$\alpha$ irradiation. 
\end{enumerate}
We attribute the lack of correlations between CN and HCN emission and UV-H$_2$ fluxes to the spatial distributions of the three molecular species: the UV-H$_2$ is concentrated in surface layers of the inner disk, the sub-mm CN emission extends from the inner to the outer disk, and the HCN emission originates in deeper layers of the inner disk. By combining UV spectra with IR and sub-mm fluxes from UV-dependent molecular gas species, we are able to investigate model predictions of molecule formation pathways and observationally confirm that the FUV continuum plays an important role in regulating CN and HCN populations in protoplanetary disks. The analysis presented here can be extended to additional species (e.g. hydrocarbons) in the era of \emph{JWST}, which will enable higher spectral resolution observations of warm molecular gas and more accurate physical-chemical models of surface reactions in planet-forming systems.      

\section{Acknowledgements}

We are thankful to the referee for their thoughtful comments that helped strengthen the analysis presented here. NA is supported by NASA Earth and Space Science Fellowship grant 80NSSC17K0531 and HST-GO-14604 (PIs: C.F. Manara, P.C. Schneider). H.M.G. was supported by program HST-GO-15204.001, which was provided by NASA through a grant from the Space Telescope Science Institute, which is operated by the Associations of Universities for Research in Astronomy, Incorporated, under NASA contract NAS5-26 555. We are grateful to M.K. McClure and C. Walsh for helpful discussions regarding the analysis presented here. This paper makes use of the following ALMA data: ADS/JAO.ALMA\#2013.1.00220.S. ALMA is a partnership of ESO (representing its member states), NSF (USA) and NINS (Japan), together with NRC (Canada), MOST and ASIAA (Taiwan), and KASI (Republic of Korea), in cooperation with the Republic of Chile. The Joint ALMA Observatory is operated by ESO, AUI/NRAO and NAOJ. This work utilized the RMACC Summit supercomputer, which is supported by the National Science Foundation (awards ACI-1532235 and ACI-1532236), the University of Colorado Boulder, and Colorado State University. The Summit supercomputer is a joint effort of the University of Colorado Boulder and Colorado State University. This research made use of Astropy,\footnote{http://www.astropy.org} a community-developed core Python package for Astronomy \citep{astropy2013, astropy2018}.    

\appendix

\renewcommand\thefigure{\thesection.\arabic{figure}} 
\section{UV-H$_2$ Spectra and Modeling Results}

We present the nine strongest observed UV-H$_2$ emission lines for RU Lupi, MY Lupi, Sz 68, and J1608-3070, with ``best-fit'' models from the 2-D radiative transfer approach described in Section 4.2 overlaid on the data. The corresponding model radial distributions of UV-H$_2$ flux for each system, with contours representing $+/-1-\sigma$ bounds, are shown as well. 

\setcounter{figure}{0}

\begin{figure}[t!]
\centering
\includegraphics[width=0.7\linewidth]{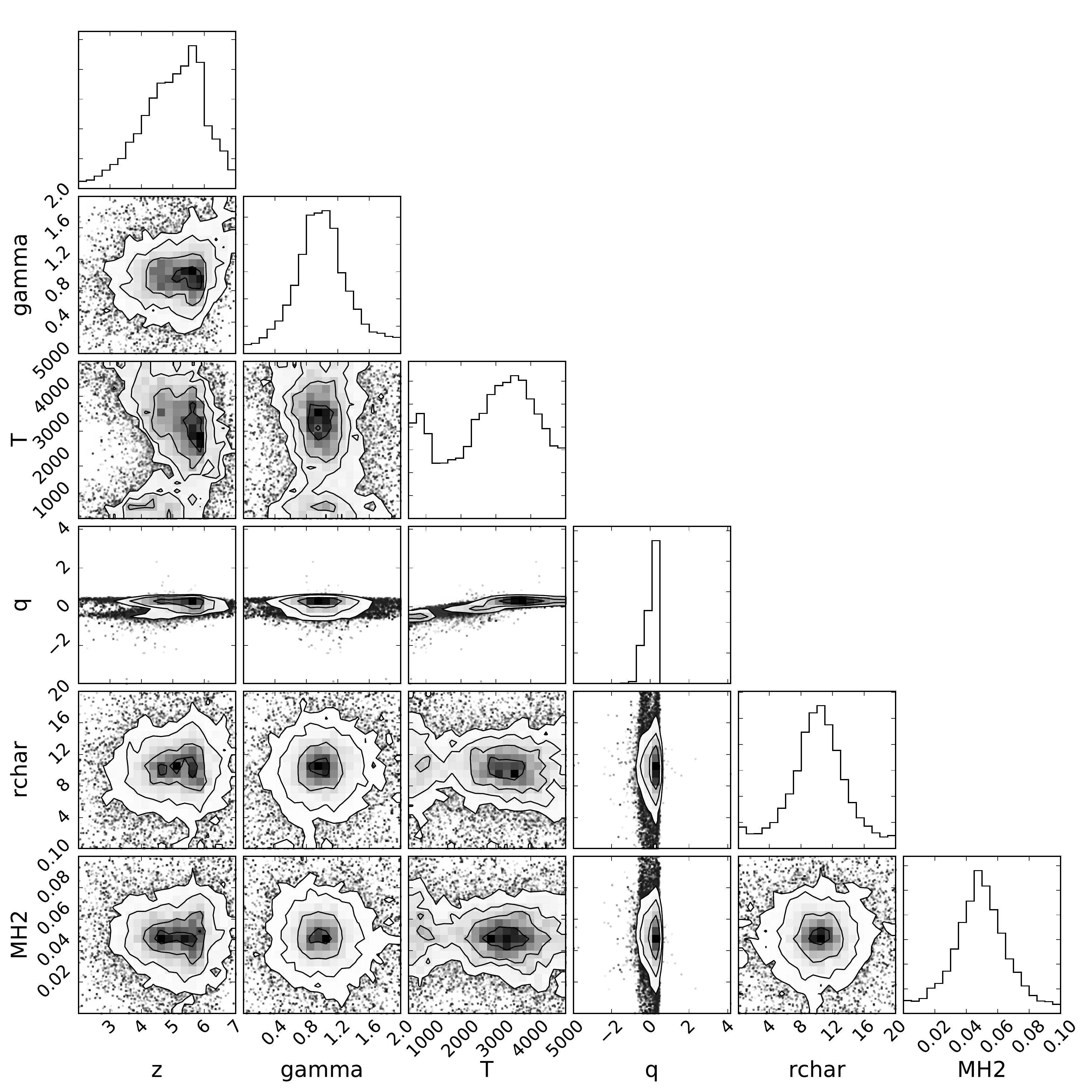}
\caption{Corner plot showing marginalized probability distributions of the variable parameters, derived from MCMC re-sampling of the 2-D radiative transfer models described in Section 4.1. This set of distributions was acquired by calculating log-likelihood values using the observed fluxes from the MY Lupi spectrum.} \label{RYLup_corner}
\end{figure}

\begin{figure}
\centering
\includegraphics[width=0.7\linewidth]{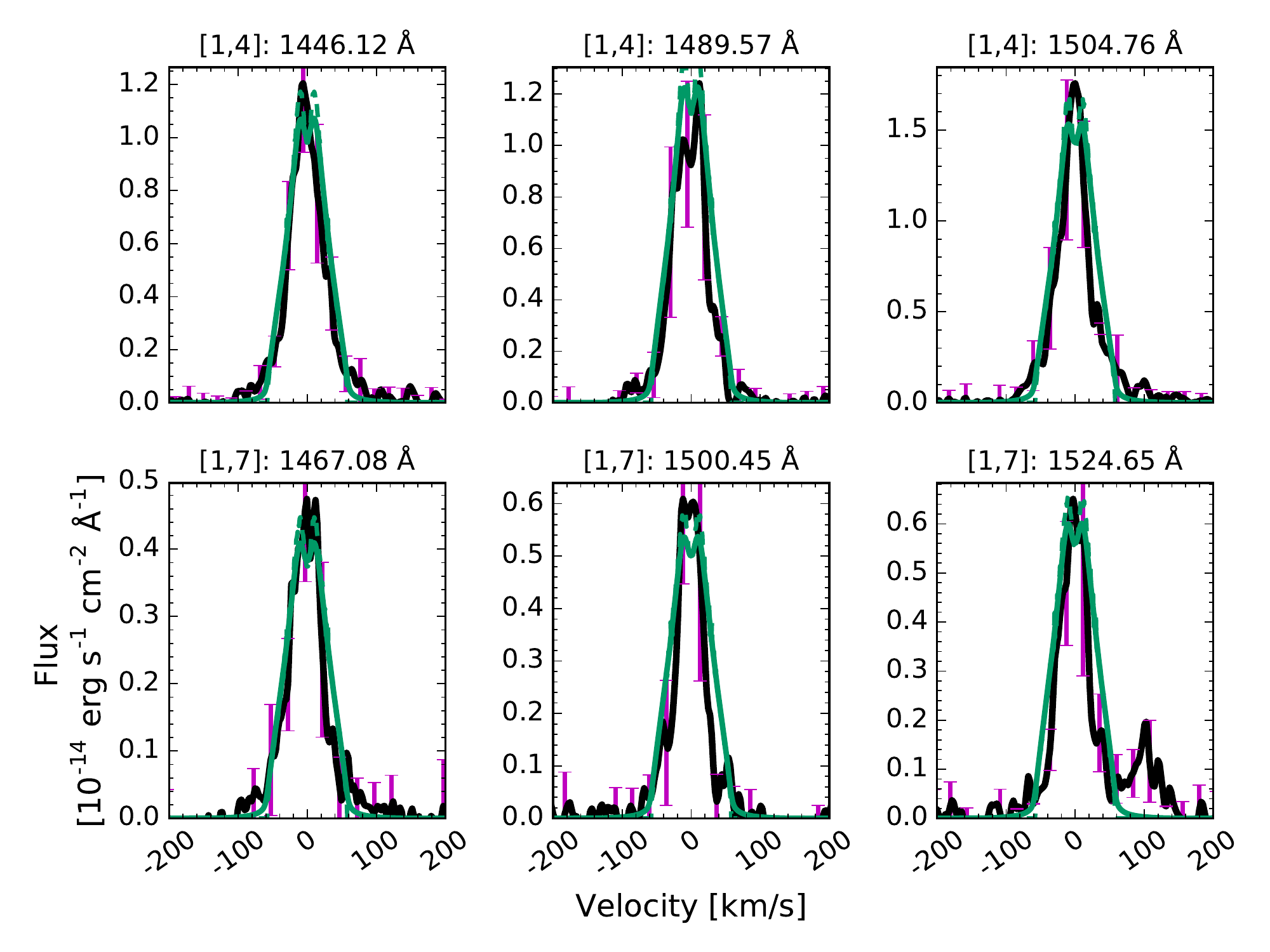}
\caption{Observed UV-H$_2$ emission lines (black) and model distributions (teal) for MY Lupi. Dashed lines show the model profile prior to convolution with the \emph{HST}-COS line spread function (LSF), while the solid lines are used to show the convolved line profile.} \label{MYLup_obsH2}
\end{figure}

\begin{figure}
\centering
\includegraphics[width=0.7\linewidth]{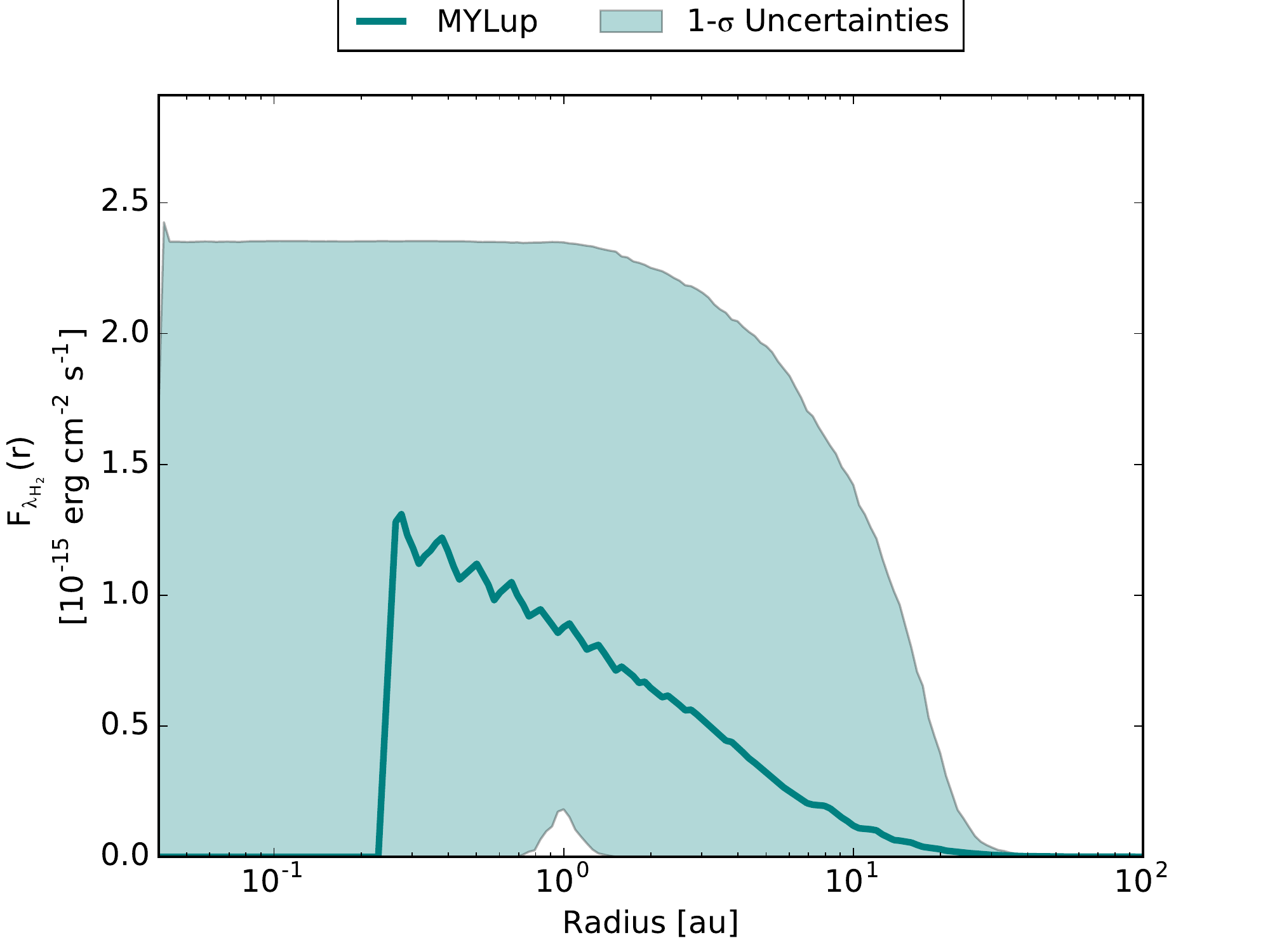}
\caption{Best-fit model radial distribution of UV-H$_2$ flux from the disk around MY Lupi, with contours marking $+/-1\sigma$ bounds on the median distribution. Uncertainties were estimated using MCMC sampling \citep{ForemanMackey2013} over the parameter space defined by \citet{Hoadley2015}.} \label{MYLup_raddistH2}
\end{figure}

\begin{figure}[t!]
\centering
\includegraphics[width=0.7\linewidth]{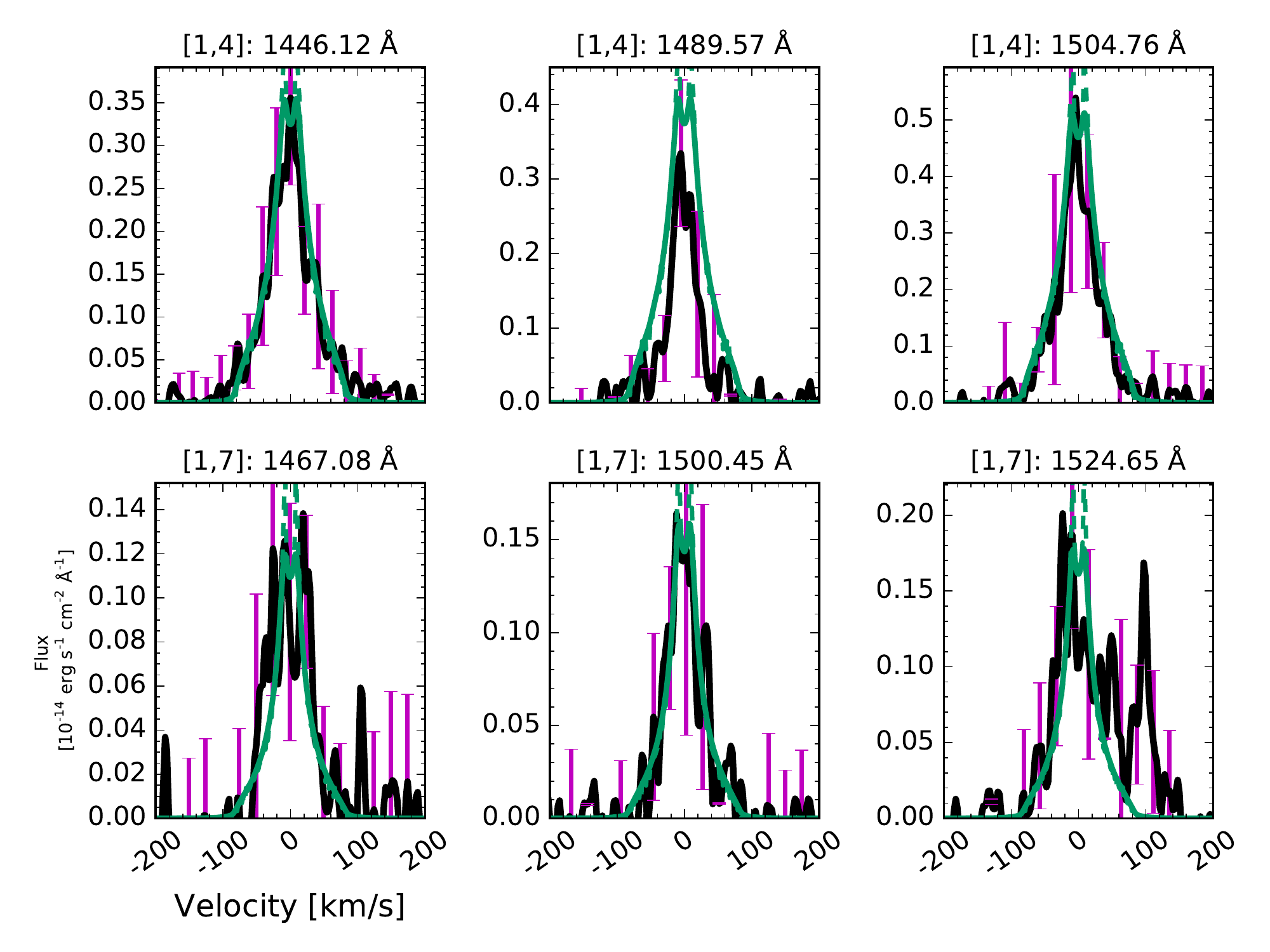}
\caption{Observed UV-H$_2$ emission lines (black) and model distributions (teal) for Sz 68}. Dashed lines show the model profile prior to convolution with the \emph{HST}-COS line spread function (LSF), while the solid lines are used to show the convolved line profile. \label{Sz68_obsH2}
\end{figure}

\begin{figure}[b!]
\centering
\includegraphics[width=0.7\linewidth]{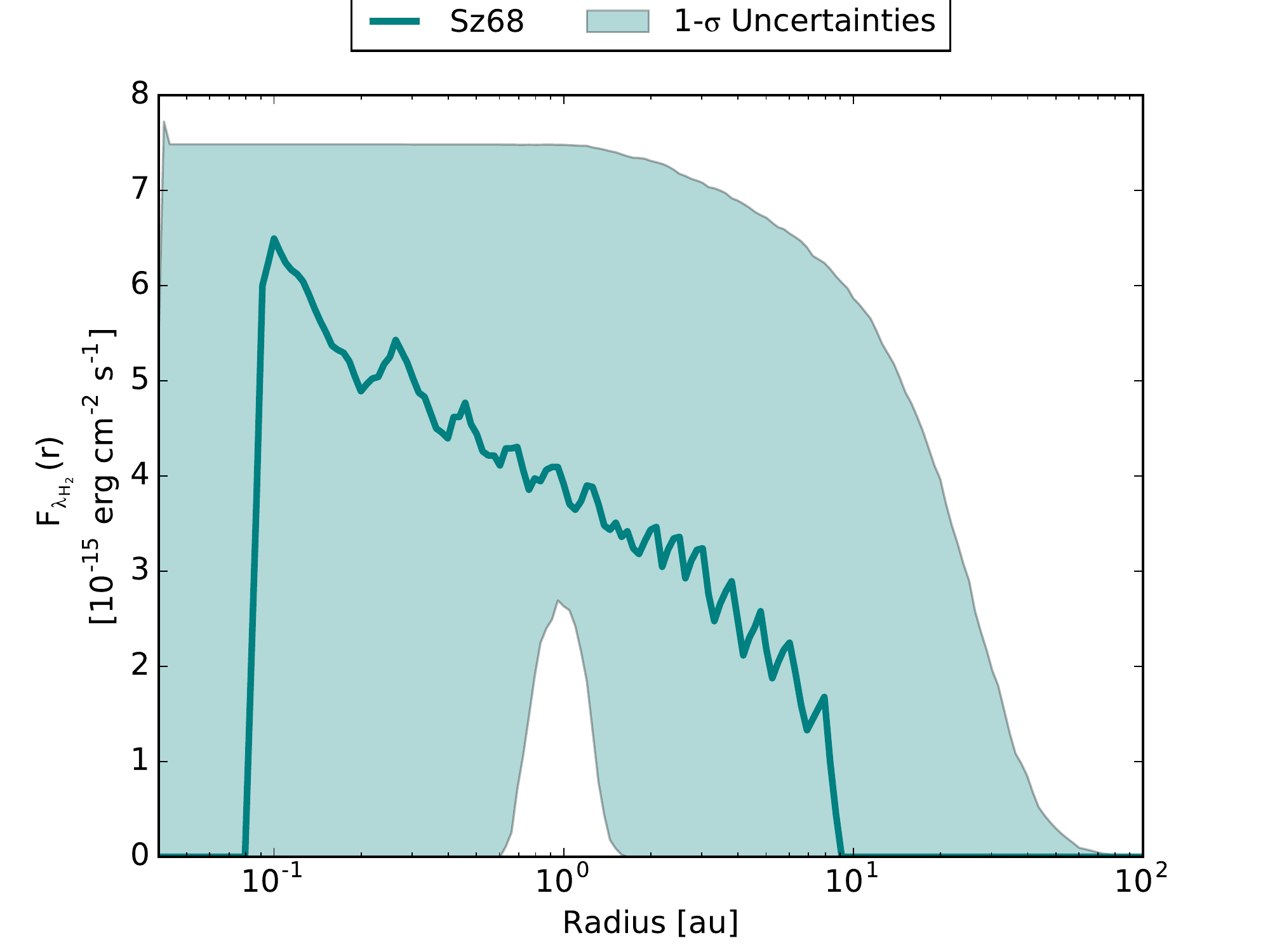}
\caption{Best-fit model radial distribution of UV-H$_2$ flux from the disk around Sz 68, with contours marking $+/-1\sigma$ bounds on the median distribution. Uncertainties were estimated using MCMC sampling \citep{ForemanMackey2013} over the parameter space defined by \citet{Hoadley2015}.} \label{Sz68_raddistH2}
\end{figure}

\begin{figure}
\centering
\includegraphics[width=0.7\linewidth]{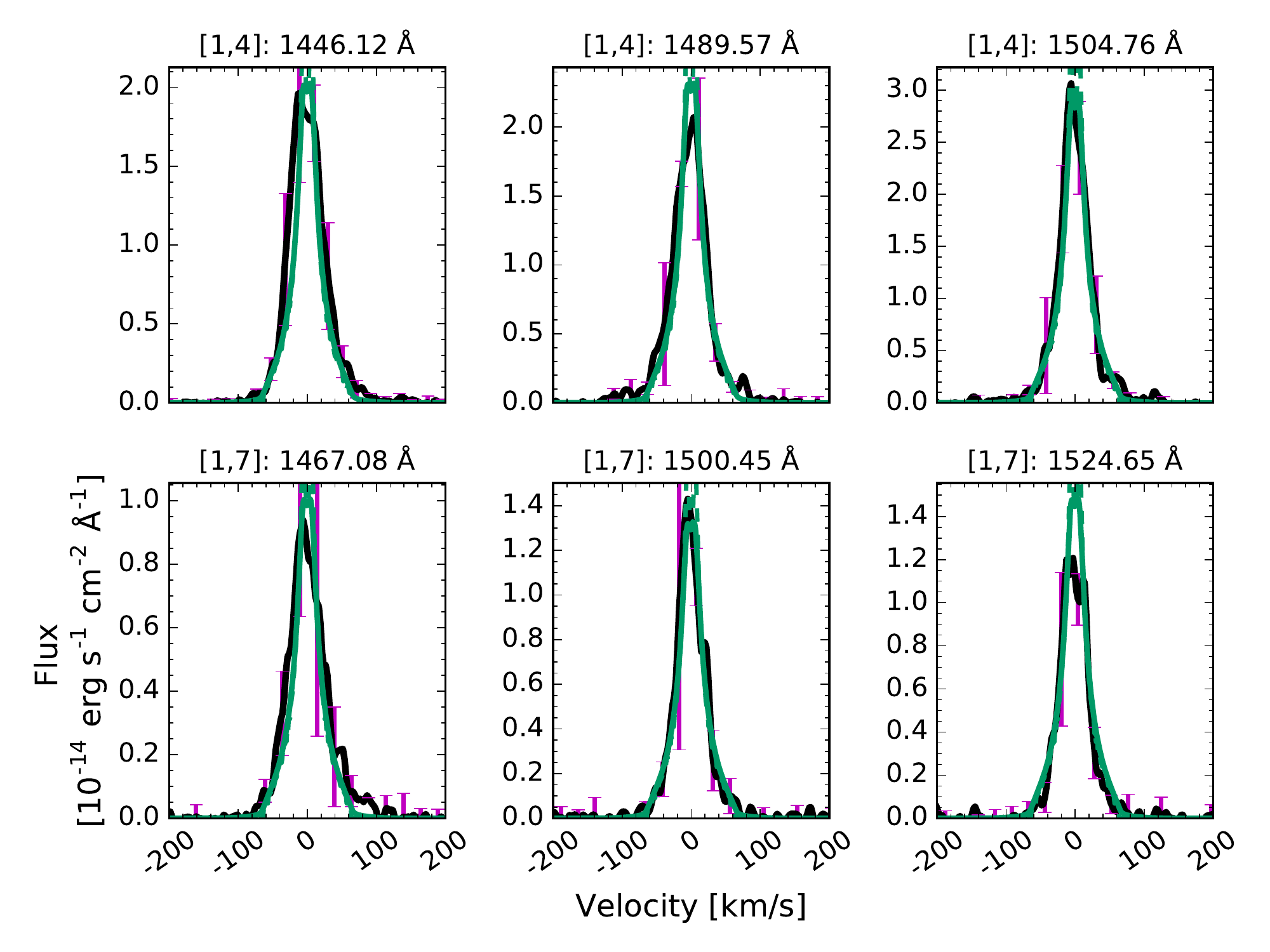}
\caption{Observed UV-H$_2$ emission lines (black) and model distributions (teal) for J1608-3070. Dashed lines show the model profile prior to convolution with the \emph{HST}-COS line spread function (LSF), while the solid lines are used to show the convolved line profile.} \label{J16083070_obsH2}
\end{figure}

\begin{figure}
\centering
\includegraphics[width=0.7\linewidth]{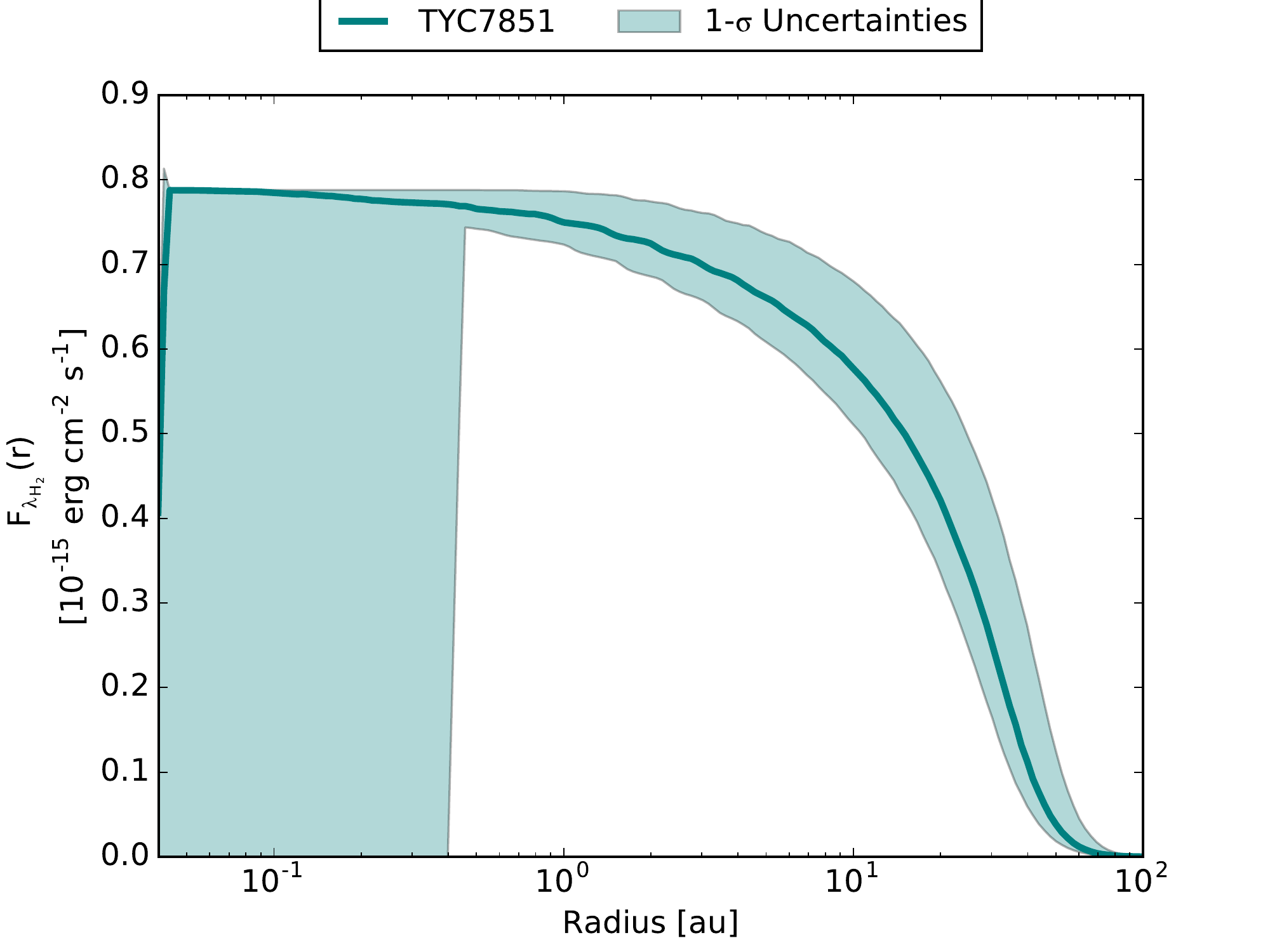}
\caption{Best-fit model radial distribution of UV-H$_2$ flux from the disk around J1608-3070, with contours marking $+/-1\sigma$ bounds on the median distribution. We note that the MCMC sampling \citep{ForemanMackey2013} used to estimate the uncertainties on the flux distribution was carried out over a tighter parameter space than for the other two targets, since the signal-to-noise in the UV-H$_2$ emission lines is lower for J1608-3070.}
\label{J16083070_raddistH2}
\end{figure}

\bibliographystyle{apj}
\bibliography{Lupus}

\begin{thebibliography}{107}
\expandafter\ifx\csname natexlab\endcsname\relax\def\natexlab#1{#1}\fi

\bibitem[{{{\'A}d{\'a}mkovics} {et~al.}(2016){{\'A}d{\'a}mkovics}, {Najita}, \&
  {Glassgold}}]{Adamkovics2016}
{{\'A}d{\'a}mkovics}, M., {Najita}, J.~R., \& {Glassgold}, A.~E. 2016, \apj,
  817, 82

\bibitem[{{Ag{\'u}ndez} {et~al.}(2008){Ag{\'u}ndez}, {Cernicharo}, \&
  {Goicoechea}}]{Agundez2008}
{Ag{\'u}ndez}, M., {Cernicharo}, J., \& {Goicoechea}, J.~R. 2008, \aap, 483,
  831

\bibitem[{{Ag{\'u}ndez} {et~al.}(2018){Ag{\'u}ndez}, {Roueff}, {Le Petit}, \&
  {Le Bourlot}}]{Agundez2018}
{Ag{\'u}ndez}, M., {Roueff}, E., {Le Petit}, F., \& {Le Bourlot}, J. 2018,
  \aap, 616, A19

\bibitem[{{Aikawa} {et~al.}(2002){Aikawa}, {van Zadelhoff}, {van Dishoeck}, \&
  {Herbst}}]{Aikawa2002}
{Aikawa}, Y., {van Zadelhoff}, G.~J., {van Dishoeck}, E.~F., \& {Herbst}, E.
  2002, \aap, 386, 622

\bibitem[{{Alcala'} {et~al.}(2019){Alcala'}, {Manara}, {France}, {Schneider},
  {Arulanantham}, {Miotello}, {Guenther}, \& {Brown}}]{Alcala2019}
{Alcala'}, J.~M., {Manara}, C.~F., {France}, K., {Schneider}, C.~P.,
  {Arulanantham}, N., {Miotello}, A., {Guenther}, H.~M., \& {Brown}, A. 2019,
  arXiv e-prints

\bibitem[{{Alcal{\'a}} {et~al.}(2017){Alcal{\'a}}, {Manara}, {Natta}, {Frasca},
  {Testi}, {Nisini}, {Stelzer}, {Williams}, {Antoniucci}, {Biazzo}, {Covino},
  {Esposito}, {Getman}, \& {Rigliaco}}]{Alcala2017}
{Alcal{\'a}}, J.~M., {et~al.} 2017, \aap, 600, A20

\bibitem[{{Alcal{\'a}} {et~al.}(2014){Alcal{\'a}}, {Natta}, {Manara}, {Spezzi},
  {Stelzer}, {Frasca}, {Biazzo}, {Covino}, {Randich}, {Rigliaco}, {Testi},
  {Comer{\'o}n}, {Cupani}, \& {D'Elia}}]{Alcala2014}
---. 2014, \aap, 561, A2

\bibitem[{{Andrews} {et~al.}(2018){Andrews}, {Huang}, {P{\'e}rez}, {Isella},
  {Dullemond}, {Kurtovic}, {Guzm{\'a}n}, {Carpenter}, {Wilner}, {Zhang}, {Zhu},
  {Birnstiel}, {Bai}, {Benisty}, {Hughes}, {{\"O}berg}, \&
  {Ricci}}]{Andrews2018}
{Andrews}, S.~M., {et~al.} 2018, \apjl, 869, L41

\bibitem[{{Andrews} {et~al.}(2013){Andrews}, {Rosenfeld}, {Kraus}, \&
  {Wilner}}]{Andrews2013}
{Andrews}, S.~M., {Rosenfeld}, K.~A., {Kraus}, A.~L., \& {Wilner}, D.~J. 2013,
  \apj, 771, 129

\bibitem[{{Ansdell} {et~al.}(2017){Ansdell}, {Williams}, {Manara}, {Miotello},
  {Facchini}, {van der Marel}, {Testi}, \& {van Dishoeck}}]{Ansdell2017}
{Ansdell}, M., {Williams}, J.~P., {Manara}, C.~F., {Miotello}, A., {Facchini},
  S., {van der Marel}, N., {Testi}, L., \& {van Dishoeck}, E.~F. 2017, \aj,
  153, 240

\bibitem[{{Ansdell} {et~al.}(2018){Ansdell}, {Williams}, {Trapman}, {van
  Terwisga}, {Facchini}, {Manara}, {van der Marel}, {Miotello}, {Tazzari},
  {Hogerheijde}, {Guidi}, {Testi}, \& {van Dishoeck}}]{Ansdell2018}
{Ansdell}, M., {et~al.} 2018, \apj, 859, 21

\bibitem[{{Ansdell} {et~al.}(2016){Ansdell}, {Williams}, {van der Marel},
  {Carpenter}, {Guidi}, {Hogerheijde}, {Mathews}, {Manara}, {Miotello},
  {Natta}, {Oliveira}, {Tazzari}, {Testi}, {van Dishoeck}, \& {van
  Terwisga}}]{Ansdell2016}
---. 2016, \apj, 828, 46

\bibitem[{{Ardila} {et~al.}(2013){Ardila}, {Herczeg}, {Gregory}, {Ingleby},
  {France}, {Brown}, {Edwards}, {Johns-Krull}, {Linsky}, {Yang}, {Valenti},
  {Abgrall}, {Alexander}, {Bergin}, {Bethell}, {Brown}, {Calvet}, {Espaillat},
  {Hillenbrand}, {Hussain}, {Roueff}, {Schindhelm}, \& {Walter}}]{Ardila2013}
{Ardila}, D.~R., {et~al.} 2013, \apjs, 207, 1

\bibitem[{{Arulanantham} {et~al.}(2018){Arulanantham}, {France}, {Hoadley},
  {Manara}, {Schneider}, {Alcal{\'a}}, {Banzatti}, {G{\"u}nther}, {Miotello},
  {van der Marel}, {van Dishoeck}, {Walsh}, \& {Williams}}]{Arulanantham2018}
{Arulanantham}, N., {et~al.} 2018, \apj, 855, 98

\bibitem[{{Astropy Collaboration} {et~al.}(2013){Astropy Collaboration},
  {Robitaille}, {Tollerud}, {Greenfield}, {Droettboom}, {Bray}, {Aldcroft},
  {Davis}, {Ginsburg}, {Price-Whelan}, {Kerzendorf}, {Conley}, {Crighton},
  {Barbary}, {Muna}, {Ferguson}, {Grollier}, {Parikh}, {Nair}, {Unther},
  {Deil}, {Woillez}, {Conseil}, {Kramer}, {Turner}, {Singer}, {Fox}, {Weaver},
  {Zabalza}, {Edwards}, {Azalee Bostroem}, {Burke}, {Casey}, {Crawford},
  {Dencheva}, {Ely}, {Jenness}, {Labrie}, {Lian Lim}, {Pierfederici},
  {Pontzen}, {Ptak}, {Refsdal}, {Servillat}, \& {Streicher}}]{astropy2013}
{Astropy Collaboration} {et~al.} 2013, \aap, 558, A33

\bibitem[{{Bailer-Jones} {et~al.}(2018){Bailer-Jones}, {Rybizki}, {Fouesneau},
  {Mantelet}, \& {Andrae}}]{Gaia}
{Bailer-Jones}, C.~A.~L., {Rybizki}, J., {Fouesneau}, M., {Mantelet}, G., \&
  {Andrae}, R. 2018, \aj, 156, 58

\bibitem[{{Banzatti} {et~al.}(2019){Banzatti}, {Pascucci}, {Edwards}, {Fang},
  {Gorti}, \& {Flock}}]{Banzatti2019}
{Banzatti}, A., {Pascucci}, I., {Edwards}, S., {Fang}, M., {Gorti}, U., \&
  {Flock}, M. 2019, \apj, 870, 76

\bibitem[{{Barenfeld} {et~al.}(2016){Barenfeld}, {Carpenter}, {Ricci}, \&
  {Isella}}]{Barenfeld2016}
{Barenfeld}, S.~A., {Carpenter}, J.~M., {Ricci}, L., \& {Isella}, A. 2016,
  \apj, 827, 142

\bibitem[{{Baulch} {et~al.}(1994){Baulch}, {Cobos}, {Cox}, {Frank}, {Hayman},
  {Just}, {Kerr}, {Murrells}, {Pilling}, {Troe}, {Walker}, \&
  {Warnatz}}]{Baulch1994}
{Baulch}, D.~L., {et~al.} 1994, Journal of Physical and Chemical Reference
  Data, 23, 847

\bibitem[{{Bergin} {et~al.}(2003){Bergin}, {Calvet}, {D'Alessio}, \&
  {Herczeg}}]{Bergin2003}
{Bergin}, E., {Calvet}, N., {D'Alessio}, P., \& {Herczeg}, G.~J. 2003, \apjl,
  591, L159

\bibitem[{{Bergin} {et~al.}(2004){Bergin}, {Calvet}, {Sitko}, {Abgrall},
  {D'Alessio}, {Herczeg}, {Roueff}, {Qi}, {Lynch}, {Russell}, {Brafford}, \&
  {Perry}}]{Bergin2004}
{Bergin}, E., {et~al.} 2004, \apjl, 614, L133

\bibitem[{{Bergner} {et~al.}(2019){Bergner}, {{\"O}berg}, {Bergin}, {Loomis},
  {Pegues}, \& {Qi}}]{Bergner2019}
{Bergner}, J.~B., {{\"O}berg}, K.~I., {Bergin}, E.~A., {Loomis}, R.~A.,
  {Pegues}, J., \& {Qi}, C. 2019, \apj, 876, 25

\bibitem[{{Bethell} \& {Bergin}(2011)}]{Bethell2011}
{Bethell}, T.~J., \& {Bergin}, E.~A. 2011, \apj, 739, 78

\bibitem[{{Black} \& {van Dishoeck}(1987)}]{Black1987}
{Black}, J.~H., \& {van Dishoeck}, E.~F. 1987, \apj, 322, 412

\bibitem[{{Bohlin} {et~al.}(1978){Bohlin}, {Savage}, \& {Drake}}]{Bohlin1978}
{Bohlin}, R.~C., {Savage}, B.~D., \& {Drake}, J.~F. 1978, \apj, 224, 132

\bibitem[{{Bottinelli} {et~al.}(2010){Bottinelli}, {Boogert}, {Bouwman},
  {Beckwith}, {van Dishoeck}, {{\"O}berg}, {Pontoppidan}, {Linnartz}, {Blake},
  {Evans}, \& {Lahuis}}]{Bottinelli2010}
{Bottinelli}, S., {et~al.} 2010, \apj, 718, 1100

\bibitem[{{Cazzoletti} {et~al.}(2019){Cazzoletti}, {Manara}, {Liu}, {van
  Dishoeck}, {Facchini}, {Alcal{\`a}}, {Ansdell}, {Testi}, {Williams},
  {Carrasco-Gonz{\'a}lez}, {Dong}, {Forbrich}, {Fukagawa}, {Galv{\'a}n-Madrid},
  {Hirano}, {Hogerheijde}, {Hasegawa}, {Muto}, {Pinilla}, {Takami}, {Tamura},
  {Tazzari}, \& {Wisniewski}}]{Cazzoletti2019}
{Cazzoletti}, P., {et~al.} 2019, arXiv e-prints

\bibitem[{{Cazzoletti} {et~al.}(2018){Cazzoletti}, {van Dishoeck}, {Visser},
  {Facchini}, \& {Bruderer}}]{Cazzoletti2018}
{Cazzoletti}, P., {van Dishoeck}, E.~F., {Visser}, R., {Facchini}, S., \&
  {Bruderer}, S. 2018, \aap, 609, A93

\bibitem[{{Chapillon} {et~al.}(2012){Chapillon}, {Guilloteau}, {Dutrey},
  {Pi{\'e}tu}, \& {Gu{\'e}lin}}]{Chapillon2012}
{Chapillon}, E., {Guilloteau}, S., {Dutrey}, A., {Pi{\'e}tu}, V., \&
  {Gu{\'e}lin}, M. 2012, \aap, 537, A60

\bibitem[{{Cleeves} {et~al.}(2016){Cleeves}, {{\"O}berg}, {Wilner}, {Huang},
  {Loomis}, {Andrews}, \& {Czekala}}]{Cleeves2016}
{Cleeves}, L.~I., {{\"O}berg}, K.~I., {Wilner}, D.~J., {Huang}, J., {Loomis},
  R.~A., {Andrews}, S.~M., \& {Czekala}, I. 2016, \apj, 832, 110

\bibitem[{{Cleeves} {et~al.}(2018){Cleeves}, {{\"O}berg}, {Wilner}, {Huang},
  {Loomis}, {Andrews}, \& {Guzman}}]{Cleeves2018}
{Cleeves}, L.~I., {{\"O}berg}, K.~I., {Wilner}, D.~J., {Huang}, J., {Loomis},
  R.~A., {Andrews}, S.~M., \& {Guzman}, V.~V. 2018, \apj, 865, 155

\bibitem[{{Dent} {et~al.}(2013){Dent}, {Thi}, {Kamp}, {Williams}, {Menard},
  {Andrews}, {Ardila}, {Aresu}, {Augereau}, {Barrado y Navascues}, {Brittain},
  {Carmona}, {Ciardi}, {Danchi}, {Donaldson}, {Duchene}, {Eiroa}, {Fedele},
  {Grady}, {de Gregorio-Molsalvo}, {Howard}, {Hu{\'e}lamo}, {Krivov},
  {Lebreton}, {Liseau}, {Martin-Zaidi}, {Mathews}, {Meeus},
  {Mendigut{\'{\i}}a}, {Montesinos}, {Morales-Calderon}, {Mora}, {Nomura},
  {Pantin}, {Pascucci}, {Phillips}, {Pinte}, {Podio}, {Ramsay}, {Riaz},
  {Riviere-Marichalar}, {Roberge}, {Sandell}, {Solano}, {Tilling}, {Torrelles},
  {Vandenbusche}, {Vicente}, {White}, \& {Woitke}}]{Dent2013}
{Dent}, W.~R.~F., {et~al.} 2013, \pasp, 125, 477

\bibitem[{{Espaillat} {et~al.}(2019){Espaillat}, {Robinson}, {Grant}, \&
  {Reynolds}}]{Espaillat2019}
{Espaillat}, C.~C., {Robinson}, C., {Grant}, S., \& {Reynolds}, M. 2019, arXiv
  e-prints

\bibitem[{{Foreman-Mackey} {et~al.}(2013){Foreman-Mackey}, {Hogg}, {Lang}, \&
  {Goodman}}]{ForemanMackey2013}
{Foreman-Mackey}, D., {Hogg}, D.~W., {Lang}, D., \& {Goodman}, J. 2013, \pasp,
  125, 306

\bibitem[{{France} {et~al.}(2017){France}, {Roueff}, \& {Abgrall}}]{France2017}
{France}, K., {Roueff}, E., \& {Abgrall}, H. 2017, \apj, 844, 169

\bibitem[{{France} {et~al.}(2014){France}, {Schindhelm}, {Bergin}, {Roueff}, \&
  {Abgrall}}]{France2014}
{France}, K., {Schindhelm}, E., {Bergin}, E.~A., {Roueff}, E., \& {Abgrall}, H.
  2014, \apj, 784, 127

\bibitem[{{France} {et~al.}(2011{\natexlab{a}}){France}, {Schindhelm}, {Burgh},
  {Herczeg}, {Harper}, {Brown}, {Green}, {Linsky}, {Yang}, {Abgrall}, {Ardila},
  {Bergin}, {Bethell}, {Brown}, {Calvet}, {Espaillat}, {Gregory},
  {Hillenbrand}, {Hussain}, {Ingleby}, {Johns-Krull}, {Roueff}, {Valenti}, \&
  {Walter}}]{France2011_FUVII}
{France}, K., {et~al.} 2011{\natexlab{a}}, \apj, 734, 31

\bibitem[{{France} {et~al.}(2012){France}, {Schindhelm}, {Herczeg}, {Brown},
  {Abgrall}, {Alexander}, {Bergin}, {Brown}, {Linsky}, {Roueff}, \&
  {Yang}}]{France2012}
---. 2012, \apj, 756, 171

\bibitem[{{France} {et~al.}(2011{\natexlab{b}}){France}, {Yang}, \&
  {Linsky}}]{France2011_FUVI}
{France}, K., {Yang}, H., \& {Linsky}, J.~L. 2011{\natexlab{b}}, \apj, 729, 7

\bibitem[{{Ghez} {et~al.}(1997){Ghez}, {McCarthy}, {Patience}, \&
  {Beck}}]{Ghez1997}
{Ghez}, A.~M., {McCarthy}, D.~W., {Patience}, J.~L., \& {Beck}, T.~L. 1997,
  \apj, 481, 378

\bibitem[{{G{\'o}mez de Castro} \& {Ferro-Font{\'a}n}(2005)}]{GdC2005}
{G{\'o}mez de Castro}, A.~I., \& {Ferro-Font{\'a}n}, C. 2005, \mnras, 362, 569

\bibitem[{{Green} {et~al.}(2012){Green}, {Froning}, {Osterman}, {Ebbets},
  {Heap}, {Leitherer}, {Linsky}, {Savage}, {Sembach}, {Shull}, {Siegmund},
  {Snow}, {Spencer}, {Stern}, {Stocke}, {Welsh}, {B{\'e}land}, {Burgh},
  {Danforth}, {France}, {Keeney}, {McPhate}, {Penton}, {Andrews},
  {Brownsberger}, {Morse}, \& {Wilkinson}}]{Green2012}
{Green}, J.~C., {et~al.} 2012, \apj, 744, 60

\bibitem[{{Greenwood} {et~al.}(2019){Greenwood}, {Kamp}, {Waters}, {Woitke}, \&
  {Thi}}]{Greenwood2019}
{Greenwood}, A.~J., {Kamp}, I., {Waters}, L.~B.~F.~M., {Woitke}, P., \& {Thi},
  W.-F. 2019, \aap, 626, A6

\bibitem[{{Guilloteau} {et~al.}(2013){Guilloteau}, {Di Folco}, {Dutrey},
  {Simon}, {Grosso}, \& {Pi{\'e}tu}}]{Guilloteau2013}
{Guilloteau}, S., {Di Folco}, E., {Dutrey}, A., {Simon}, M., {Grosso}, N., \&
  {Pi{\'e}tu}, V. 2013, \aap, 549, A92

\bibitem[{{Hartigan} \& {Kenyon}(2003)}]{Hartigan2003}
{Hartigan}, P., \& {Kenyon}, S.~J. 2003, \apj, 583, 334

\bibitem[{{Heays} {et~al.}(2017){Heays}, {Bosman}, \& {van
  Dishoeck}}]{Heays2017}
{Heays}, A.~N., {Bosman}, A.~D., \& {van Dishoeck}, E.~F. 2017, \aap, 602, A105

\bibitem[{{Henning} {et~al.}(2010){Henning}, {Semenov}, {Guilloteau}, {Dutrey},
  {Hersant}, {Wakelam}, {Chapillon}, {Launhardt}, {Pi{\'e}tu}, \&
  {Schreyer}}]{Henning2010}
{Henning}, T., {et~al.} 2010, \apj, 714, 1511

\bibitem[{{Herczeg} {et~al.}(2002){Herczeg}, {Linsky}, {Valenti},
  {Johns-Krull}, \& {Wood}}]{Herczeg2002}
{Herczeg}, G.~J., {Linsky}, J.~L., {Valenti}, J.~A., {Johns-Krull}, C.~M., \&
  {Wood}, B.~E. 2002, \apj, 572, 310

\bibitem[{{Herczeg} {et~al.}(2005){Herczeg}, {Walter}, {Linsky}, {Gahm},
  {Ardila}, {Brown}, {Johns-Krull}, {Simon}, \& {Valenti}}]{Herczeg2005}
{Herczeg}, G.~J., {et~al.} 2005, \aj, 129, 2777

\bibitem[{{Herczeg} {et~al.}(2004){Herczeg}, {Wood}, {Linsky}, {Valenti}, \&
  {Johns-Krull}}]{Herczeg2004}
{Herczeg}, G.~J., {Wood}, B.~E., {Linsky}, J.~L., {Valenti}, J.~A., \&
  {Johns-Krull}, C.~M. 2004, \apj, 607, 369

\bibitem[{{Hoadley} {et~al.}(2015){Hoadley}, {France}, {Alexander}, {McJunkin},
  \& {Schneider}}]{Hoadley2015}
{Hoadley}, K., {France}, K., {Alexander}, R.~D., {McJunkin}, M., \&
  {Schneider}, P.~C. 2015, \apj, 812, 41

\bibitem[{{Hoadley} {et~al.}(2017){Hoadley}, {France}, {Arulanantham}, {Loyd},
  \& {Kruczek}}]{Hoadley2017}
{Hoadley}, K., {France}, K., {Arulanantham}, N., {Loyd}, R.~O.~P., \&
  {Kruczek}, N. 2017, \apj, 846, 6

\bibitem[{{Houck} {et~al.}(2004){Houck}, {Roellig}, {van Cleve}, {Forrest},
  {Herter}, {Lawrence}, {Matthews}, {Reitsema}, {Soifer}, {Watson}, {Weedman},
  {Huisjen}, {Troeltzsch}, {Barry}, {Bernard-Salas}, {Blacken}, {Brandl},
  {Charmandaris}, {Devost}, {Gull}, {Hall}, {Henderson}, {Higdon}, {Pirger},
  {Schoenwald}, {Sloan}, {Uchida}, {Appleton}, {Armus}, {Burgdorf},
  {Fajardo-Acosta}, {Grillmair}, {Ingalls}, {Morris}, \& {Teplitz}}]{Houck2004}
{Houck}, J.~R., {et~al.} 2004, \apjs, 154, 18

\bibitem[{{Huang} {et~al.}(2018){Huang}, {Andrews}, {Dullemond}, {Isella},
  {P{\'e}rez}, {Guzm{\'a}n}, {{\"O}berg}, {Zhu}, {Zhang}, {Bai}, {Benisty},
  {Birnstiel}, {Carpenter}, {Hughes}, {Ricci}, {Weaver}, \&
  {Wilner}}]{Huang2018_annular}
{Huang}, J., {et~al.} 2018, \apjl, 869, L42

\bibitem[{{Ingleby} {et~al.}(2013){Ingleby}, {Calvet}, {Herczeg}, {Blaty},
  {Walter}, {Ardila}, {Alexander}, {Edwards}, {Espaillat}, {Gregory},
  {Hillenbrand}, \& {Brown}}]{Ingleby2013}
{Ingleby}, L., {et~al.} 2013, \apj, 767, 112

\bibitem[{{Johns-Krull} \& {Herczeg}(2007)}]{JKH2007}
{Johns-Krull}, C.~M., \& {Herczeg}, G.~J. 2007, \apj, 655, 345

\bibitem[{{Kenyon} \& {Hartmann}(1995)}]{Kenyon1995}
{Kenyon}, S.~J., \& {Hartmann}, L. 1995, \apjs, 101, 117

\bibitem[{{Kessler-Silacci} {et~al.}(2006){Kessler-Silacci}, {Augereau},
  {Dullemond}, {Geers}, {Lahuis}, {Evans}, {van Dishoeck}, {Blake}, {Boogert},
  {Brown}, {J{\o}rgensen}, {Knez}, \& {Pontoppidan}}]{KS2006}
{Kessler-Silacci}, J., {et~al.} 2006, \apj, 639, 275

\bibitem[{{Kurtovic} {et~al.}(2018){Kurtovic}, {P{\'e}rez}, {Benisty}, {Zhu},
  {Zhang}, {Huang}, {Andrews}, {Dullemond}, {Isella}, {Bai}, {Carpenter},
  {Guzm{\'a}n}, {Ricci}, \& {Wilner}}]{Kurtovic2018}
{Kurtovic}, N.~T., {et~al.} 2018, \apjl, 869, L44

\bibitem[{{Lebouteiller} {et~al.}(2015){Lebouteiller}, {Barry}, {Goes},
  {Sloan}, {Spoon}, {Weedman}, {Bernard-Salas}, \& {Houck}}]{Lebouteiller2015}
{Lebouteiller}, V., {Barry}, D.~J., {Goes}, C., {Sloan}, G.~C., {Spoon},
  H.~W.~W., {Weedman}, D.~W., {Bernard-Salas}, J., \& {Houck}, J.~R. 2015, The
  Astrophysical Journal Supplement Series, 218, 21

\bibitem[{{Lebouteiller} {et~al.}(2011){Lebouteiller}, {Barry}, {Spoon},
  {Bernard-Salas}, {Sloan}, {Houck}, \& {Weedman}}]{Lebouteiller2011}
{Lebouteiller}, V., {Barry}, D.~J., {Spoon}, H.~W.~W., {Bernard-Salas}, J.,
  {Sloan}, G.~C., {Houck}, J.~R., \& {Weedman}, D.~W. 2011, \apjs, 196, 8

\bibitem[{{Li} {et~al.}(2013){Li}, {Heays}, {Visser}, {Ubachs}, {Lewis},
  {Gibson}, \& {van Dishoeck}}]{Li2013}
{Li}, X., {Heays}, A.~N., {Visser}, R., {Ubachs}, W., {Lewis}, B.~R., {Gibson},
  S.~T., \& {van Dishoeck}, E.~F. 2013, \aap, 555, A14

\bibitem[{{Long} {et~al.}(2019){Long}, {Herczeg}, {Harsono}, {Pinilla},
  {Tazzari}, {Manara}, {Pascucci}, {Cabrit}, {Nisini}, {Johnstone}, {Edwards},
  {Salyk}, {Menard}, {Lodato}, {Boehler}, {Mace}, {Liu}, {Mulders}, {Hendler},
  {Ragusa}, {Fischer}, {Banzatti}, {Rigliaco}, {van de Plas}, {Dipierro},
  {Gully-Santiago}, \& {Lopez-Valdivia}}]{Long2019_Taurus}
{Long}, F., {et~al.} 2019, \apj, 882, 49

\bibitem[{{Long} {et~al.}(2018){Long}, {Herczeg}, {Pascucci}, {Apai},
  {Henning}, {Manara}, {Mulders}, {Sz{\H u}cs}, \&
  {Hendler}}]{Long2018_Chamaeleon}
---. 2018, \apj, 863, 61

\bibitem[{{Loomis} {et~al.}(2017){Loomis}, {{\"O}berg}, {Andrews}, \&
  {MacGregor}}]{Loomis2017}
{Loomis}, R.~A., {{\"O}berg}, K.~I., {Andrews}, S.~M., \& {MacGregor}, M.~A.
  2017, \apj, 840, 23

\bibitem[{{Loyd} {et~al.}(2018){Loyd}, {France}, {Youngblood}, {Schneider},
  {Brown}, {Hu}, {Segura}, {Linsky}, {Redfield}, {Tian}, {Rugheimer}, {Miguel},
  \& {Froning}}]{Loyd2018}
{Loyd}, R.~O.~P., {et~al.} 2018, \apj, 867, 71

\bibitem[{{Madhusudhan} {et~al.}(2011){Madhusudhan}, {Mousis}, {Johnson}, \&
  {Lunine}}]{Madhusudhan2011}
{Madhusudhan}, N., {Mousis}, O., {Johnson}, T.~V., \& {Lunine}, J.~I. 2011,
  \apj, 743, 191

\bibitem[{{Manset} {et~al.}(2009){Manset}, {Bastien}, {M{\'e}nard}, {Bertout},
  {Le van Suu}, \& {Boivin}}]{Manset2009}
{Manset}, N., {Bastien}, P., {M{\'e}nard}, F., {Bertout}, C., {Le van Suu}, A.,
  \& {Boivin}, L. 2009, \aap, 499, 137

\bibitem[{{McClure}(2019)}]{McClure2019}
{McClure}, M.~K. 2019, arXiv e-prints

\bibitem[{{McElroy} {et~al.}(2013){McElroy}, {Walsh}, {Markwick}, {Cordiner},
  {Smith}, \& {Millar}}]{UMIST2013}
{McElroy}, D., {Walsh}, C., {Markwick}, A.~J., {Cordiner}, M.~A., {Smith}, K.,
  \& {Millar}, T.~J. 2013, \aap, 550, A36

\bibitem[{{McJunkin} {et~al.}(2014){McJunkin}, {France}, {Schneider},
  {Herczeg}, {Brown}, {Hillenbrand}, {Schindhelm}, \& {Edwards}}]{McJunkin2014}
{McJunkin}, M., {France}, K., {Schneider}, P.~C., {Herczeg}, G.~J., {Brown},
  A., {Hillenbrand}, L., {Schindhelm}, E., \& {Edwards}, S. 2014, \apj, 780,
  150

\bibitem[{{Miotello} {et~al.}(2019){Miotello}, {Facchini}, {van Dishoeck},
  {Cazzoletti}, {Testi}, {Williams}, {Ansdell}, {van Terwisga}, \& {van der
  Marel}}]{Miotello2019}
{Miotello}, A., {et~al.} 2019, arXiv e-prints

\bibitem[{{Miotello} {et~al.}(2016){Miotello}, {van Dishoeck}, {Kama}, \&
  {Bruderer}}]{Miotello2016}
{Miotello}, A., {van Dishoeck}, E.~F., {Kama}, M., \& {Bruderer}, S. 2016,
  \aap, 594, A85

\bibitem[{{Miotello} {et~al.}(2017){Miotello}, {van Dishoeck}, {Williams},
  {Ansdell}, {Guidi}, {Hogerheijde}, {Manara}, {Tazzari}, {Testi}, {van der
  Marel}, \& {van Terwisga}}]{Miotello2017}
{Miotello}, A., {et~al.} 2017, \aap, 599, A113

\bibitem[{{Najita} {et~al.}(2013){Najita}, {Carr}, {Pontoppidan}, {Salyk}, {van
  Dishoeck}, \& {Blake}}]{Najita2013}
{Najita}, J.~R., {Carr}, J.~S., {Pontoppidan}, K.~M., {Salyk}, C., {van
  Dishoeck}, E.~F., \& {Blake}, G.~A. 2013, \apj, 766, 134

\bibitem[{{Nuth} \& {Glicker}(1982)}]{Nuth1982}
{Nuth}, J.~A., \& {Glicker}, S. 1982, \jqsrt, 28, 223

\bibitem[{{{\"O}berg} {et~al.}(2008){{\"O}berg}, {Boogert}, {Pontoppidan},
  {Blake}, {Evans}, {Lahuis}, \& {van Dishoeck}}]{Oberg2008}
{{\"O}berg}, K.~I., {Boogert}, A.~C.~A., {Pontoppidan}, K.~M., {Blake}, G.~A.,
  {Evans}, N.~J., {Lahuis}, F., \& {van Dishoeck}, E.~F. 2008, \apj, 678, 1032

\bibitem[{{{\"O}berg} {et~al.}(2010){{\"O}berg}, {Qi}, {Fogel}, {Bergin},
  {Andrews}, {Espaillat}, {van Kempen}, {Wilner}, \& {Pascucci}}]{Oberg2010}
{{\"O}berg}, K.~I., {et~al.} 2010, \apj, 720, 480

\bibitem[{{{\"O}berg} {et~al.}(2011){{\"O}berg}, {Qi}, {Fogel}, {Bergin},
  {Andrews}, {Espaillat}, {Wilner}, {Pascucci}, \& {Kastner}}]{Oberg2011}
---. 2011, \apj, 734, 98

\bibitem[{{Pascucci} {et~al.}(2009){Pascucci}, {Apai}, {Luhman}, {Henning},
  {Bouwman}, {Meyer}, {Lahuis}, \& {Natta}}]{Pascucci2009}
{Pascucci}, I., {Apai}, D., {Luhman}, K., {Henning}, T., {Bouwman}, J.,
  {Meyer}, M.~R., {Lahuis}, F., \& {Natta}, A. 2009, \apj, 696, 143

\bibitem[{{Pascucci} {et~al.}(2013){Pascucci}, {Herczeg}, {Carr}, \&
  {Bruderer}}]{Pascucci2013}
{Pascucci}, I., {Herczeg}, G., {Carr}, J.~S., \& {Bruderer}, S. 2013, \apj,
  779, 178

\bibitem[{{Pascucci} {et~al.}(2016){Pascucci}, {Testi}, {Herczeg}, {Long},
  {Manara}, {Hendler}, {Mulders}, {Krijt}, {Ciesla}, {Henning}, {Mohanty},
  {Drabek-Maunder}, {Apai}, {Sz{\H u}cs}, {Sacco}, \&
  {Olofsson}}]{Pascucci2016}
{Pascucci}, I., {et~al.} 2016, \apj, 831, 125

\bibitem[{{Pontoppidan} {et~al.}(2019){Pontoppidan}, {Salyk}, {Banzatti},
  {Blake}, {Walsh}, {Lacy}, \& {Richter}}]{Pontoppidan2019}
{Pontoppidan}, K.~M., {Salyk}, C., {Banzatti}, A., {Blake}, G.~A., {Walsh}, C.,
  {Lacy}, J.~H., \& {Richter}, M.~J. 2019, \apj, 874, 92

\bibitem[{{Pontoppidan} {et~al.}(2010){Pontoppidan}, {Salyk}, {Blake},
  {Meijerink}, {Carr}, \& {Najita}}]{Pontoppidan2010}
{Pontoppidan}, K.~M., {Salyk}, C., {Blake}, G.~A., {Meijerink}, R., {Carr},
  J.~S., \& {Najita}, J. 2010, \apj, 720, 887

\bibitem[{{Salyk} {et~al.}(2011{\natexlab{a}}){Salyk}, {Blake}, {Boogert}, \&
  {Brown}}]{Salyk2011CO}
{Salyk}, C., {Blake}, G.~A., {Boogert}, A.~C.~A., \& {Brown}, J.~M.
  2011{\natexlab{a}}, \apj, 743, 112

\bibitem[{{Salyk} {et~al.}(2011{\natexlab{b}}){Salyk}, {Pontoppidan}, {Blake},
  {Najita}, \& {Carr}}]{Salyk2011Spitzer}
{Salyk}, C., {Pontoppidan}, K.~M., {Blake}, G.~A., {Najita}, J.~R., \& {Carr},
  J.~S. 2011{\natexlab{b}}, \apj, 731, 130

\bibitem[{{Schindhelm} {et~al.}(2012{\natexlab{a}}){Schindhelm}, {France},
  {Burgh}, {Herczeg}, {Green}, {Brown}, {Brown}, \&
  {Valenti}}]{Schindhelm2012_CO}
{Schindhelm}, E., {France}, K., {Burgh}, E.~B., {Herczeg}, G.~J., {Green},
  J.~C., {Brown}, A., {Brown}, J.~M., \& {Valenti}, J.~A. 2012{\natexlab{a}},
  \apj, 746, 97

\bibitem[{{Schindhelm} {et~al.}(2012{\natexlab{b}}){Schindhelm}, {France},
  {Herczeg}, {Bergin}, {Yang}, {Brown}, {Brown}, {Linsky}, \&
  {Valenti}}]{Schindhelm2012}
{Schindhelm}, E., {et~al.} 2012{\natexlab{b}}, \apjl, 756, L23

\bibitem[{{Schneider} {et~al.}(2015){Schneider}, {France}, {G{\"u}nther},
  {Herczeg}, {Robrade}, {Bouvier}, {McJunkin}, \& {Schmitt}}]{Schneider2015}
{Schneider}, P.~C., {France}, K., {G{\"u}nther}, H.~M., {Herczeg}, G.,
  {Robrade}, J., {Bouvier}, J., {McJunkin}, M., \& {Schmitt}, J.~H.~M.~M. 2015,
  \aap, 584, A51

\bibitem[{{Schwarz} {et~al.}(2018){Schwarz}, {Bergin}, {Cleeves}, {Zhang},
  {{\"O}berg}, {Blake}, \& {Anderson}}]{Schwarz2018}
{Schwarz}, K.~R., {Bergin}, E.~A., {Cleeves}, L.~I., {Zhang}, K., {{\"O}berg},
  K.~I., {Blake}, G.~A., \& {Anderson}, D. 2018, \apj, 856, 85

\bibitem[{{Simon} {et~al.}(2016){Simon}, {Pascucci}, {Edwards}, {Feng},
  {Gorti}, {Hollenbach}, {Rigliaco}, \& {Keane}}]{Simon2016}
{Simon}, M.~N., {Pascucci}, I., {Edwards}, S., {Feng}, W., {Gorti}, U.,
  {Hollenbach}, D., {Rigliaco}, E., \& {Keane}, J.~T. 2016, \apj, 831, 169

\bibitem[{{Skrutskie} {et~al.}(1990){Skrutskie}, {Dutkevitch}, {Strom},
  {Edwards}, {Strom}, \& {Shure}}]{Skrutskie1990}
{Skrutskie}, M.~F., {Dutkevitch}, D., {Strom}, S.~E., {Edwards}, S., {Strom},
  K.~M., \& {Shure}, M.~A. 1990, \aj, 99, 1187

\bibitem[{{Strom} {et~al.}(1989){Strom}, {Strom}, {Edwards}, {Cabrit}, \&
  {Skrutskie}}]{Strom1989}
{Strom}, K.~M., {Strom}, S.~E., {Edwards}, S., {Cabrit}, S., \& {Skrutskie},
  M.~F. 1989, \aj, 97, 1451

\bibitem[{{Tazzari} {et~al.}(2017){Tazzari}, {Testi}, {Natta}, {Ansdell},
  {Carpenter}, {Guidi}, {Hogerheijde}, {Manara}, {Miotello}, {van der Marel},
  {van Dishoeck}, \& {Williams}}]{Tazzari2017}
{Tazzari}, M., {et~al.} 2017, \aap, 606, A88

\bibitem[{{Teske} {et~al.}(2011){Teske}, {Najita}, {Carr}, {Pascucci}, {Apai},
  \& {Henning}}]{Teske2011}
{Teske}, J.~K., {Najita}, J.~R., {Carr}, J.~S., {Pascucci}, I., {Apai}, D., \&
  {Henning}, T. 2011, \apj, 734, 27

\bibitem[{{The Astropy Collaboration} {et~al.}(2018){The Astropy
  Collaboration}, {Price-Whelan}, {Sip{\H o}cz}, {G{\"u}nther}, {Lim},
  {Crawford}, {Conseil}, {Shupe}, {Craig}, {Dencheva}, {Ginsburg},
  {VanderPlas}, {Bradley}, {P{\'e}rez-Su{\'a}rez}, {de Val-Borro}, {Aldcroft},
  {Cruz}, {Robitaille}, {Tollerud}, {Ardelean}, {Babej}, {Bachetti}, {Bakanov},
  {Bamford}, {Barentsen}, {Barmby}, {Baumbach}, {Berry}, {Biscani}, {Boquien},
  {Bostroem}, {Bouma}, {Brammer}, {Bray}, {Breytenbach}, {Buddelmeijer},
  {Burke}, {Calderone}, {Cano Rodr{\'{\i}}guez}, {Cara}, {Cardoso},
  {Cheedella}, {Copin}, {Crichton}, {D{\'A}vella}, {Deil}, {Depagne},
  {Dietrich}, {Donath}, {Droettboom}, {Earl}, {Erben}, {Fabbro}, {Ferreira},
  {Finethy}, {Fox}, {Garrison}, {Gibbons}, {Goldstein}, {Gommers}, {Greco},
  {Greenfield}, {Groener}, {Grollier}, {Hagen}, {Hirst}, {Homeier}, {Horton},
  {Hosseinzadeh}, {Hu}, {Hunkeler}, {Ivezi{\'c}}, {Jain}, {Jenness}, {Kanarek},
  {Kendrew}, {Kern}, {Kerzendorf}, {Khvalko}, {King}, {Kirkby}, {Kulkarni},
  {Kumar}, {Lee}, {Lenz}, {Littlefair}, {Ma}, {Macleod}, {Mastropietro},
  {McCully}, {Montagnac}, {Morris}, {Mueller}, {Mumford}, {Muna}, {Murphy},
  {Nelson}, {Nguyen}, {Ninan}, {N{\"o}the}, {Ogaz}, {Oh}, {Parejko}, {Parley},
  {Pascual}, {Patil}, {Patil}, {Plunkett}, {Prochaska}, {Rastogi}, {Reddy
  Janga}, {Sabater}, {Sakurikar}, {Seifert}, {Sherbert}, {Sherwood-Taylor},
  {Shih}, {Sick}, {Silbiger}, {Singanamalla}, {Singer}, {Sladen}, {Sooley},
  {Sornarajah}, {Streicher}, {Teuben}, {Thomas}, {Tremblay}, {Turner},
  {Terr{\'o}n}, {van Kerkwijk}, {de la Vega}, {Watkins}, {Weaver}, {Whitmore},
  {Woillez}, \& {Zabalza}}]{astropy2018}
{The Astropy Collaboration} {et~al.} 2018, ArXiv e-prints

\bibitem[{{van der Marel} {et~al.}(2018){van der Marel}, {Williams}, {Ansdell},
  {Manara}, {Miotello}, {Tazzari}, {Testi}, {Hogerheijde}, {Bruderer}, {van
  Terwisga}, \& {van Dishoeck}}]{vanderMarel2018}
{van der Marel}, N., {et~al.} 2018, \apj, 854, 177

\bibitem[{{van Dishoeck} {et~al.}(2006){van Dishoeck}, {Jonkheid}, \& {van
  Hemert}}]{vanDishoeck2006}
{van Dishoeck}, E.~F., {Jonkheid}, B., \& {van Hemert}, M.~C. 2006, Faraday
  Discussions, 133, 231

\bibitem[{{van Terwisga} {et~al.}(2019){van Terwisga}, {van Dishoeck},
  {Cazzoletti}, {Facchini}, {Trapman}, {Williams}, {Manara}, {Miotello}, {van
  der Marel}, {Ansdell}, {Hogerheijde}, {Tazzari}, \&
  {Testi}}]{vanTerwisga2019}
{van Terwisga}, S.~E., {et~al.} 2019, \aap, 623, A150

\bibitem[{{van Zadelhoff} {et~al.}(2003){van Zadelhoff}, {Aikawa},
  {Hogerheijde}, \& {van Dishoeck}}]{vanZadelhoff2003}
{van Zadelhoff}, G.~J., {Aikawa}, Y., {Hogerheijde}, M.~R., \& {van Dishoeck},
  E.~F. 2003, in SFChem 2002: Chemistry as a Diagnostic of Star Formation, ed.
  C.~L. {Curry} \& M.~{Fich}, 440

\bibitem[{{Visser} {et~al.}(2018){Visser}, {Bruderer}, {Cazzoletti},
  {Facchini}, {Heays}, \& {van Dishoeck}}]{Visser2018}
{Visser}, R., {Bruderer}, S., {Cazzoletti}, P., {Facchini}, S., {Heays}, A.~N.,
  \& {van Dishoeck}, E.~F. 2018, \aap, 615, A75

\bibitem[{{Walsh} {et~al.}(2012){Walsh}, {Nomura}, {Millar}, \&
  {Aikawa}}]{Walsh2012}
{Walsh}, C., {Nomura}, H., {Millar}, T.~J., \& {Aikawa}, Y. 2012, \apj, 747,
  114

\bibitem[{{Walsh} {et~al.}(2015){Walsh}, {Nomura}, \& {van
  Dishoeck}}]{Walsh2015}
{Walsh}, C., {Nomura}, H., \& {van Dishoeck}, E. 2015, \aap, 582, A88

\bibitem[{{Williams} {et~al.}(2019){Williams}, {Cieza}, {Hales}, {Ansdell},
  {Ruiz-Rodriguez}, {Casassus}, {Perez}, \& {Zurlo}}]{Williams2019}
{Williams}, J.~P., {Cieza}, L., {Hales}, A., {Ansdell}, M., {Ruiz-Rodriguez},
  D., {Casassus}, S., {Perez}, S., \& {Zurlo}, A. 2019, \apjl, 875, L9

\bibitem[{{Woodgate} {et~al.}(1997{\natexlab{a}}){Woodgate}, {Kimble},
  {Bowers}, {Kraemer}, {Kaiser}, {Gull}, {Danks}, {Grady}, {Loiacono},
  {Brumfield}, {Feinberg}, {Hood}, {Meyer}, {Vanhouten}, {Argabright}, {Bybee},
  {Timothy}, {Blouke}, {Dorn}, {Bottema}, {Woodruff}, {Michika}, {Sullivan},
  {Hetlinger}, {Stocker}, {Ludtke}, {Ebbets}, {Delamere}, {Rose}, {Gardner},
  {Breyer}, {Lindler}, {Content}, {Standley}, {Hartig}, {Heap}, {Joseph},
  {Green}, {Jenkins}, {Linsky}, {Hutchings}, {Moos}, {Boggess}, {Maran},
  {Roesler}, \& {Weistrop}}]{Woodgate1997_overview}
{Woodgate}, B., {et~al.} 1997{\natexlab{a}}, in Bulletin of the American
  Astronomical Society, Vol.~29, American Astronomical Society Meeting
  Abstracts \#190, 836

\bibitem[{{Woodgate} {et~al.}(1997{\natexlab{b}}){Woodgate}, {Kimble},
  {Bowers}, {Kraemer}, {Kaiser}, {Danks}, {Grady}, {Loiacono}, {Hood}, {Meyer},
  {van Houten}, {Argabright}, {Bybee}, {Timothy}, {Blouke}, {Dorn}, {Bottema},
  {Woodruff}, {Michika}, {Sullivan}, {Hetlinger}, {Stocker}, {Brumfield},
  {Feinberg}, {Delamere}, {Rose}, {Garner}, {Lindler}, {Gull}, {Heap},
  {Joseph}, {Green}, {Jenkins}, {Linsky}, {Hutchings}, {Moos}, {Boggess},
  {Maran}, {Roesler}, {Weistrop}, {Sonneborn}, {Bower}, \&
  {Gardner}}]{Woodgate1997_firstresults}
{Woodgate}, B.~E., {et~al.} 1997{\natexlab{b}}, in \procspie, Vol. 3118,
  Imaging Spectrometry III, ed. M.~R. {Descour} \& S.~S. {Shen}, 2--12

\bibitem[{{Yang} {et~al.}(2012){Yang}, {Herczeg}, {Linsky}, {Brown},
  {Johns-Krull}, {Ingleby}, {Calvet}, {Bergin}, \& {Valenti}}]{Yang2012}
{Yang}, H., {et~al.} 2012, \apj, 744, 121

\end{thebibliography}

\end{document}